\documentclass[review,3p]{elsarticle}

\usepackage{graphicx}
\usepackage{dcolumn}
\usepackage{bm}
\usepackage{color}
\usepackage{epstopdf}
\usepackage{ulem}

\usepackage{amsmath}
\usepackage{amssymb}

\usepackage[T1]{fontenc}
\usepackage[latin9]{inputenc}
\usepackage{amsmath,amssymb,amsthm,amsfonts}
\usepackage{slashed}
\usepackage{graphicx}
\usepackage{color,rotating}
\usepackage{dsfont}
\usepackage{setspace}
\usepackage{verbatim}
\usepackage{fancyhdr}

\usepackage{ulem}

\newcommand{\nn}{\nonumber\\}

\def\di{\displaystyle}

\def\bg{\begin{eqnarray}\begin{array}{rcl}\displaystyle}
\def\eg{\end{array} &\di    &\di   \end{eqnarray}}
\def\bm#1{\begin{eqnarray}\begin{array}{#1}\di}
\def\bmo#1{\begin{eqnarray*}\begin{array}{#1}\di}
\def\bml#1#2{\begin{eqnarray}\begin{array}{#1}\label{#2}\di}
\def\bgo{\begin{eqnarray*}\begin{array}{rcl}\displaystyle}
\def\ego{\end{array} &\di    &\di \nonumber  \end{eqnarray*}}
\def\btensor#1#2{\renew\left#1\begin{array}{#2}\di}
\def\brtensor#1#2#3{\ren#3\left#1\begin{array}{#2}}
\def\botensor#1#2{\renew\left#1\begin{array}{#2}}
\def\etensor#1{\end{array}\right#1}

\def\eq#1{(\ref{#1})}

\def\s0#1#2{\mbox{\small{$ \frac{#1}{#2} $}}}
\def\0#1#2{\frac{#1}{#2}}

\newcommand{\rmd}{{\rm d}}
\newcommand{\rme}{{\rm e}}

\def\ren#1{\renewcommand{\arraystretch}{#1}}

\def\renew{\renewcommand{\arraystretch}{1}}

\def\simle{\mathrel{\rlap{\raise 0.511ex \hbox{$<$}}{\lower 0.511ex \hbox{$\sim$}}}}
\def\simge{\mathrel{ \rlap{\raise 0.511ex \hbox{$>$}}{\lower 0.511ex \hbox{$\sim$}}}}

\newcommand \beq{\begin{eqnarray}}
\newcommand \eeq{\end{eqnarray}}
\def\simle{\mathrel{\rlap{\raise 0.511ex \hbox{$<$}}{\lower 0.511ex 
\hbox{$\sim$}}}}
\def\simge{\mathrel{ \rlap{\raise 0.511ex 
\hbox{$>$}}{\lower 0.511ex \hbox{$\sim$}}}}
\newcommand{\del}{\partial}

\newcommand{\kp}{\kappa}
\newcommand{\pk}{\partial_\kappa}

\journal{Annals of Physics}
\begin{document}

\begin{frontmatter}

\title{Functional renormalization group and 2PI effective action formalism}

\author[1]{Jean-Paul Blaizot}
\ead{jean-paul.blaizot@ipht.fr}

\author[2]{Jan M. Pawlowski}
\ead{pawlowski@thphys.uni-heidelberg.de}

\author[3]{Urko Reinosa}
\ead{reinosa@cpht.polytechnique.fr}

\address[1]{Institut de Physique Th\'eorique, Universit\'e Paris-Saclay, CEA, CNRS, 91191 Gif-sur-Yvette Cedex, France}
\address[2]{Institut f\"ur Theoretische Physik, University of Heidelberg, Philosophenweg 16, 69120 Heidelberg, Germany}
\address[3]{Centre de Physique Th\'eorique (CPHT), CNRS, Ecole Polytechnique,\\ Institut Polytechnique de Paris,  Route de Saclay, F-91128 Palaiseau, France.}

\date{\today}

\begin{abstract}
We combine two non-perturbative approaches, one based on the two-particle-irreducible (2PI) action, the other on the functional renormalization group (fRG), in an effort to develop new non-perturbative approximations for the field theoretical description of strongly coupled systems. In particular, we exploit the exact 2PI relations between the two-point and four-point functions in order to truncate the infinite hierarchy of equations of the functional renormalization group. The truncation is ``exact'' in two ways. First, the solution of the resulting flow equation is independent of the choice of the regulator. Second, this solution coincides with that of the 2PI equations for the two-point and the four-point functions, for any selection of two-skeleton diagrams characterizing a so-called $\Phi$-derivable approximation. The transformation of the equations of the 2PI formalism into flow equations offers new ways to solve these equations in practice, and provides new insight on certain aspects of their renormalization. It also opens the possibility to develop approximation schemes going  beyond the strict $\Phi$-derivable ones, as well as new truncation schemes for the fRG hierarchy.
  \end{abstract}
\begin{keyword}
key words \sep key words \sep key words
 PACS codes here, in the form: 
\end{keyword}

\end{frontmatter}


\section{Introduction}

There is an obvious need to develop non-perturbative methods in quantum field theory in order to deal with systems, both in-equilibrium and out-of-equilibrium, that are strongly coupled. A natural starting point for discussing such methods is the generating functional of connected Green's functions $W[J]$, or equivalently, the one-particle-irreducible (1PI) effective action $\Gamma[\phi]$, a functional of the field expectation value. This functional can be obtained as the Legendre transform of $W[J]$ with respect to a source $J$.  Alternatively, it can be derived from more general functionals that involve Legendre transforms with respect to additional sources. Such functionals are commonly referred to as $n$-particle-irreducible effective actions ($n$PI in short) and depend not only on the field expectation value, but also on the propagator, and possibly on a number of  dressed $n$-point vertices. This paper  focuses on the 2PI effective action, denoted $\Gamma[\phi,G]$, which depends both on the field expectation value $\phi$ and on the propagator $G$. It coincides with the 1PI effective action $\Gamma[\phi]$ once $G$ is fixed to its stationary value determined by the condition $\delta\Gamma[\phi,G]/\delta G=0$. When $\phi$ also assumes its stationary value, the 1PI and 2PI functionals coincide with the free energy of the system.

Our goal in this paper is to explore the connections between techniques based on the 2PI formalism, and the functional renormalization group (fRG). 
The fRG is usually formulated in terms of the 1PI effective action, \cite{Wetterich:1992yh, Ellwanger:1993mw, Morris:1993qb}. It has led to many applications in different areas (for reviews see
\cite{Berges:2000ew, Bagnuls:2000ae, Pawlowski:2005xe, Gies:2006wv, Delamotte:2007pf, Rosten:2010vm, Kopietz:2010zz, Braun:2011pp, 2012RvMP...84..299M, Reuter:2019byg, Dupuis:2020fhh}.  One of its major advantages is to allow for the formulation of approximations at the level of the effective action itself, a prominent example being the local potential approximation \cite{Wetterich:1992yh}. A drawback of the method is, however, that the flow for the effective action yields naturally an infinite hierarchy of equations for the $n$-point functions, analog to the Dyson-Schwinger hierarchy,  whose solution
requires in practice some truncation.  Such truncations are usually justified on the basis of physical considerations, or the ease of their implementation, which may introduce uncontrollable elements. One example of such uncontrollable elements relates to the presence of a regulator that controls the flow of the various $n$-point functions. Within a given truncation, final values of these $n$-point functions may depend on the choice of this regulator, which introduces uncertainties that are not easy to control a priori.\footnote{One possibility is to try to select appropriate (optimal) regulators that minimize the dependence upon small variations, see e.g.\ \cite{Litim:2000ci, Pawlowski:2015mlf, Balog:2019rrg}.}

The 2PI formalism has been developed initially in the context of the non-relativistic many-body problem 
\cite{Luttinger:1960ua,DeDominicis:1964a}, and formulated in \cite{Cornwall:1974vz} in terms of a 2PI effective action more suited to applications to relativistic field theories. An essential element of the 2PI effective action is a specific functional of the single-particle propagator, $\Phi[G]$, often referred to as the Luttinger-Ward (LW) functional, which is the sum of all two-particle irreducible skeleton diagrams evaluated with the full propagator $G$. The LW functional is also the generating functional for the self-energy, and other 2PI $n$-point functions that are obtained as functional derivatives of $\Phi[G]$ with respect to $G$. Selecting a specific class of skeleton diagrams in $\Phi[G]$ yields so-called $\Phi$-derivable approximations which have special symmetry conserving properties \cite{Baym:1961zz,Baym:1962sx} (see however the discussion in \cite{vanHees:2002bv}, and references therein). In the context of relativistic field theories, the 2PI formalism has been applied to the study of systems both in-equilibrium (see e.g. \cite{Blaizot:1999ip}) and out-of-equilibrium (see e.g. \cite{Tsutsui:2017uzd} and references therein), including the calculation of transport coefficients (see e.g. \cite{Berges:2000ur,Aarts:2004sd,Carrington:2009kh}). One may also mention a recent application in the more formal context of the SYK model \cite{Benedetti:2018goh}. There has also been much effort to extend their application to gauge theories (see e.g. 
\cite{Arrizabalaga:2002hn,Carrington:2003ut,Berges:2004pu,Reinosa:2009tc}). The 2PI formalism leads to non-linear equations for the self-consistent propagator that are often difficult to solve, and may have unphysical solutions that can annihilate with physical ones in some particular cases (see e.g. \cite{Kozik:2014,Marko:2015gpa}). In the context of relativistic quantum field theories, a further difficulty concerns their renormalizability  \cite{vanHees:2001ik,Blaizot:2003an,Berges:2005hc}.  

The purpose of this paper, as already mentioned, is to foster the connections between the fRG and the 2PI formalism, and get better insight into their non-perturbative properties. We shall do so by considering a simple scalar $\varphi^4$ theory in four dimensions. We shall in particular 
analyze further the truncation of the fRG flow equations that has been proposed in Ref.~\cite{Blaizot:2010zx}, and 
that exploits the relation between the four-point function and the
two-point function in the 2PI formalism. 
The truncation that we use is based on an exact relation. It leads on the one hand to an exact reformulation of the flow equations for the two-point and four-point functions, and on the other hand to an exact reformulation of the 2PI equations in terms of flow equations. Approximations are formulated in terms of the selection of skeletons in the LW functional, in the spirit of $\Phi$-derivable approximations. Thus, aside from providing a possible truncation of the usual flow equations of the fRG, conversely, the fRG provides an alternative way to solve the 2PI equations. It also greatly clarifies some issues concerning the renormalization of $\Phi$-derivable approximations. Ref.~\cite{Blaizot:2010zx} was limited to the case of a super-renomalizable theory ($\varphi^4$ theory in three dimensions). Here we consider the more general case of renormalizable theories on the example of $\varphi^4 $ scalar theory in four dimensions.\footnote{The discussion could also be extended to other systems, in particular fermionic ones where truncations at the level of four-point functions are often employed.} This allows us to analyze more fully the specificities of the renormalization of $\Phi$-derivable approximations from the point of view of the exact renormalization group.

We emphasize  that our use of 2PI relations is distinct from that developed in \cite{Dupuis:2005ij, Dupuis:2013vda} or \cite{Wetterich:2002ky,Rentrop_2015}.  In these works, one writes flow equations for the 2PI effective action, which in turn translates into a hierarchy of equations for the 2PI $n$-point functions. What we use in this paper is rather a hybrid scheme, where we close the 1PI hierarchy of flow equations by using the 2PI relation between the two-point and four-point functions \cite{Blaizot:2010zx}. While the latter relation can  be  seen as part of the 2PI hierarchy, we do not focus on the 2PI $n$-point functions, as done for instance in \cite{Carrington:2017lry} (see also \cite{Carrington:2012ea,Carrington:2014lba,Carrington:2019fw0p}, and similarly \cite{Alexander:2019cgw}), but rather on different objects, identified as loop truncations of the four-point function. These objects emerge naturally in the flow formulation of $\Phi$-derivable approximations and turn out to play an important role in  their renormalization, whether this is achieved through the flow equations or via the more standard diagrammatic approach. As we shall see, the renormalization of $\Phi$-derivable approximations via the flow equations brings both insight and flexibility. This flexibility can be exploited to extend standard $\Phi$-derivable approximations beyond their standard diagrammatic formulations. 

The outline of the paper is as follows. In Sect.~\ref{sect:general}, we provide a short summary of both the functional renormalization group based on the 1PI effective action,  and the 2PI formalism. We show how the central equations of the 2PI formalism, the gap equation that defines the two-point function self-consistently, as well as the equation that relates this two-point function to the free-energy, can be transformed into flow equations: one that relates the self-energy to the four-point function,  and another that relates the free-energy to the two-point function. These equations are identical to the flow equations for the two- and zero-point functions that can be deduced from the fRG. In Sect.~\ref{Phitruncations}, we show how the relation between the two-point function and the four-point function of the 2PI formalism allows us to close the infinite fRG hierarchy of equations for the 1PI $n$-point functions at the level of the four-point function. At this point we have obtained a possible calculation scheme where a given $\Phi$-derivable approximation is exactly reformulated in terms of flow equations. To this point, all the relevant equations depend on an ultraviolet cutoff. The following sections will be concerned with the elimination of this cutoff dependence through renormalization. In Sect.~\ref{sec:flowI} we develop further the analysis of the flow equation for the four-point functions, in preparation for the renormalization proper, which is addressed in the following two sections. In particular, we emphasize the role of auxiliary four-point functions truncated at a given loop order corresponding to the  loop order of the $\Phi$-derivable approximation considered. In Sect.~\ref{sec:42}, we show that the equations derived in Sect.~\ref{sec:flowI} are finite and can be made independent of the bare parameters. One can then extend the strategy  of the standard 1PI flow equations to renormalize the 2PI approximation,  without having to introduce the counterterms of the diagrammatic approach. In Sect.~\ref{sec:2PIren}, we use these same flow equations to make contact with the diagrammatic approach. The flow equations  clarify how the divergences are distributed among the various $n$-point functions involved in a given  $\Phi$-derivable approximation, and  provide clear prescriptions for the explicit determination of the counterterms. Overall, the flow equations bring new insights on the whole renormalization procedure of the diagrammatic approach. In Sect.~7, we take a general view on what has been achieved in previous sections and emphasize features that could potentially be generalized to other approaches, beyond the $\Phi$-derivable framework. We also propose yet a third, more practical flow reformulation of $\Phi$-derivable approximations that combines the benefits of the flow equations derived in Secs.~\ref{Phitruncations} and \ref{sec:flowI}. In the last section we discuss possible extensions of $\Phi$-derivable approximations, as well as new possible truncations of the fRG.  \ref{sec:twoloop} illustrates with the simple two-loop example many of the concepts developed in the main text. The remaining appendices gather technical material that complements the developments of the main text. In particular,  we show in \ref{app:great}  how, for certain generic collections of diagrams, it is possible to hide any reference to the quartic coupling (of the considered $\varphi^4$ theory) using the exact four-point function. This property is crucial to the strategy followed in the present work and it may have applications elsewhere.

\section{A connection between the 2PI formalism and the functional renormalization group}\label{sect:general}

We consider in this paper a  scalar field theory  with Euclidean action 
\beq\label{eactON} 
S[\varphi] =\int \rmd^{d}x\,\left\lbrace{ \frac{1}{2}}\left(\del\varphi(x)\right)^2  + \frac{m_{\rm b}^2}{2} \, \varphi^2(x) + \frac{\lambda_{\rm b}}{4!}\,\varphi^4(x) \right\rbrace,
\eeq     
in $d=4$ dimensions. When discussing systems at finite temperature, the integration over the (imaginary) time is restricted to the interval $\smash{[0,\beta=1/T]}$.  For clarity, we put a subscript b on the mass and the coupling constant, and refer to $m_{\rm b}$ and $\lambda_{\rm b}$ as to bare parameters, although that terminology takes its full (and standard) meaning only when we discuss renormalization issues, which will come later.\footnote{We keep the notation simple for the field $\varphi$, which here is also to be understood as the bare field. We shall switch to the renormalized field when discussing renormalization in later sections (see e.g. Sect.~\ref{renormdiag}).}   

Our goal in this section is  to relate two non-perturbative formulations of the same cut off theory, leaving aside the issues of ultraviolet divergences and renormalization  which will be addressed in later  sections. Thus, all the momentum integrals that we shall introduce will be assumed to be evaluated with an ultraviolet  regulator characterized by a cutoff scale $\Lambda_{\rm uv}$. For instance, this can be a sharp cutoff, in which case all the loop integrals are limited to momenta smaller than $\Lambda_{\rm uv}$. Except in a few cases where explicit calculations are done, this ultraviolet  regulator will be left implicit.  

\subsection{The 1PI effective action and the functional renormalization group}\label{frgbasics}

 The one-particle-irreducible (1PI) effective action $\Gamma[\phi]$ is obtained from the action $S[\varphi]$ through a Legendre transform, with $\phi$ denoting the expectation value of the field in the presence of an external source:  we   introduce  a source $J(x)$ coupled  to the field $ \varphi(x)$, and write the  generating functional of connected Green's functions  as\footnote{Throughout this paper we use a shorthand notation for the integrations over spatial and momentum integrals, viz.
 $$
 \int_x \equiv \int {\rmd }^d x\,,\qquad \int_q\equiv \int\frac{\rmd^d q}{(2\pi)^d}\,.
 $$
 }
\beq\label{eq:WJ}
W[J]=\ln\int {\cal D}\varphi \,\rme^{-S[\varphi]+\int_x J(x)\varphi(x)},\qquad 
\frac{\delta W[J]}{\delta J(x)}=\langle\varphi(x)\rangle\equiv \phi(x)\,.
\eeq
The Legendre transform of $W[J]$ yields $\Gamma[\phi]$
\beq\label{1PIeffaction}
\Gamma[\phi]+W[J]=\int_x J(x) \phi(x)\,,\qquad \frac{\delta \Gamma[\phi]}{\delta\phi(x)}=J(x)\,.
\eeq
In thermal equilibrium, the effective action $\Gamma[\phi]$ is the free energy for a given expectation value $\phi$ of the field  (to within a factor $1/\beta$). 

The 1PI $n$-point functions are obtained from $\Gamma[\phi]$ by functional differentiation with respect to $\phi(x)$. More precisely, we define
\beq\label{eq:1PI_vertex}
\Gamma^{(n)}(x_1,\dots,x_n;\phi)\equiv \frac{\delta^n\Gamma}{\delta\phi(x_1) \cdots
\delta\phi(x_n)}\,,
\eeq
where we leave explicit the functional dependence on the background field $\phi$. By construction, the $n$-point functions are invariant under any permutation of their arguments, a property known as crossing symmetry which we will discuss further in later sections. It is usually convenient to work with the Fourier transform of these functions. With an obvious abuse of notation we set
 \beq\label{eq:FT}
\Gamma^{(n)}(p_1,\dots,p_n;\phi)\equiv\int_{x_1}\cdots\int_{x_{n}} \,e^{i\sum_{j=1}^n
p_jx_j}\,\Gamma^{(n)}(x_1,\dots,x_n;\phi)\,.
\eeq
Our choice of a common sign for all the exponential factors corresponds to a convention of all incoming or all outgoing momenta. For the diagrammatic representation to be used in this work, we choose incoming momenta. It is easily seen that\footnote{We absorb a factor of $(2\pi)^d$ in the definition of the functional derivative $\delta/\delta\varphi(p)$ in momentum space. We shall absorb a similar factor in the definition of the $\delta$-function in momentum space.}
\beq
\Gamma^{(n)}(p_1,\dots,p_n;\phi)=\frac{\delta^n\Gamma}{\delta\phi(-p_1) \cdots
\delta\phi(-p_n)}\,,
\eeq
which makes the crossing symmetry explicit in terms of the momentum variables. 

In the case of a constant  background field $\smash{\phi(x)=\phi}$, the $n$-point functions $\Gamma^{(n)}(x_1,\dots,x_n;\phi)$ are invariant under translations of the coordinates, and it is convenient to factor out of the definition of their Fourier transform $\Gamma^{(n)}(p_1,\dots,p_n;\phi)$ the $\delta$-function that expresses the conservation of the total momentum:
 \beq\label{gamman} 
 \Gamma^{(n)}(p_1,\dots,p_n;\phi)\to 
\delta^{(d)}\left(p_1+\cdots
+p_n\right)\Gamma^{(n)}(p_1,\dots,p_n;\phi)\,.
\eeq
We shall refer to the function $\Gamma^{(n)}(p_1,\dots,p_n)$ in the right-hand side of (\ref{gamman}) as to the ``reduced'' Fourier transform. Note that we shall use the same notation for both the reduced and the full Fourier transforms, unless confusion may arise, in which case we shall specify which one is used. It should be stressed that this reduced Fourier transform is  actually a function of $n-1$ independent variables. The reason for using a redundant notation  is that it makes it simpler to track down the crossing symmetry of $\Gamma^{(n)}(p_1,\dots,p_n)$.\footnote{One could decide to denote the reduced Fourier transform as a function of $n-1$ variables among $p_1,\dots p_n$, for instance as $\Gamma^{(n)}(p_1,\dots,p_{n-1})$. In this case, however, crossing symmetry, which just amounts to permutation invariance of $\Gamma^{(n)}(p_1,\dots,p_n)$, translates into permutation invariance of $\Gamma^{(n)}(p_1,\dots,p_{n-1})$ plus invariance under the substitution of any momentum $p_i$ among $p_1,\dots,p_{n-1}$ by $-\sum_{j=1}^{n-1} p_j$.}  Finally, we mention that a similar factorization of the momentum conserving delta-function applies in a sense to the effective action itself. Indeed, for a constant background configuration, the effective action is proportional to the space time volume $\smash{\int_x 1 = \delta^{(d)}(0)}$ (recall our convention for $\delta$-functions in momentum space) 
\beq\label{effpot}
\Gamma[\phi]\to \delta^{(d)}(0)\,\Gamma(\phi)\,,
\eeq 
 where  $\Gamma(\phi)$ in the right-hand side defines the so-called effective potential.\\

The 1PI effective action (\ref{1PIeffaction}) plays a central role in the  functional renormalization group. Let us recall that in one of its popular implementations, the fRG consists in adding to the original action
$S[\varphi]$ a non-local regulator term $\Delta S_\kappa[\varphi]$ of the form \cite{Dupuis:2020fhh} 
\beq \label{eq:deformation}
\Delta S_\kappa[\varphi]=\frac{1}{2}\int_x\int_y R_\kappa(x-y)\varphi(y)\varphi(x)= \frac{1}{2}\int_q\:
R_\kappa(q)\varphi(q)\varphi(-q)\,, 
\eeq where the parameter $\kappa$
runs continuously from a high momentum scale $\Lambda$ (to be
specified below) down to 0 where the original theory is recovered. The regulator $\Delta S_\kappa$ modifies the free propagator. Its role is that of a
mass term that suppresses the fluctuations with momenta lower than
$\kappa$, while leaving unaffected those with momenta greater than
$\kappa$. This is usually achieved with a smooth cutoff function
$R_\kappa(q)$ such that, for small momenta $q$,  $R_\kappa(q\ll \kappa)\simeq \kappa^2$,
while at large momenta,  $R_\kappa(q)$ goes sufficiently rapidly to 0 as $q\simge \kappa$ so that
$\partial_\kappa R_\kappa(q)$ can play the role of an ultraviolet
cutoff in the flow equations. A convenient choice for the regulator, to which we shall occasionally refer to, is one which substitutes  in the propagator the scale $\kappa$ for the momentum $q$ when $q\le \kappa$, and vanishes for $q>\kappa$, that is, $R_\kappa(q)=(\kappa^2-q^2)\theta(\kappa-q)$, \cite{Litim:2000ci}. We shall also use at some places a sharp cutoff that completely eliminates  all contributions  to loop integrals coming from momenta below the scale $\kappa$: in this case the loop momenta run strictly from $\kappa$ to $\Lambda_{\rm uv}$.

We may consider the addition  of the regulator term as a (non local) continous ``deformation'' of the original $\varphi^4$ theory. The generating functional of 1PI $n$-point functions in this deformed theory, $\Gamma_\kappa[\phi]$,\footnote{The later is defined as in Eqs.~(\ref{eq:WJ})-(\ref{1PIeffaction}) with $S$ replaced by $S+\Delta S_\kp$ and with the additional convenient subtraction of $\frac{1}{2}\int_x\int_y R_\kappa(x-y)\phi(y)\phi(x)$.}  obeys the following flow equation \cite{Wetterich:1992yh, Ellwanger:1993mw, Morris:1993qb}
 \beq \label{NPRGeq0}
\partial_\kappa  \Gamma_\kappa[\phi]=\frac{1}{2} \int_x \int_y
\,\partial_\kappa R_\kappa(x-y)\,
G_\kappa(y,x;\phi)\,,
\eeq
where $G_\kappa(x,y;\phi)$ denotes the full propagator in the deformed theory, related to the two-point function $\Gamma^{(2)}_\kappa(z,y;\phi)$ by
\beq\label{inv}
\int_z G_\kappa(x,z;\phi)\,\left[\Gamma^{(2)}_\kappa(z,y;\phi)+R_\kappa(z-y)\right]=\delta^{(d)}(x-y)\,.
\eeq
In Fourier space, this becomes
 \beq\label{NPRGeq}
\partial_\kappa  \Gamma_\kappa[\phi]=\frac{1}{2} \int_q
\,\partial_\kappa R_\kappa(q)\,
G_\kappa(q,-q;\phi)\,,
\eeq
with
\beq\label{eq:Ginv}
\int_r G_\kappa(p,-r;\phi)\,\left[\Gamma^{(2)}_\kappa(r,-q;\phi)+\delta^{(d)}(r-q)R_\kappa(r)\right]=\delta^{(d)}(p-q)\,,
\eeq
and where $G_\kappa(p,q)$ and $\Gamma^{(2)}_\kappa(p,q)$ denote the full Fourier transforms (see above). In the case of a constant background, Eq.~(\ref{NPRGeq}) retains its form if we replace $\Gamma_\kappa[\phi]$ by the effective potential (\ref{effpot}) and $G_\kappa(q,-q)$ by the reduced propagator (\ref{gamman}). This is because the same volume factor $\delta^{(d)}(0)=\int_x 1$ can be factored out on both sides of the equation. In contrast,  Eq.~(\ref{eq:Ginv}) becomes 
\beq\label{Ginverse}
G_\kappa^{-1}(q,-q;\phi)=\Gamma_\kappa^{(2)}(q,-q;\phi)+R_\kappa(q)\,,
\eeq
with $G_\kappa(q,-q;\phi)$ and $\Gamma_\kappa^{(2)}(q,-q;\phi)$ the reduced Fourier transforms. In what follows, and unlike our choice of notation for the higher reduced $n$-point functions, we denote $G_\kappa(q,-q;\phi)$ and $\Gamma_\kappa^{(2)}(q,-q;\phi)$ simply as $G_\kappa(q;\phi)$ and $\Gamma_\kappa^{(2)}(q;\phi)$, the crossing symmetry implying in this case that $G_\kappa(q;\phi)=G_\kappa(-q;\phi)$ and $\Gamma_\kappa^{(2)}(-q;\phi)=\Gamma_\kappa^{(2)}(q;\phi)$.

\begin{figure}
\begin{center}
\includegraphics[angle=0,scale=.20]{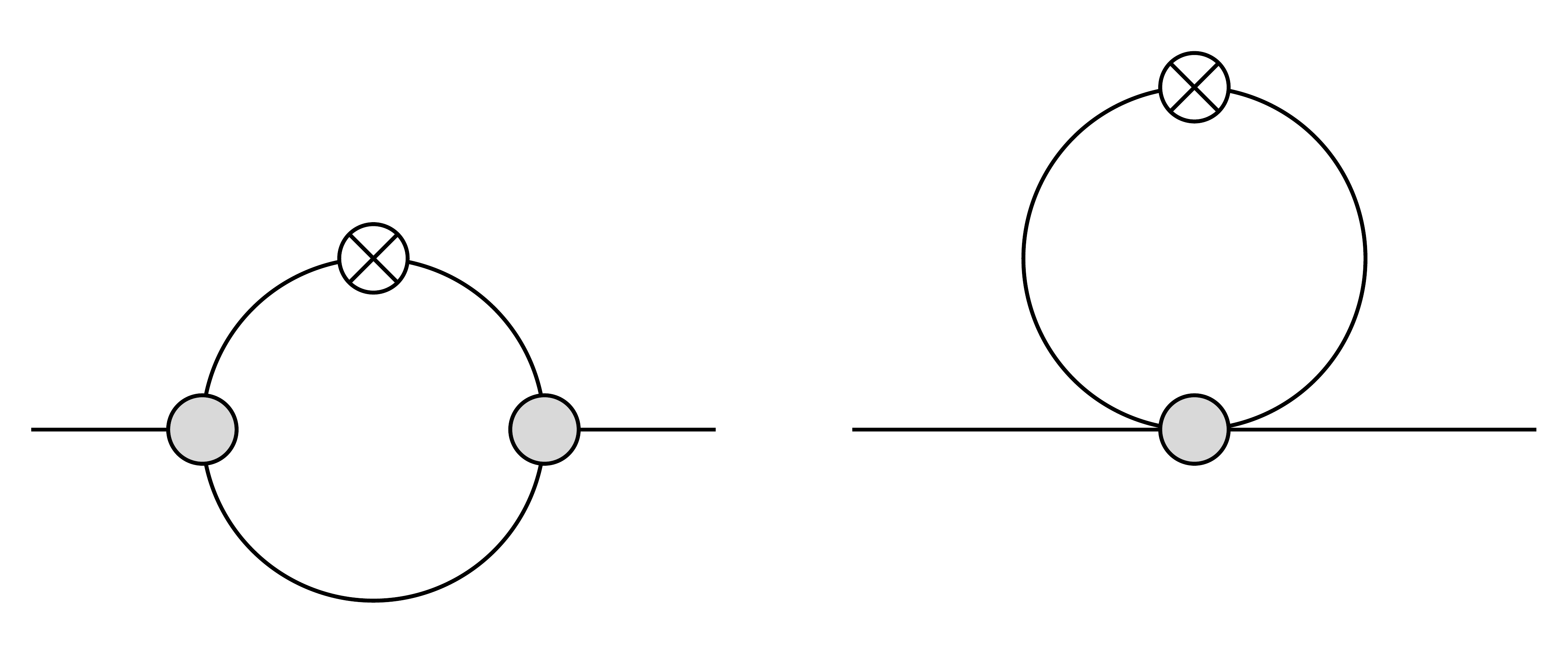} 
\end{center}
\caption{The two diagrams contributing to the flow of the two-point function, Eq.~(\ref{gamma2champnonnul}). The internal  lines represent dressed propagators, $G_\kappa$. The circled
 cross represents an insertion of $\partial_\kappa R_k$. The vertices denoted by   grey dots are respectively the three-point functions $\Gamma^{(3)}_\kappa$ (left) and the four-point function $\Gamma^{(4)}_\kappa$ (right). Note that $\Gamma^{(3)}_\kappa$ vanishes when $\phi$ vanishes, leaving then the right diagram as the only contribution to the flow of the two-point function. \label{two-point-diagrams}}
\end{figure}

 By  taking two derivatives of  Eq.~(\ref{NPRGeq0}) with respect to $\phi$, exploiting Eq.~(\ref{inv}), and Fourier transforming while restricting to a constant background field, one obtains the flow equation for the two-point function:\footnote{Alternatively, one could also start from Eq.~(\ref{NPRGeq}), take two field derivatives in the momentum representation while exploiting Eq.~(\ref{eq:Ginv}), and then restrict oneself  to a constant background field configuration.}
\beq
\label{gamma2champnonnul}
\partial_\kappa\Gamma_{\kappa}^{(2)}(p;\phi)&=&\int_q
\,\partial_\kappa R_\kappa(q)\,G_{\kappa}^2(q,\phi)\times\nonumber\\
&\times&\left\{\Gamma_{\kappa}^{(3)}(q,p,-p-q;\phi) G_{\kappa}(q+p,\phi)\Gamma_{\kappa}^{(3)}(p+q,-p,-q;\phi)
-\frac{1}{2}\Gamma_{\kappa}^{(4)}
(q,-q,p,-p;\phi)\right\},\nonumber \\
\eeq
where it is understood that here all the $n$-point functions are the reduced ones.
A diagrammatic illustration of this equation is given in Fig.~\ref{two-point-diagrams}. Flow equations for other $n$-point function are obtained similarly by taking further functional derivatives of $\Gamma[\phi]$.

In this paper, we shall be considering only systems with $\smash{\phi=0}$. External background fields will be introduced occasionally  only as intermediate tools to derive equations of motion. When $\phi=0$, the equation for the two-point function, Eq.~(\ref{gamma2champnonnul}), simplifies since $\Gamma_{\kappa}^{(3)}(p,q,-p-q;\phi=0)=0$.  We shall also use the simplified notation $\Gamma^{(n)}(p_1,\dots,p_n;\phi=0)=\Gamma^{(n)}(p_1,\dots,p_n)$ and similarly $G_\kappa(p;\phi=0)=G_\kappa(p)$.

\subsection{The 2PI formalism}\label{sec:2PI}

The 2PI formalism involves an additional Legendre transform with respect to a source $K(x,y)$ coupled to the bilinear $\varphi(x)\varphi(y)$. One defines the functional
\beq\label{eq:WJK}
W[J,K]=\ln\int {\cal D}\varphi \,\rme^{-S[\varphi]+\int_x J(x)\varphi(x)+\int_x\int_y K(x,y)\varphi(x)\varphi(y)}\,,
\eeq
together with the fields $\phi(x)$ and $G(x,y)$ conjugated to the sources $J(x)$ and $K(x,y)$:
\beq\label{2PIconj}
\frac{\delta W[J,K]}{\delta J(x)}=\langle\varphi(x)\rangle\equiv \phi(x)\,, \quad \frac{\delta W[J,K]}{\delta K(x,y)}=\langle\varphi(x)\varphi(y)\rangle\equiv G(x,y)+\phi(x)\phi(y)\,.
\eeq
The Legendre transform of $W[J,K]$ yields the 2PI effective action $\Gamma[\phi,G]$
\beq\label{2PIeffaction}
\Gamma[\phi,G]+W[J,K]=\int_x J(x) \phi(x)+\int_x\int_y K(x,y)\big[G(x,y)+\phi(x)\phi(y)\big]\,,
\eeq
with
\beq\label{2PIsources}
\frac{\delta \Gamma[\phi,G]}{\delta\phi(x)}=J(x)+2\int_y K(x,y)\phi(y)\,, \qquad \frac{\delta \Gamma[\phi,G]}{\delta G(x,y)}=K(x,y)\,.
\eeq
The relation between the 1PI effective action $\Gamma[\phi]$ and the 2PI effective action $\Gamma[\phi,G]$ is easily unveiled by noticing that $W[J]$ as defined in Eq.~(\ref{eq:WJ}) is obtained from $W[J,K]$ in the limit $\smash{K\to 0}$. Taking the same limit in Eqs.~(\ref{2PIeffaction})-(\ref{2PIsources}) and comparing to Eq.~(\ref{1PIeffaction}) one finds that $\Gamma[\phi]$ and $\Gamma[\phi,G]$ coincide when $G$ is chosen to obey its equation of motion:
\beq\label{eq:1PI2PI}
\Gamma[\phi]=\Gamma[\phi,G]\,, \qquad \frac{\delta\Gamma[\phi,G]}{\delta G(x,y)}=0\,.
\eeq 
We mention that functional derivatives with respect to $G(x,y)$ need to be taken in a symmetrical sense:
\beq\label{eq:sym_der}
\frac{\delta}{\delta G(x,y)}\to \frac{1}{2}\left[\frac{\delta}{\delta G(x,y)}+\frac{\delta}{\delta G(y,x)}\right].
\eeq
Indeed, it follows from Eq.~(\ref{2PIconj}) that $G$ is crossing symmetric for any choice of sources, and, therefore, the variations of the propagator need to satisfy this constraint. This does not play any crucial role for the derivative that appears in Eq.~(\ref{eq:1PI2PI}) but becomes crucial once one considers higher derivatives of $\Gamma[\phi,G]$ with respect to $G$.\footnote{Similarly, because $K(x,y)$ only enters $W[J,K]$ in the combination $K(x,y)\varphi(x)\varphi(y)$, we can restrict ourself to symmetrical sources such that $\smash{K(x,y)=K(y,x)}$. In fact, this is a necessary requirement if one wants to ensure the invertibility of the Legendre transform.}

\begin{figure}[htbp]
\begin{center}
\includegraphics[scale=0.2]{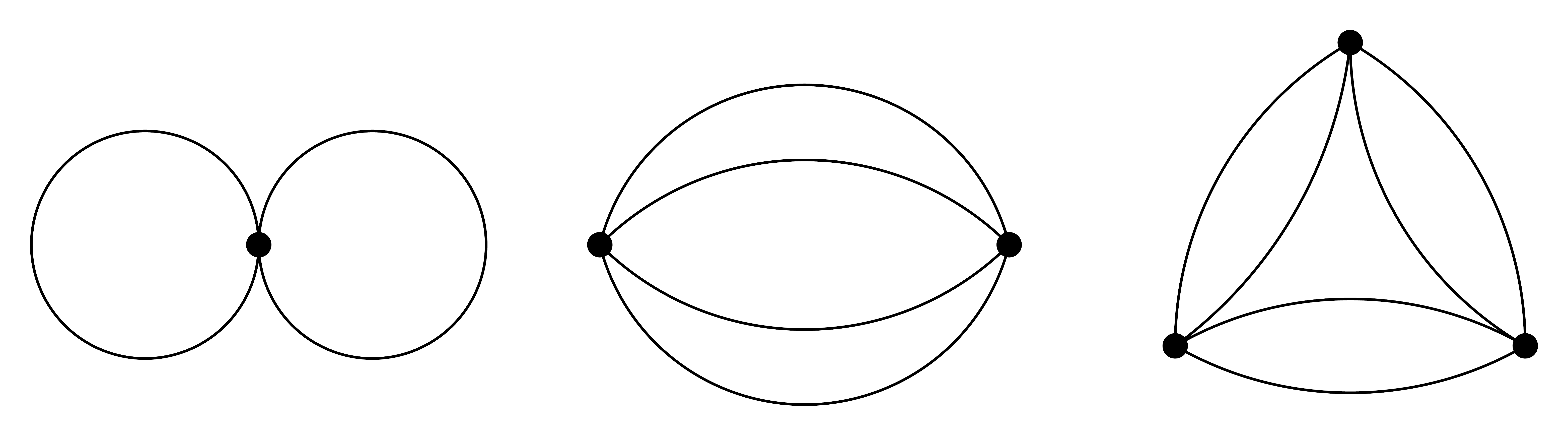}
\caption{The three skeleton diagrams that contribute to $\Phi[G]$ at order four-loop. The black dots represent bare vertices $\lambda_{\rm b}$.}\label{fig:skeletonsPhi}
\end{center}
\end{figure}

Just as the 1PI effective action can be expressed in terms of 1PI diagrams, the 2PI effective action admits an expansion in terms of 2PI diagrams \cite{Luttinger:1960ua,DeDominicis:1964b,Cornwall:1974vz}. In the case of a vanishing field expectation value $\phi=0$, $\Gamma[G]$ takes the form
\beq\label{Omegafunctional}
\Gamma[G]=\frac{1}{2} {\rm Tr} \log G^{-1}+\frac{1}{2} {\rm Tr} \,G_0^{-1} G +\Phi[G]\,,
\eeq
where ${\rm Tr}\,{\cal O}$ denotes the trace of the operator ${\cal O}$ that can be expressed either as an integral over space-time coordinates or as an integral over 4-momenta:\footnote{At finite temperature, integration over 4-momenta is replaced by integration over 3-momenta and summation over Matsubara  frequencies $\omega_n=2\pi n T$.}
\beq
{\rm Tr}\,{\cal O}\equiv\int_x {\cal O}(x,x)=\int_p {\cal O}(p,-p)\,.
\eeq
The functional $\Phi[G]$ in Eq.~(\ref{Omegafunctional}), often referred to as the Luttinger-Ward (LW) functional, plays a central role in the 2PI formalism. It is built as the sum of the ``two-particle-irreducible'' (2PI) diagrams of ordinary perturbation theory, with no external lines,  but evaluated with the full propagator $G$ rather than the free propagator $G_0$. These diagrams, sometimes also called ``two-line-irreducible'' diagrams, are diagrams that cannot be split apart by cutting two lines. In the absence of external lines, this notion coincides with that of ``skeleton'' diagrams (or two-skeletons), that is diagrams in which one cannot isolate self-energy insertions. In the absence of any external field, the case treated in this paper, $\Phi[G]$ is a  functional of the full propagator $G$. The  skeletons diagrams contributing to $\Phi$ up to order four-loop are displayed in Fig.~\ref{fig:skeletonsPhi}.
One  defines  similarly skeleton diagrams with external lines as obtained by taking successive functional derivatives of $\Phi[G]$ with respect to $G$. For two or more such derivatives, the corresponding skeleton diagrams are not completely two-particle-irreducible: they are 2PI only with respect to cuts of two lines that leave the two external legs associated to a given derivative $\delta/\delta G$ on the same side of the cut.

\begin{figure}[htbp]
\begin{center}
\includegraphics[scale=0.2]{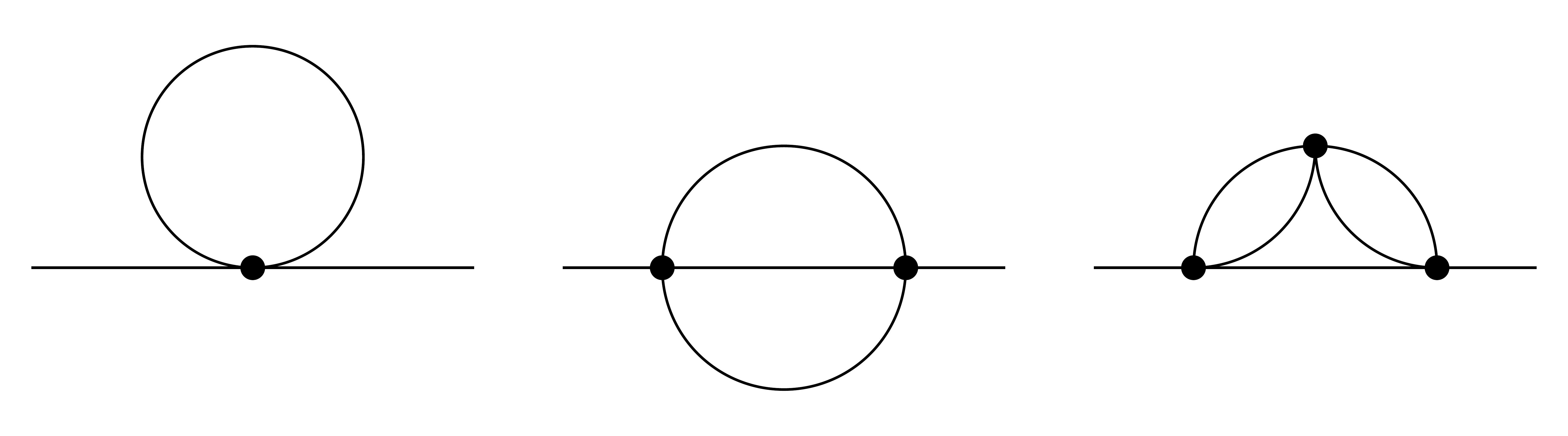}
\caption{ The skeleton diagrams  that contribute to $\Sigma[G]$ up to order three-loop. These are obtained from the skeletons of $\Phi[G]$ shown in Fig.~\ref{fig:skeletonsPhi} by taking a functional derivative with respect to $G$, Eq.~(\ref{PhiPi}).\label{fig:skeletons_Sigma}}
\end{center}
\end{figure}

Let us emphasize that $G(x,y)$ in Eq.~(\ref{Omegafunctional}) is a priori a general symmetric function of the two position variables $x$ and $y$. In contrast, the equation of motion for $G$ given in Eq.~(\ref{eq:1PI2PI}) defines a specific value for $G$ that may be restricted by symmetry. For instance, in equilibrium and in the presence of a constant $\phi$ (such as the vanishing background case considered here), the propagator determined by the equation of motion is translation invariant: $G(x,y)=G(x-y,0)\equiv G(x-y)$. For many purposes, it is enough to restrict the functional $\Gamma[G]$ to such propagators.\footnote{Below, we will discuss examples where the most general $G(x,y)$ needs to be used.} In this case, it is readily seen that a trivial volume factor $\delta^{(d)}(0)$ factorizes from all terms in Eq.~(\ref{Omegafunctional}). By dropping this factor, one is left with the reduced effective action
\beq\label{eq:2PIpot}
\Gamma[G]=\frac{1}{2} \int_p \log G^{-1}(p)+\frac{1}{2} \int_p \,G_0^{-1}(p)\,G(p) +\Phi[G]\,,
\eeq
in terms of a reduced Luttinger-Ward functional. Here, $G_0^{-1}(p)$ stands for the inverse free propagator, $G_0^{-1}(p)=p^2+m^2_{\rm b}$. The full propagator is obtained by extremizing  the functional $\Gamma[G]$. This yields a self-consistent Dyson equation, commonly called a ``gap equation''
\beq\label{Dyson}
G^{-1}(p)=G_0^{-1}(p)+\Sigma(p),
\eeq
where the self-energy functional $\Sigma[G]$ is obtained from $\Phi[G]$ via a functional derivative, viz.\footnote{We recall that our convention for the functional derivative in momentum space includes an implicit factor $(2\pi)^d$.}
\beq\label{PhiPi}
\Sigma(p)=2 \frac{\delta \Phi[G]}{\delta G(p)}.
\eeq
The skeleton diagrams contributing to $\Sigma$ up to order three-loop\footnote{The counting of loops depends on the quantity that one is looking at. Thus for instance a four-loop contribution to $\Phi$ generates three-loop contributions to $\Sigma$. Note however that the functional derivative does not change the dependence on the coupling constant, and both the four-loop contribution to $\Phi$ and the three-loop contributions to $\Sigma$ are of order $\lambda^3_{\rm b}$.} are displayed in Fig.~\ref{fig:skeletons_Sigma}. By construction, these diagrams do not contain subdiagrams that can be identified as self-energy insertions. Solving the gap equation (\ref{Dyson}) represents a major task in the 2PI formalism. After substituting the solution of this equation into Eq.~(\ref{Omegafunctional}), one obtains the free energy as the sum of all the Feynman diagrams of ordinary perturbation theory (i.e., irreducible and reducible and evaluated with propagators $G_0$).\footnote{ Note that when substituting $G$ for $G_0$ in $\Phi$, one ends up with an over-counting of the perturbative diagrams. This over-counting is precisely corrected by the first two terms in Eq.~(\ref{Omegafunctional}) (see e.g. \cite{Blaizot:2003an}).  However, in contrast to what happens for $\Phi$, the substitution $G_0\to G$ in skeletons with external lines does not generate any over-counting.}

\begin{figure}[htbp]
\begin{center}
\includegraphics[scale=0.17]{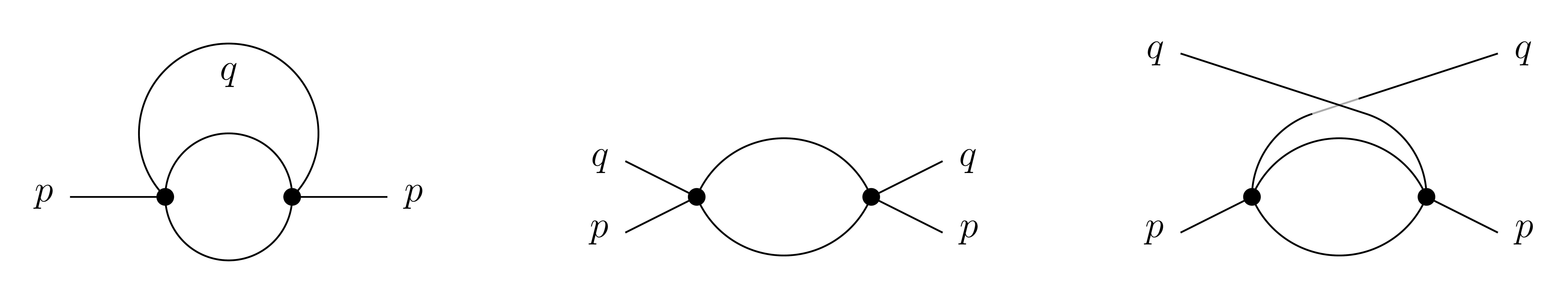}
\caption{The symmetric derivative  (see text) of $\Sigma(p)$ (left diagram) generates the two contributions to ${\cal I}(q,p)$ displayed on the right. These two contributions, of order $\lambda^2_{\rm b}$, correspond to the $t$ and the $u$ channels respectively.  \label{fig:kernel}}
\end{center}
\end{figure}

A further differentiation of $\Phi[G]$ with respect to $G$ yields the two-particle-irreducible kernel 
\beq\label{eq:Lambda}
{\cal I}(q,p)=2\frac{\delta \Sigma(p)}{\delta G(q)}=4\frac{\delta^2\Phi}{\delta G(q)\delta G(p)}={\cal I}(p,q).
\eeq
This kernel can be used to construct the full four-point function $\Gamma^{(4)}(q,p)$, via  a Bethe-Salpeter (BS) equation
\beq\label{BS1}
\Gamma^{(4)}(q,p) & = & {\cal I}(q,p)-\frac{1}{2}\int_r \,\,\,\Gamma^{(4)}(q,r)\,G^2(r)\,{\cal I}(r,p)\nonumber\\
& = & {\cal I}(q,p)-\frac{1}{2}\int_r \,\,\,{\cal I}(q,r)\,G^2(r)\,\Gamma^{(4)}(r,p).
\eeq
This equation  allows the calculation of the four-point function at specific values of the external momenta, namely $\Gamma^{(4)}(q,p)\equiv\Gamma^{(4)}(q,-q,p,-p)$, where it is here understood that $\Gamma^{(4)}(q,-q,p,-p)$ denotes the reduced Fourier transform.  As follows from Eq.~(\ref{eq:sym_der}), the functional derivative with respect to the propagator $G(q)$ needs to be taken in a symmetrical sense
 \beq
 {\cal I}(q,p)=2\frac{\delta \Sigma(p)}{\delta G(q)}\longrightarrow \frac{\delta \Sigma(p)}{\delta G(q)}+\frac{\delta \Sigma(p)}{\delta G(-q)}\,,
 \eeq
 reflecting the crossing symmetry $\smash{G(-q)=G(q)}$.  The diagrammatic interpretation of this operation is illustrated in Fig.~\ref{fig:kernel}, and the skeletons contributing to ${\cal I}$ up to two-loop order are given in Fig.~\ref{fig:skeletons_kernel} below. The kernel ${\cal I}(q,p)$ is the two-line irreducible contribution to $\Gamma^{(4)}(q,p)$ in one particular channel. Borrowing from  the particle physics terminology, we  refer to this particular channel as the $s$-channel, with momenta $p,-p$ entering and $q,-q$ leaving.\footnote{Recall that we use a convention of ``incoming'' momenta: a momentum $p$ is counted as $+p$ if it enters a vertex, and $-p$ if it leaves it.}  The other two channels are  the $t$ and $u$ channels.

It may not be immediately obvious why we need to consider the four-point function when calculating  the free energy and the self energy according to Eqs.~(\ref{Omegafunctional}) and (\ref{Dyson}), since these are functionals of the propagator. However, we shall see later that the four-point function and the corresponding BS equation play an important role in the renormalization, as they serve in particular to eliminate subdivergences both in the self-energy and in the free energy. They also emerge naturally when reformulating the 2PI formalism in terms of flow equations as we show below.

Not unrelated to this, the four-point function appears when one looks at how the solution of the gap equation, i.e. the self-consistent propagator $G$ or the self-energy $\Sigma$,  depends on parameters, in particular on the mass $m_{\rm b}$.  To see that, let us take the derivative of the self-energy, Eq.~(\ref{PhiPi}), with respect to $m_{\rm b}^2$.  We obtain
\beq\label{delSigma2}
\frac{\del\Sigma(p)}{\del m_{\rm b}^2}=2\int_q  \frac{\del G(q)}{\del m_{\rm b}^2}\,\frac{\delta^2 \Phi[G]}{\delta G(q)\delta G(p)}= \frac{1}{2}\int_q \frac{\del G(q)}{\del m_{\rm b}^2} \,{\cal I}(q,p)\,.
\eeq
Observe next that,  according to Eq.~(\ref{Dyson}), $G^{-1}=p^2+m_{\rm b}^2+\Sigma[G]$, so that
\beq\label{deltakappaG2}
\frac{\del G(q)}{\del m_{\rm b}^2}=-G^2(q)\,\frac{\del G^{-1}(q)}{\del m_{\rm b}^2}=-G^2(q) \left(  1+\frac{\del \Sigma(q)}{\del m_{\rm b}^2}\right).
\eeq
It follows  that Eq.~(\ref{delSigma2}) can be written as 
\beq\label{eqforSigma1b}
\int_q \frac{\del \Sigma(q)}{\del m_{\rm b}^2}\left(\delta^{(d)}(q-p)+\frac{1}{2}\,G^2(q)\,{\cal I}(q,p)\right)= -\frac{1}{2}\int_q \,G^2(q)\,{\cal I}(q,p)\,.
\eeq
\begin{figure}[h]
\begin{center}
\includegraphics[angle=0,scale=0.25]{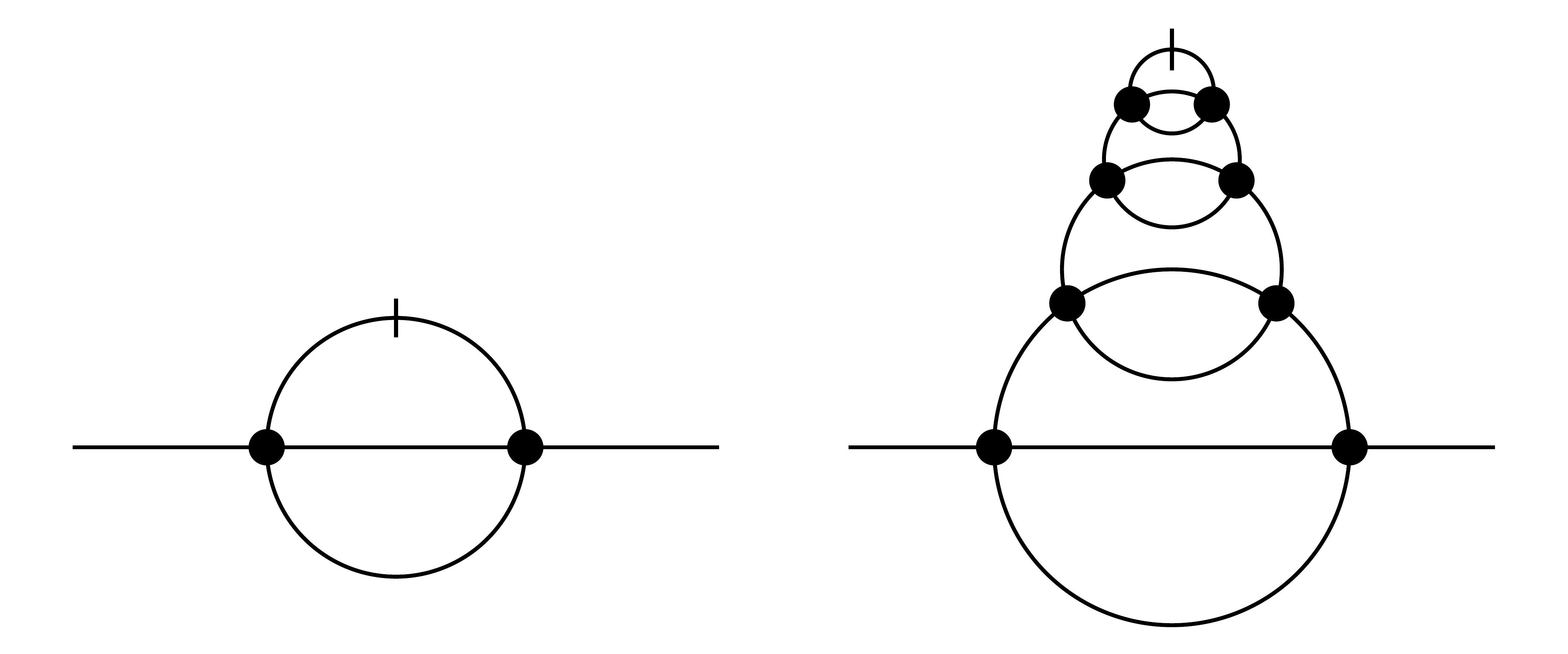}
\caption{Graphical illustration of Eq.~(\ref{dSigmadm2}) for the two-loop skeleton. The derivative with respect to $m_{\rm b}$ is denoted by a slash. When acting on a self-consistent propagator, this derivative generates the infinite tower of one-loop kernels  ${\cal I}$ that are resummed by the BS equation, Eq.~(\ref{BS1}).} 
\label{fig:delSigmadelm}
\end{center}
\end{figure}
This equation can be read as a (continuous) matrix equation. Let us then define the matrix  $M(q,p)\equiv \delta^{(d)}(q-p)+\frac{1}{2}\,G^2(q)\,{\cal I}(q,p)$. It is easily checked, using Eq.~(\ref{BS1}), that the inverse matrix is given by $M^{-1}(q,p)=\delta^{(d)}(q-p)-\frac{1}{2}\,G^2(q)\,\Gamma^{(4)}(q,p)$. Multiplying now both sides of Eq.~(\ref{eqforSigma1b}) by this inverse matrix and using Eq.~(\ref{BS1}) once more, one arrives finally at
\beq\label{dSigmadm2}
\frac{\del\Sigma(p)}{\del m_{\rm b}^2}=-\frac{1}{2}\int_q \,G^2(q)\,\Gamma^{(4)}(q,p)\label{eq:flow2b}
\eeq
where, as anticipated, the four-point function explicitly appears. 
 Note that this result has taken into account the fact that the dressed propagator depends on $m_{\rm b}$ both explicitly and implicitly, through its dependence on the self-energy, itself a functional of the propagator. It is this self-consistency which is at the origin of the appearance of the four-point function in the final result (see also Fig.~\ref{fig:delSigmadelm}). The flow equation (\ref{eq:BS2b}) discussed in the next subsection  can be seen as a generalization of this equation (\ref{dSigmadm2}).\\

If all skeletons are kept in $\Phi[G]$, all the functional relations written above are exact relations in the cutoff theory. But, of course,  the main interest of the LW functional is to lend itself to particular approximations.  The so-called ``$\Phi$-derivable''
approximations \cite{Baym:1962sx}  consist in selecting a class of skeletons in $\Phi[G]$ and calculating $\Sigma$ and $\Gamma^{(4)}$ from the equations above. It is easy to verify that all the relations mentioned above remain valid then,  whichever group of skeletons is selected.  It will be convenient in the foregoing discussion to refer to a systematic expansion of the skeleton diagrams. One such expansion  consists in selecting skeletons with an increasing number of loops\footnote{Another systematic expansion is that in number of components of the field, in the case of a scalar theory with O(N) symmetry \cite{Berges:2001fi}. Most of the results of this paper extend to the corresponding $1/N$ expansion.}. Since we  shall constantly refer to this loop expansion in the present discussion, we introduce a specific notation for the various objects that appear in this expansion. We denote by $\Phi_L$ the contribution to the LW functional of all the skeletons that contains up to $L$ loops.  For instance, the four-loop approximation to $\Phi[G]$ is the following functional of $G$:
\beq\label{Phi3loop}
\Phi_{L=4}[G] &=&\frac{\lambda_{\rm b}}{8}\left(\int_q G(q)\right)^2-\frac{\lambda^2_{\rm b}}{48}\int_p\int_q\int_r G(p)G(q)G(r)G(r+q+p)\nn
&+& \frac{\lambda^3_{\rm b}}{48}\int_p\int_q\int_r\int_k\, G(p)G(q)G(r) G(k)G(r+q+p)G(k+q+p)\,.
\eeq
We shall call $\Phi^{(l)}$ the $l$-loop contribution to $\Phi$, with the factor $\lambda^{l-1}_{\rm b}$ left out, that is, we write\footnote{A $l$-loop diagram that contributes to $\Phi$ contains $l-1$ vertices. It is tacitly assumed that the counting in loops is here made with respect to a given propagator $G$. Of course, when such a  propagator is replaced by the solution of the gap equation, each of the finite loop contributions that we are considering contains infinitely many ``perturbative'' loops with respect to the free propagator $G_0$.} 
\beq\label{eq:Phi_exp}
\Phi_L=\lambda_{\rm b} \Phi^{(2)}-\lambda^2_{\rm b} \Phi^{(3)}+\cdots +\lambda^{L-1}_{\rm b} (-)^L\Phi^{(L)}.
\eeq
 Similar expansions will be used for other objects, such as the self-energy $\Sigma$ ($\Sigma_L,\Sigma^{(l)}$), the irreducible kernel ${\cal I}$ (${\cal I}_L, {\cal I}^{(l)}$) or the four-point function $\Gamma^{(4)}$ ($\Gamma_L^{(4)}, \Gamma^{(4,l)}$), viz.
\beq\label{eq:bare_exp}
\Gamma^{(4)}_L=\lambda_{\rm b} -\lambda^2_{\rm b}\Gamma^{(4,1)} +\cdots +\lambda^{L+1}_{\rm b} (-)^L\Gamma^{(4,L)}.
\eeq
The  skeletons corresponding to $\Phi_{L=4}[G]$ are shown in Fig.~\ref{fig:skeletonsPhi}, and those which contribute  to $\Sigma_{L=3}[G]$ and to ${\cal I}_{L=2}[G]$ in this approximation are displayed respectively  in  Figs.~\ref{fig:skeletons_Sigma} and \ref{fig:skeletons_kernel}.

\begin{figure}[h]
\begin{center}
\includegraphics[angle=0,scale=0.15]{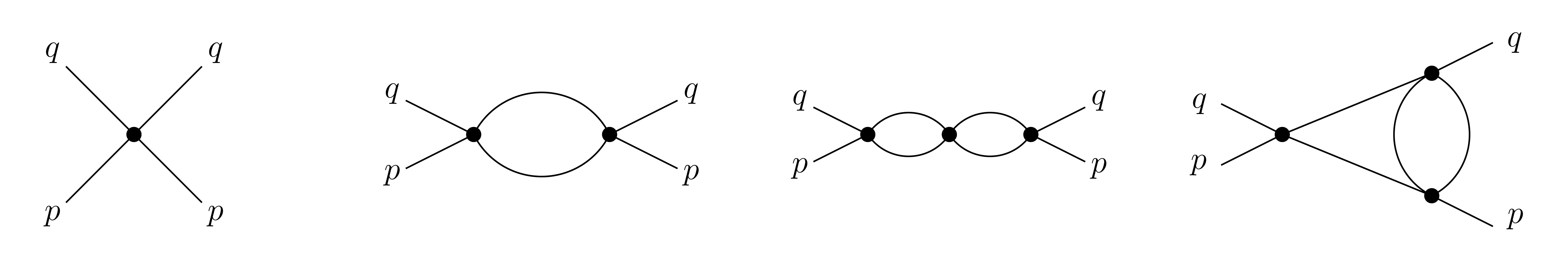}
\caption{ The skeleton diagrams   that contribute to the kernel ${\cal I}(q,p)$ up to order two-loop, i.e., the contributions to ${\cal I}_2(q,p)$. These are obtained from the skeletons of $\Phi_4[G]$ by taking two functional derivatives with respect to $G$, which amounts to cutting two lines in the diagrams of $\Phi$, that are subsequently labelled  respectively $p$ and $q$. These diagrams are two-line irreducible in the $s$ channel, defined by the external lines that carry momentum $p$ (or equivalently $q$). Only the $t$ channel contributions of the one-loop and two-loop diagrams are shown (the $u$ channel diagrams are obtained from the $t$ channel ones by exchanging the $q$ lines, as in Fig.~\ref{fig:kernel}). \label{fig:skeletons_kernel}}
\end{center}
\end{figure}

\begin{figure}[htbp]
\begin{center}
\includegraphics[scale=0.3]{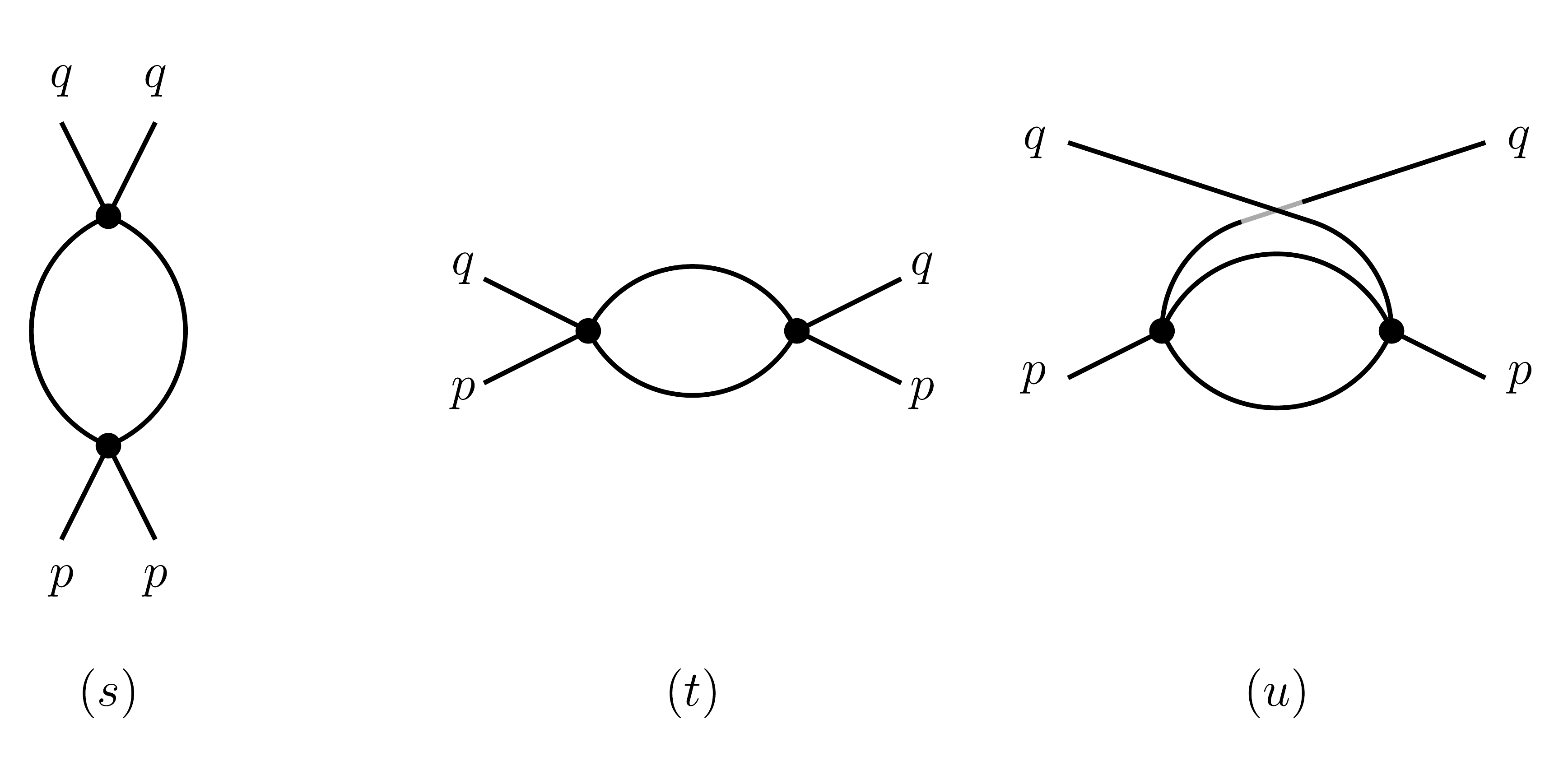}
\caption{ The   diagrams   that contribute to the four-point function $\Gamma^{(4)}(p,q)$ at order one-loop.  The first diagram, which is reducible in the $s$ channel, is obtained by iterating once the four-point vertex (the lowest order contribution to ${\cal I}$) in the Bethe-Salpeter equation. The other two diagrams are two-line irreducible in the $s$ channel (but reducible in the $t$ or the $u$ channel), and correspond respectively to the one-loop contributions of the $t$ and $u$ channels to the kernel ${\cal I}^{(l=1)}(p,q)$.\label{fig:three_channels}}
\end{center}
\end{figure}

It is useful for the foregoing discussion to observe  how the successive loop contributions to $\Gamma^{(4)}(q,p)$ build up as one takes successive skeletons into account in the  loop expansion. At a given order in the loop expansion, the diagrams of $\Gamma^{(4,l)}(q,p)$ that are two-line-irreducible in the $s$-channel,  are  included into the kernel ${\cal I}^{(l)}$, while those which are reducible are obtained from iterations in the Bethe-Salpeter equation of contributions  ${\cal I}^{(l')}$ with  $l'<l$.  An illustration  is provided in Fig.~\ref{fig:three_channels} which displays the three contributions to $\Gamma^{(4,l=1)}(q,p)$. This illustration makes apparent a well-known feature of the $\Phi$-derivable approximations: by focussing on a specific channel, one looses the crossing symmetry which is present in each order of perturbation theory. Let us stress that, within the approximation defined by the replacement $\Phi\to\Phi_L$, the solution to the BS equation (with ${\cal I}\to {\cal I}_{L-2}$), which we shall denote throughout $\Gamma^{(4)}_{\Phi_L}$,   contains of course $\Gamma^{(4)}_L$ but also infinitely many higher order contributions that make it distinct from $\Gamma^{(4)}_L$. In particular, while the various channels are always treated in a symmetrical way in $\Gamma^{(4)}_L$, this is not so for $\Gamma^{(4)}_{\Phi_L}$. Both $\Gamma^{(4)}_{\Phi_L}$ and the $\Gamma^{(4)}_{L'}$ with $L'<L-3$ play a role in the foregoing analysis.

\subsection{Making contact with the exact renormalization group}
\label{sec:contactRG}

We return now to the  functional  renormalization group, and observe that all the functional relations discussed in the previous subsection hold for the deformed theory, with the obvious modifications that are needed in order to take into account the presence of the regulator in the free propagator. Consider in particular the  gap equation in the deformed theory:
\beq\label{renormgapeqn}
G^{-1}_\kappa(p)=p^2+m^2_{\rm b}+\Sigma_\kappa(p)+R_\kappa(p),
\eeq
where the $\kappa$ dependence of  the self-energy comes entirely from its dependence on the propagator $G_\kappa$, i.e.,  $\Sigma_\kappa\equiv \Sigma[G_\kappa]$. For $\kappa=0$,  since the regulator then vanishes, this equation reduces trivially to the gap equation  of  the original theory, Eq.~(\ref{Dyson}). 
Now, by taking a derivative of $\Sigma_\kappa$ with respect to $\kappa$ and using the extension of Eq.~(\ref{PhiPi}) to the deformed theory, we obtain  
\beq\label{delSigma}
\del_\kappa \Sigma_\kappa(p)=2\int_q \del_\kappa G_\kappa(q)\left. \frac{\delta^2 \Phi[G]}{\delta G(q)\delta G(p)}\right|_{G_\kappa}= \frac{1}{2}\int_q \del_\kappa G_\kappa(q) \,{\cal I}_\kappa(q,p),
\eeq
where the irreducible kernel ${\cal I}_\kappa(q,p)$ appears: this is given by the same skeletons as in the original  theory, with  the propagators replaced by $G_\kappa$. Again, $G_\kappa$ is the sole source of $\kappa$ dependence of the kernel, i.e., $\smash{{\cal I}_\kappa={\cal I}[G_\kappa]}$.
\begin{figure}[h]
\begin{center}
\includegraphics[scale=0.080]{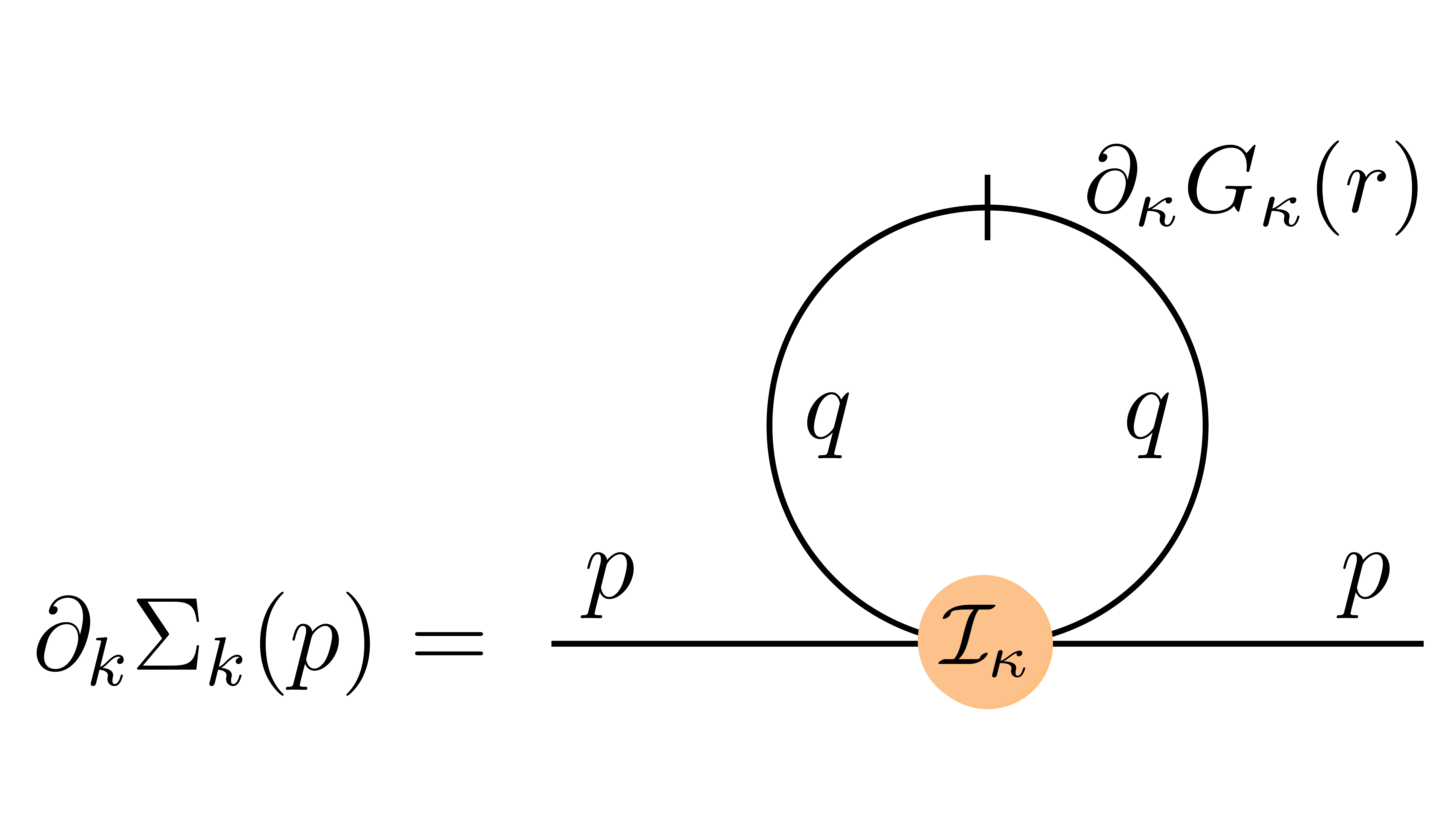}
\caption{Graphical illustration of Eq.~(\ref{delSigma}).}\label{fig:dSigmadG}
\end{center}
\end{figure}
A graphical illustration of this equation, analogous to Eq.~(\ref{delSigma2}),  is presented in Fig.~\ref{fig:dSigmadG}.
The derivatice $\del_\kappa G_\kappa(q)$ in Eq.~(\ref{delSigma}) can be calculated from Eq.~(\ref{renormgapeqn}):\beq\label{deltakappaG}
\del_\kappa G_\kappa(q)=-G_\kappa^2(q)\,\del_\kappa G_\kappa^{-1}(q)=-G^2_\kappa(q) \left (\del_\kappa \Sigma_\kappa(q)+\del_\kappa R_\kappa(q)\right).
\eeq
This allows us to rewrite Eq.~(\ref{delSigma}) as
\beq\label{eqforSigma1}
\int_q \pk \Sigma_\kappa(q)\left(\delta^{(d)}(q-p)+\frac{1}{2}\,G_\kappa^2(q)\,{\cal I}_\kappa(q,p)\right)= -\frac{1}{2}\int_q \pk R_\kappa(q)\,G_\kappa^2(q)\,{\cal I}_\kappa(q,p)\,.
\eeq
At this point we consider the extension of Eq.~(\ref{BS1}) to the deformed theory, viz. 
\beq\label{BSkappa}
\Gamma^{(4)}_\kp(q,p) & = & {\cal I}_\kp(q,p)-\frac{1}{2}\int_r\,\,\,\Gamma^{(4)}_\kp(q,r)\,G^2_\kp(r)\,{\cal I}_\kappa(r,p)\nonumber\\
& = & {\cal I}_\kp(q,p)-\frac{1}{2}\int_r\,\,\,{\cal I}_\kappa(q,r)\,G^2_\kp(r)\,\Gamma^{(4)}_\kp(r,p)\,,\label{eq:BS2k}
\eeq
and a simple calculation, analogous to that leading to Eq.~(\ref{dSigmadm2}), allows us to rewrite Eq.~(\ref{eqforSigma1}) in the form
\beq
\pk \Sigma_\kappa(p)=-\frac{1}{2}\int_q \,\pk R_\kappa(q)\,G^2_\kp(q)\,\Gamma^{(4)}_\kp(q,p)\,,\label{eq:flow2}\label{eq:BS2b}
\eeq
where now the four-point function $\Gamma_\kappa^{(4)}$, solution of Eq.~(\ref{eq:BS2k}), explicitly appears on the right hand side. See Fig.~\ref{fig:dSigmadR} for a graphical illustration.
\begin{figure}[htbp]
\begin{center}
\includegraphics[angle=0,scale=0.17]{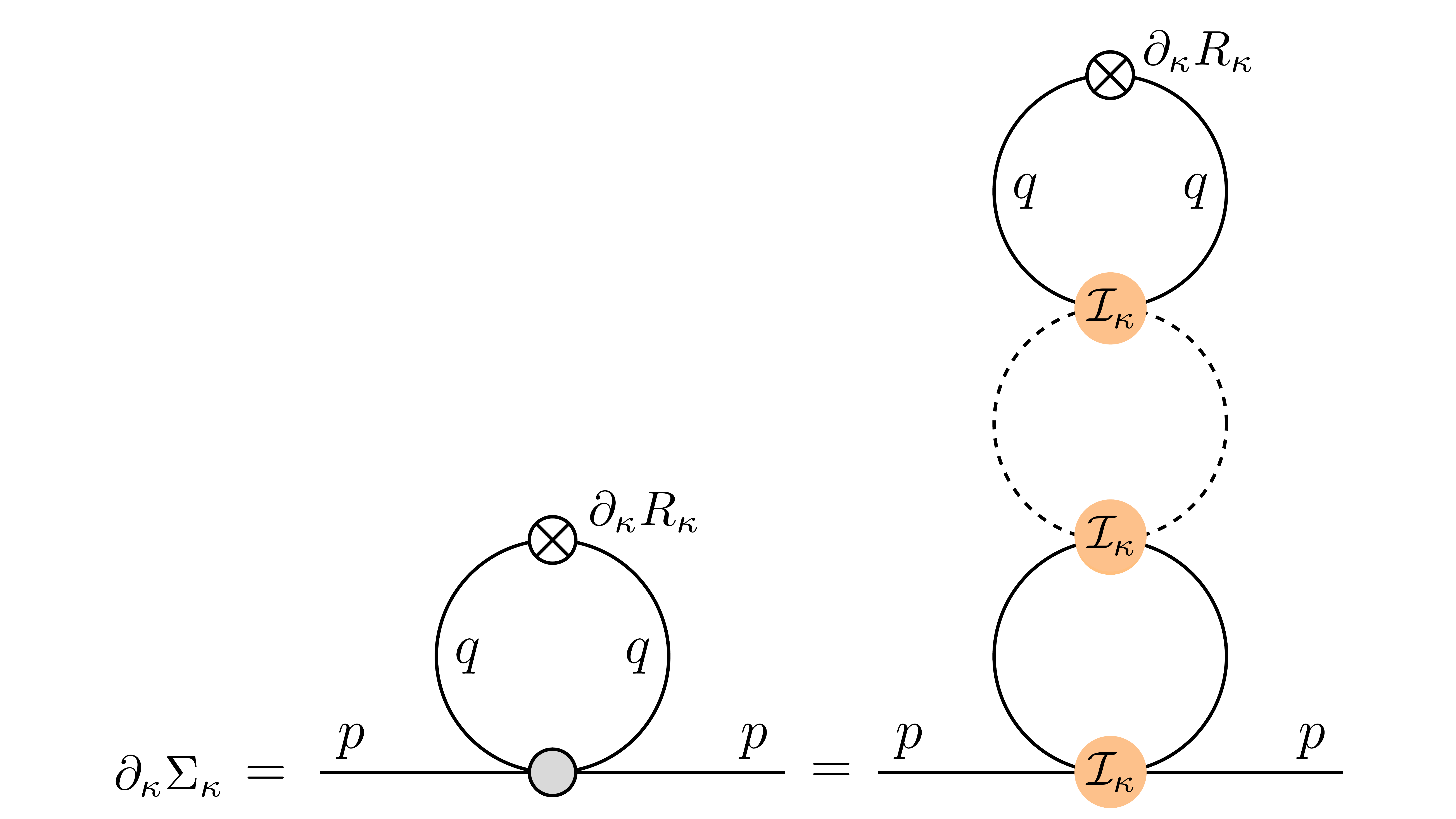}
\caption{Graphical illustration of Eq.~(\ref{eq:BS2b}). The grey blob on the left diagram represents the four-point function $\Gamma^{(4)}_\kappa$, and the chain of bubbles in the right diagram represents the expression of $\Gamma^{(4)}_\kappa$ in terms of the irreducible kernel ${\cal I}_\kappa$ as given by the BS equation (\ref{BSkappa}).}\label{fig:dSigmadR}
\end{center}    
\end{figure}

Since $\Gamma_\kappa^{(2)}(p)=p^2+m_{\rm b}^2+\Sigma_\kappa(p)$, this equation coincides formally with that deduced from the flow equation for the 1PI effective action, after taking two functional derivatives, i.e., Eq.~(\ref{gamma2champnonnul}) where we set $\phi=0$. It establishes a connection between the 2PI formalism and the functional renormalization group. A similar connection can also be made one step earlier in the  hierarchy of flow equations. Indeed, the free-energy density $f_\kp\equiv\Gamma_\kappa[\phi=0]$ can be obtained by evaluating the 2PI effective action $\Gamma_\kappa[G]$ at its stationary point $G=G_\kappa$:
\beq
f_\kp=\frac{1}{2} \int_p \log G^{-1}_\kappa(p)+\frac{1}{2} \int_p \,G_{0,\kappa}^{-1}(p)\,G_\kappa(p) +\Phi[G_\kappa]\,, \quad 0=\left.\frac{\delta \Gamma_\kappa[G]}{\delta G}\right|_{G=G_\kappa}\,.
\eeq
Because of the stationary condition, the only source of $\kappa$-dependence comes from $G_{0,\kp}^{-1}$. Using $\partial_\kappa G_{0,\kp}^{-1}=\partial_\kappa R_\kappa$, we find
\beq\label{eq:free}
\partial_\kappa f_\kp=\frac{1}{2} \int_p \,G_\kappa(p)\,\partial_\kappa R_\kappa(p)\,,
\eeq
which coincides with Eq.~(\ref{NPRGeq}) for $\phi=0$.

In contrast, moving up in the hierarchy, a major difference occurs. Indeed, the usual 1PI hierarchy continues with an equation for the four-point function which involves the six-point function $\Gamma^{(6)}_\kappa$ (see Eq.~(\ref{eq:flow4}) below), and so on. As already mentioned, a difficulty in the practical implementation of the fRG is that of closing this hierarchy in order to get a finite set of equations that one can solve. The equations that we have obtained so far realize such a truncation. Indeed the four-point function is directly calculated by solving a BS equation with the irreducible kernel that corresponds to the considered $\Phi$-derivable approximation, with no information being needed about the higher point functions. In the next section, we analyze further this truncation, and discuss an alternative to solving the BS equation that consists in writing a flow equation for the four-point function.

\section{$\Phi$-derivable approximations as truncations of the fRG equations}\label{Phitruncations}

The truncation that we have just discussed consists  in using the expression of $\Gamma^{(4)}_\kappa$ obtained from Eq.~(\ref{eq:BS2k}), in place of the usual 1PI flow equation (\ref{eq:flow4}). From the way Eq.~(\ref{eq:BS2b}) was obtained, it is clear that solving the coupled set of equations (\ref{eq:BS2k})-(\ref{eq:BS2b}) is equivalent to solving the gap equation (\ref{renormgapeqn}) for each value of $\kappa$. Thus, aside from providing a truncation of the hierarchy of flow equations, these coupled equations  constitute also an alternative formulation of a given $\Phi$-derivable approximation: a flow equation for $\Sigma_\kappa(p)$, coupled to a BS equation for $\Gamma_\kappa^{(4)}$ that needs to be solved at each step in $\kappa$. The approximation involves explicitly only the kernel ${\cal I}_\kp$, a known functional of $G_\kappa$ once a choice of skeletons has been made for $\Phi$. In this sense the system of equations is closed.  In fact, if we were to use the entire set of skeleton diagrams for $\Phi$, the relation between $\Gamma^{(4)}_\kappa$ and $\Sigma_\kappa$ would be exact. So the only approximation that we use here is that of a typical $\Phi$-derivable approximation, namely a selection of a finite class of skeleton diagrams.

From the point of view of the fRG, we note that the present truncation does not treat the two-point function and the four-point function on the same footing: the two-point function is indeed the solution of a flow equation, Eq.~(\ref{eq:BS2b}), while the four-point function is obtained, for each value of $\kappa$, from the solution of a BS equation, Eq.~(\ref{eq:BS2k}). However a simple alternative to solving the BS equation for each value of $\kappa$  is to write a flow equation for $\Gamma_\kappa^{(4)}$. This is what we do in the next subsection.  

\subsection{Flow equation for $\Gamma^{(4)}_\kappa$}\label{sec:gamma4flow}
Let us then take the  derivative of  Eq.~(\ref{eq:BS2k}) with respect to $\kappa$. We get
\beq
\int_r\partial_\kappa\Gamma^{(4)}_\kappa(q,r)\,M_\kappa(r,p)=\partial_\kappa{\cal I}_\kappa(q,p) & - & \frac{1}{2}\int_r\Gamma^{(4)}_\kappa(q,r)\,\partial_\kappa G_\kappa^2(r)\,{\cal I}_\kappa(r,p)\nonumber\\
& - & \frac{1}{2}\int_r\Gamma^{(4)}_\kappa(q,r)\,G_\kappa^2(r)\,\partial_\kappa {\cal I}_\kappa(r,p)
\eeq
where $M_\kappa$  is defined in terms of $G_\kp$ and ${\cal I}_\kp$ in the same way as $M$ in terms of $G$ and ${\cal I}$ at $\kappa=0$ (see the text after Eq.~(\ref{eqforSigma1b})). By multiplying both sides of this equation by $M^{-1}_\kappa$ (whose explicit expression follows also from that of $M^{-1}$ given  after Eq.~(\ref{eqforSigma1b})), and using again  Eq.~(\ref{eq:BS2k}), one ends up with
\begin{eqnarray}\label{eq:flowGamma4gen}
\partial_\kappa\Gamma^{(4)}_\kappa(p,q)=\partial_\kappa{\cal I}_\kappa(p,q) & - & \frac{1}{2}\int_r\Gamma^{(4)}_\kappa(p,r)\,\partial_\kappa G_\kappa^2(r)\,\Gamma^{(4)}_\kappa(r,q)\nonumber\\
& - & \frac{1}{2}\int_r\partial_\kappa{\cal I}_\kappa(p,r)\,G_\kappa^2(r)\,\Gamma^{(4)}_\kappa(r,q)\nonumber\\
& - & \frac{1}{2}\int_r\Gamma^{(4)}_\kappa(p,r)\,G_\kappa^2(r)\,\partial_\kappa {\cal I}_\kappa(r,q)\nonumber\\
& + & \frac{1}{4}\int_r\int_s \Gamma^{(4)}_\kappa(p,r)\,G_\kappa^2(r)\,\partial_\kappa {\cal I}_\kappa(r,s)\,G_\kappa^2(s)\,\Gamma^{(4)}_\kp(s,q). 
\end{eqnarray} 
Note that ${\cal I}_\kp$ has dropped out of the equation for $\pk\Gamma^{(4)}_\kp$, which depends now only  on $\pk{\cal I}_\kp$, that is, on the flow of the kernel rather than on the kernel itself. 

Solving Eq.~(\ref{eq:flowGamma4gen}) requires  the evaluation of $\partial_\kappa{\cal I}_\kappa$. In principle, this is a known functional of the propagator, as is ${\cal I}_\kappa$ itself.  Indeed, using the property that  the $\kappa$-dependence of ${\cal I}_\kappa$ comes entirely from its functional dependence on the propagator $G_\kappa$, one finds
\beq\label{eq:chain_rule}
\partial_\kappa {\cal I}_\kappa(q,p)=\int_r \partial_\kappa G_\kappa(r)\left.\frac{\delta{\cal I}(q,p)}{\delta G(r)}\right|_{G=G_\kappa},
\eeq
where, at each step in the flow, $\partial_\kappa G_\kappa$ can be evaluated from the known flow of $\Sigma_\kappa$, viz.
\beq\label{eq:dGa}
\partial_\kappa G_\kappa(r)=-(\partial_\kappa R_\kappa(r)+\partial_\kappa\Sigma_\kappa(r))\,G_\kappa^2(r)\,.
\eeq
We emphasize that Eq.~(\ref{eq:chain_rule}) should not be considered as an additional flow equation, on the same level as Eq.~(\ref{eq:flowGamma4gen}) for instance. Indeed this is not an equation to be solved to get ${\cal I}_\kappa$, since only  $\del_\kappa{\cal I}_\kappa$ is needed and this can be calculated explicitly in terms of the propagator $G_\kappa$. 
Equations~(\ref{eq:BS2b}), (\ref{eq:flowGamma4gen}), (\ref{eq:chain_rule}) and (\ref{eq:dGa}) define therefore a practical computational scheme, that we shall return to shortly. At this point, however, a few remarks are in order. 

These remarks concern the nature of the truncation realized by the present scheme. To better appreciate what it does, let us  compare Eq.~(\ref{eq:flowGamma4gen}) with the usual fRG flow equation for the four-point function 
\beq\label{eq:flow4}
\pk\Gamma^{(4)}_\kp(q,p) &=& -\frac{1}{2}\int_r \pk R_\kp(r)\,G_\kp^2(r)\,\Gamma^{(6)}_\kp(r,q,p) 
\nn
& + & \int_r \Gamma^{(4)}_\kp(q,-q,r,-r)\,G_\kp^3(r)\pk R_\kp(r)\,\Gamma^{(4)}_\kp(r,-r,p,-p)\nn
& + &  \int_r \Gamma^{(4)}_\kp(p,q,r,-p-q-r)\,G_\kp^2(r)\pk R_\kp(r)\,G_\kp^2(p+q+r)\,\Gamma^{(4)}_\kp(p,q,-r,-p-q+r)\nn
& + & \int_r \Gamma^{(4)}_\kp(p,-q,r,-p+q-r)\,G_\kp^2(r)\pk R_\kp(r)\,G_\kp^2(p-q+r)\,\Gamma^{(4)}_\kp(q,-p,r,p-q-r).\nn
\eeq
This equation is 
obtained after taking four functional derivatives of Eq.~(\ref{NPRGeq}) with respect to $\phi$, then setting $\phi=0$,  and choosing  a particular arrangement of external momenta.

 The appearance of  the six-point function $\Gamma^{(6)}_\kappa$ in Eq.~(\ref{eq:flow4}) reflects the fact that this equation for $\Gamma^{(4)}_\kappa$ is part of an infinite hierarchy. In contrast, Eq.~(\ref{eq:flowGamma4gen}) involves only known objects, and from that point of view it closes the system of flow  equations.  Furthermore, the present truncation is ``exact'' \cite{Blaizot:2010zx}, in the sense that the two-point and four-point functions  obtained at the end of the flow, i.e. when $\kappa=0$, are independent of the choice of the regulator\footnote{However attractive this property may look, it is worth keeping in mind that such a complete independence of the regulator often signals that what one does with the fRG is equivalent to an approximation that can be formulated in other terms, as is obviously the case here. For recent studies of the regulator choices, see \cite{Pawlowski:2015mlf,Balog:2019rrg}.} (except perhaps for a small effect in the initial conditions that we shall briefly mention at the end of this section).

There is another important difference between Eq.~(\ref{eq:flow4}) and Eq.~(\ref{eq:flowGamma4gen}), related to the convergence of the loop integrals. In Eq.~(\ref{eq:flow4}) the convergence is ensured by the derivative of the regulator, $\del_\kappa R_\kappa$(r) which, as we have already noticed, plays the role of an ultraviolet cutoff. This is not the case in Eq.~(\ref{eq:flowGamma4gen}), and indeed one of the issues that we shall have to address later is precisely that of whether  Eq.~(\ref{eq:flowGamma4gen}) can be made finite as the ultraviolet cutoff $\Lambda_{\rm uv}$ is sent to infinity. 

Finally, we note that crossing symmetry is respected in Eq.~(\ref{eq:flow4}), but this is not so in Eq.~(\ref{eq:flowGamma4gen}) when approximations are performed (since these approximations privilege a particular channel). We shall return to this issue in Sect.~\ref{sec:beyond}.

\subsection{Initial conditions}

As indicated earlier, Eqs.~(\ref{eq:BS2b}) and (\ref{eq:flowGamma4gen}),  together with Eqs.~(\ref{eq:chain_rule}) and (\ref{eq:dGa}), constitute a closed set of equations that are equivalent to the original formulation of $\Phi$-derivable approximations. These equations need to be solved with some initial conditions, and  we want to choose these so as to reproduce the results of the usual 2PI calculation at $\kappa=0$. In principle, this requires the initial conditions to be obtained by solving the gap and BS equations at some initial scale $\smash{\kappa=\Lambda}$ using standard methods. However, if the initialization scale $\Lambda$ is chosen large enough, there is no need to do so. Indeed, in this case, the initial conditions for $\Gamma^{(2)}_\Lambda(p)$ and $\Gamma^{(4)}_\Lambda(p,q)$ become simple polynomials in the external momenta whose determination does not require the resolution of the gap and BS equations. 

One way to understand that is to consider a sharp infrared regulator, in which case all momenta in the loops run strictly from $\kappa$ to $\Lambda_{\rm uv}$. In the limit $\kappa\to\Lambda_{\rm uv}$ the loop integrals then trivially vanish. Thus for instance, the $r$-integral in Eq.~(\ref{eq:BS2k}) vanishes, so that  $\Gamma^{(4)}_\Lambda={\cal I}_\Lambda=\lambda_{\rm b}$ for $\Lambda>\Lambda_{\rm uv}$, with $\lambda_{\rm b}$ the parameter of the Lagrangian. The same reasoning applies to the corrections to the mass:  all fluctuations being  suppressed when $\Lambda\ge\Lambda_{\rm uv}$, the two-point function $\Gamma^{(2)}_\Lambda$ reduces to $p^2+m_{\rm b}^2$, with $m_{\rm b}$  the parameter of the Lagrangian. 

The previous considerations extend to a smooth regulator. Indeed,   in the fixed UV cutoff approach that we have been discussing so far, nothing prevents us from taking an initial scale $\Lambda$ well above the cutoff scale $\Lambda_{\rm uv}$ and thus much larger than any loop momentum $k$ that enters the $n$-point functions (since $\smash{k<\Lambda_{\rm uv}\ll\Lambda}$). It follows that the loop corrections are suppressed by the regulator $R_\Lambda(k)\sim \Lambda^2$ (recall that $R_\kappa(k)$ is chosen so that $R_\kappa(k)\sim \kappa^2$ when $k\ll \kappa$) and the initial condition takes then the form
\beq\label{eq:init}
\Gamma^{(2)}_{\kappa=\Lambda}(p)\sim m^2_{\rm b}+p^2\,, \quad \Gamma^{(4)}_{\kappa=\Lambda}(p,q)\sim \lambda_{\rm b}\,, \quad \Lambda\gg\Lambda_{\rm uv}.
\eeq
Note that the approximate determination of the initial condition entails a small dependence on the choice of the regulator. Indeed, strict regulator independence requires the gap and BS equations to be solved exactly at the initial scale $\smash{\kappa=\Lambda}$. By approximating instead $\Gamma^{(2)}_\Lambda(p)$ and $\Gamma^{(4)}_\Lambda(q,p)$ by their simpler asymptotic behaviors in Eq.~(\ref{eq:init}), one reintroduces a slight regulator dependence that gets however smaller and smaller as $\Lambda$ is taken to larger and larger values (see the two-loop example discussed in \ref{sec:twoloop} for a concrete example).\\

One may perhaps wonder at this point what is the benefit of the reformulation of $\Phi$-derivable approximations as flow equations,  as opposed to their standard formulation in terms of diagrams. On a practical level, the equations that we have obtained provide an alternative  method to solve  the gap equation at $\smash{\kappa=0}$ via a smooth integration of fluctuations in successive momentum-shells controlled by the regulating scale $\kappa$. This is in contrast to standard methods, such as those based on iterations of the gap and BS equations, that may exhibit instabilities and do not always converge (of course other methods exist to tame such instabilities \cite{Berges:2004hn,Baacke:2004dp}). On a more formal level, the flow equations provide much insight into the renormalization of $\Phi$-derivable approximations, while opening the way to extensions that do not hinder their renormalizability. These formal aspects are our main concern in this paper. In fact, to address these renormalization issues, we need to develop further the analysis of the present section and provide a more explicit calculation scheme for the flow of the irreducible kernel. This is the purpose of the next section.

\section{Flow of the irreducible kernel}\label{sec:flowI}
As we have seen in the previous section, solving the equation for $\Gamma^{(4)}_\kappa$ requires the computation of the flow of the irreducible kernel, $\partial_\kappa {\cal I}_\kappa(q,p)$. This can be done by using Eqs.~(\ref{eq:chain_rule}) and (\ref{eq:dGa}), and this may indeed lead to a practical strategy to solve the equations. However, this simple approach suffers from conceptual limitations when it comes to understanding how the renormalization really operates, which we shall address in the following sections. Indeed, the flow of the kernel, $\partial_\kappa {\cal I}_\kappa(q,p)$, is given in Eq.~(\ref{eq:chain_rule}) as a sum of diagrams that may be divergent when $\Lambda_{\rm uv}\to\infty$, and where, furthermore, the bare coupling constant sits at the vertices. As we shall see, these represent obstacles in the renormalization, which however can be bypassed by the approach that we develop in this section. Once renormalization is clarified, we shall return in Sect.~\ref{sec:practical} to the formulation of Sect.~\ref{Phitruncations} and exploit this formulation to provide a finite and practical computational scheme for $\Phi$-derivable approximations.

A different line of approach is based on the observation that the functional derivative $\delta {\cal I} (q,p) /\delta G(r)|_{G=G_\kappa}$ in Eq.~(\ref{eq:chain_rule}) is a contribution to a six-point function (however, not a 1PI six-point function, as in Eq.~(\ref{eq:flow4})). In fact, Eq.~(\ref{eq:chain_rule}) may be viewed as the next equation, after Eq.~(\ref{delSigma}) for $\Sigma_\kappa$, in a (finite) hierarchy of equations for the 2PI $n$-point functions \cite{Dupuis:2005ij}. Examples of these $n$-point functions are given explicitly in \ref{App:npoint2PI} for the case of the three-loop skeleton. These 2PI $n$-point functions, and their flow equations, could be used to determine $\partial_\kappa {\cal I}_\kappa(q,p)$. This strategy has indeed  been put in practice in Ref.~\cite{Carrington:2014lba}. However, the 2PI $n$-point functions are not the most natural objects to be used in the present context. For one thing, they are not 1PI $n$-point functions. As such they offer limited insight on the general structure underlying the renormalization of $\Phi$-derivable approximations, which will be our main concern later in this paper. As we shall see, in this context, a prominent role is played by the functions $\Gamma^{(4)}_{L'}$ that have been introduced in Sect.~\ref{sec:2PI} (see Eq.~(\ref{eq:bare_exp})). In fact, as discussed in more details in \ref{App:npoint2PI}, these functions also naturally appear in the proper initialization of the flows of the 2PI $n$-point functions.\footnote{We stress that the discussion in \ref{App:npoint2PI} is not exactly the one followed in Ref.~\cite{Carrington:2014lba} and  we shall point out some of the differences.}

In this paper, we shall then proceed differently. We shall see that the flow of the kernel, $\del_\kappa {\cal I}_\kappa$, can be expressed in terms of the functions $\Gamma^{(4)}_{L'}$ that we have just mentioned and for which we shall obtain a finite set of flow equations. In the next section, we shall verify that these equations are finite, and can be written in such a way that no reference to the bare coupling appears. As we shall see then, this allows for a simple discussion of the renormalisation. 

\subsection{Flow of the irreducible kernel ${\cal I}_\kappa$ in terms of the $\Gamma^{(4)}_{L',\kp}$'s}
\begin{figure}[h]
\begin{center}
\includegraphics[angle=0,scale=0.09]{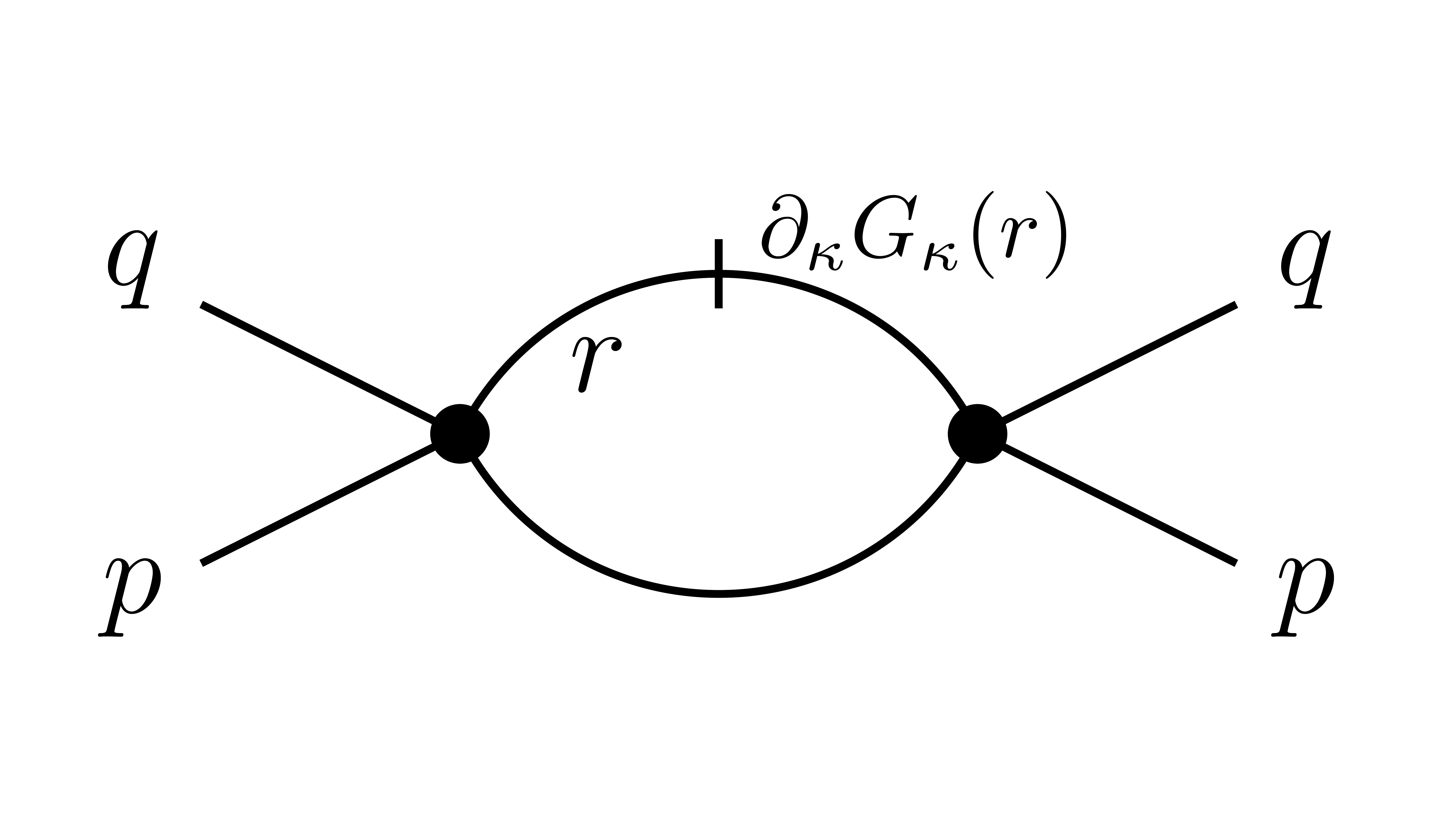}
\caption{The contributions to $\del_\kappa{\cal I}_\kappa^{(1)}$ (only the $t$-channel is drawn). The line that carries a slash represents $\del_\kappa G_\kappa(r)$. The diagram obtained by opening the slashed line is a contribution to a six-point function which is manifestly one-line reducible. \label{fig:deltaIkappaa}}
\end{center}
\end{figure}

To proceed, we start with a detailed analysis of the diagrammatic content of Eq.~(\ref{eq:chain_rule}), which is best unveiled by examining first the approximations to $\del_\kappa {\cal I}_\kappa$ up to $L=4$ loops for $\Phi$ (i.e. two-loop for ${\cal I}$). 
The two-loop approximation is trivial, since then ${\cal I}_{L=0}=\lambda_{\rm b}$, and is independent of $\kappa$. In this case only the first line of Eq.~(\ref{eq:flowGamma4gen}) contributes. Consider next the three-loop case, i.e., the one-loop approximation to ${\cal I}_\kappa$. We have
\beq
{\cal I}_{L=1,\kappa}(q,p)=\lambda_{\rm b}-\lambda_{\rm b}^2\,{\cal I}_\kappa^{(1)}(q,p)\,,\qquad {\cal I}_\kappa^{(1)}(q,p)=\frac{1}{2}\big[J_\kappa(p+q)+J_\kappa(p-q)\big],
\eeq
with 
\beq\label{Jdefp}
J_\kappa(p)\equiv \int_r  G_\kappa(r)G_\kappa(r+p)\,.
\eeq
By taking the derivative with respect to $\kappa$, one gets
\beq\label{eq:kernel_3loopa}
\partial_\kappa{\cal I}_{L=1,\kappa}(q,p) & = & -\frac{\lambda_{\rm b}^2}{2}\big[ \del_\kappa J_\kappa(p+q)+\del_\kappa J_\kappa(p-q)\big]\nn
& = & -\left[(\Gamma^{(4)}_{L=0,\kappa})^2\int_r\partial_\kappa G_\kappa(r)G_\kappa(r+p+q)\right]_{1}+(q\to -q)\,,
\eeq
where the six-point function $\delta {\cal I}^{(1)}_\kappa(q,p) /\delta G_\kappa(r)$ in the second line  has a tree structure (two vertices linked by a propagator, see the next to last diagram in Fig.~(\ref{fig:AppB})). In the second line of Eq.~(\ref{eq:kernel_3loopa}), we have used $\Gamma^{(4)}_{L=0,\kappa}=\lambda_{\rm b}$. We have also introduced a notation that will be used throughout this paper, namely $[\_]_{L}$, whose meaning is that of an operator that retains in the expression  within the square brackets all the terms up to order $L$-loop, thus extending (\ref{eq:Phi_exp}) or (\ref{eq:bare_exp}). Since the integral is already one-loop, the notation is here somewhat redundant but we shall see its usefulness below when considering the next order of approximation. We mention that, in writing Eq.~(\ref{eq:kernel_3loopa}), we have exploited the symmetries of the integrals, which correspond to the symmetries of the corresponding diagrams, see Fig.~\ref{fig:deltaIkappaa}.

\begin{figure}[htbp]
\begin{center}
\includegraphics[scale=0.20]{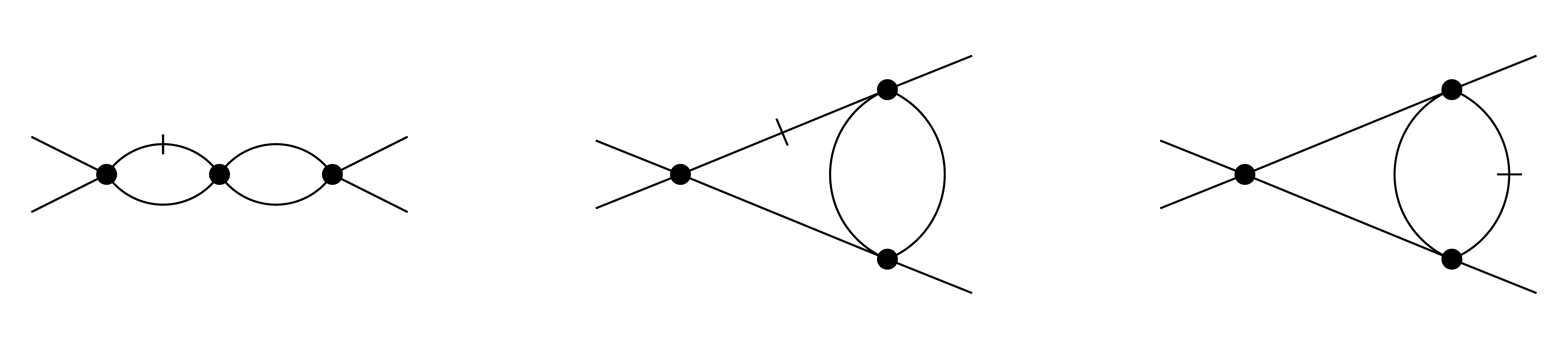}
\caption{Contributions to $\partial_\kappa{\cal I}_\kappa^{(l=2)}$. The lines that carry a dash represent $\partial_\kappa G_\kappa(r)$. Note that the one-loop subdiagrams that do not carry a dash in the first and second diagrams add up to the full one-loop contribution to $\Gamma_\kappa^{(4)}$, as shown in Eq.~(\ref{eq:dkIk}). By contracting these four-point insertions to a point, one obtains the diagram displayed in Fig.~\ref{fig:deltaIkappaa}, which is viewed as a four-skeleton diagram in the analysis of $\del_\kappa {\cal I}_\kappa$ developed in this section.} \label{fig:deltaIkappab}
\end{center}
\end{figure}

We continue with the two-loop contribution to ${\cal I}(p,q)$. This is given by
\begin{eqnarray}\label{calI2}
{\cal I}_{L=2,\kp}(q,p)=\lambda_{\rm b}-\lambda_{\rm b}^2\,{\cal I}_\kappa^{(1)}(q,p)+\lambda_{\rm b}^3\, {\cal I}_\kappa^{(2)}(q,p)\,,
\eeq
where
\beq
{\cal I}_\kappa^{(2)}(q,p)
&=&\frac{1}{4}\left[J_\kappa^2(p+q) +J_\kappa^2(p-q)\right]
\nn
 & +& \int_r G_\kappa(r)\left[G_\kappa(r+p+q)+G_\kappa(r+p-q) \right]J_\kappa(p+r)\nn
&=& \frac{1}{4}\int_r G_\kp(r)\,G_\kp(r+q+p) \int_s G_\kp(s)\,G_\kp(s+q+p)+(q\rightarrow -q)\nn
& + & \int_r G_\kp(r)\,G_\kp(r+p+q) \int_sG_\kp(s)\,G_\kp(s+r+p)+(q\rightarrow -q)\,.
\label{eq:Ika}
\end{eqnarray}
Now, in contrast to $\delta {\cal I}^{(1)}_\kappa(q,p) /\delta G_\kappa(r)$ which has a tree structure,  $\delta {\cal I}^{(2)}_\kappa(q,p) /\delta G_\kappa(r)$ contains a loop. This situation is generic in higher orders.  After taking a derivative with respect to $\kappa$, one gets
\begin{eqnarray}
\pk{\cal I}^{(2)}_\kp(q,p) & = & \int_r \pk G_\kp(r)\,G_\kp(r+p+q)\int_s G_\kp(s)\,G_\kp(s+p+q)+(q\rightarrow -q)\nonumber\\
& + & \int_r \pk G_\kp(r)\,G_\kp(r+p+q)\int_s G_\kp(s)\,G_\kp(s+r+p)+(q\rightarrow -q)\nonumber\\
& + & \int_r \pk G_\kp(r)\,G_\kp(r+p+q)\int_s G_\kp(s)\,G_\kp(s+r+q)+(q\rightarrow -q)\nonumber\\
& + & 2\int_r G_\kp(r)\,G_\kp(r+p+q)\int_s \pk G_\kp(s)\,G_\kp(s+r+p)+(q\rightarrow -q)\,,
\label{eq:Ika}
\end{eqnarray}
where we have again exploited the symmetries of the integrals. An illustration of  various contributions to the right-hand side of this equation is given in Fig.~\ref{fig:deltaIkappab}. 

An important observation is that, in the first three lines of Eq.~(\ref{eq:Ika}), one can identify the three independent contributions to the one-loop four-point function. This is illustrated in Fig.~\ref{fig:deltaIkappab}, where the first two diagrams starting from the left can be obtained form the diagram in Fig.~\ref{fig:deltaIkappaa}, by replacing one of the vertices by the one-loop four point function. As we shall argue shortly, this diagram plays the role of a skeleton from which higher order contributions to $\pk{\cal I}^{(1)}_\kp(q,p)$ can be obtained by replacing the vertices by four-point functions. We shall call these four-skeletons, in order not to confuse them with the two-skeletons introduced earlier. Combining with the result above for $\pk{\cal I}^{(1)}_\kp(q,p)$, Eq.~(\ref{eq:kernel_3loopa}), one can indeed rewrite $\pk{\cal I}_{L=2,\kp}(q,p)$ as follows
\beq\label{eq:dkIk}
\partial_\kappa{\cal I}_{L=2,\kappa}(q,p) & = & \Bigg\{-\left[\int_r \Gamma^{(4)}_{L=1,\kp}(r,q,p)\,\pk G_\kp(r)\,G_\kp(r+q+p)\,\Gamma^{(4)}_{L=1,\kp}(r,q,p)\right]_{2}\\
&  & \hspace{0.2cm}+\,2\left[(\Gamma^{(4)}_{L=0,\kp})^3\int_r G_\kp(r)\,G_\kp(r+q+p)\int_s \pk G_\kp(s)\,G_\kp(s+r+p)\right]_{2}\Bigg\}+(q\rightarrow -q)\,,\nonumber
\eeq
where
\beq\label{Gamma4rqp}
\Gamma^{(4)}_{L=1,\kp}(r,q,p) = \lambda_{\rm b}-\frac{\lambda_{\rm b}^2}{2}\big[J_\kappa(p+q)+J_\kappa(q+r)+J_\kappa(r+p)\big]
\eeq
is  the complete one-loop four-point function, and only terms up to two-loop order (that is to order $\lambda_{\rm b}^3$) have to be kept in each line. The first line has the same structure as that in Eq.~(\ref{eq:kernel_3loopa}), reflecting the skeleton structure mentioned above. It contains the contributions of the three independent channels and carries accordingly a full momentum dependence. The second line in Eq.~(\ref{eq:dkIk}) involves a new four-skeleton diagram (right most diagram in Fig.~(\ref{fig:deltaIkappab})) which is the seed for higher order contributions to $\partial_\kappa{\cal I}_\kappa$.

Equations such as Eq.~(\ref{eq:dkIk}) for $\partial_\kappa{\cal I}_{L=2,\kappa}(q,p)$ reveal an emerging structure that generalizes to all orders. That is,  $\partial_\kappa {\cal I}_\kappa$ can be written in terms of four-skeleton diagrams in which the vertices are  four-point functions. In particular, once written in terms of $\Gamma^{(4)}_{L=0,\kp}$ and $\Gamma^{(4)}_{L=1,\kp}$, the right-hand side of Eq.~(\ref{eq:dkIk}) does not contain any explicit non-trivial four-point subdiagrams. As we discuss in \ref{app:great}, this property extends to higher order truncations of $\partial_\kappa {\cal I}_\kappa$. More precisely, one defines four-skeletons as diagrams that contribute to $\partial_\kappa {\cal I}_\kappa$ and which contain no non-trivial four-point vertex insertions (a trivial insertion being just a tree-level vertex). All the diagrams contributing to the exact $\partial_\kappa {\cal I}_\kappa$ can then be obtained by replacing the trivial vertices by the exact four-point function $\Gamma^{(4)}$, with appropriate rooting of the momenta. Thus, we can write
\beq\label{eq:dec_great}
\partial_\kappa{\cal I}_{\kappa}(q,p) & = & \Bigg\{-\int_r \Gamma^{(4)}_{\kp}(r,q,p)\,\pk G_\kp(r)\,G_\kp(r+q+p)\,\Gamma^{(4)}_{\kp}(r,q,p)\nonumber\\
&  & \hspace{0.2cm}+\,2\int_s \pk G_\kp(s)\int_r \Gamma^{(4)}_{\kappa}(r,q,p)\,G_\kp(r)\,G_\kp(r+q+p)\nonumber\\
& & \hspace{1.2cm}\times\,\Gamma^{(4)}_{\kp}(-r,-s,-p)\,G_\kp(r+s+p)\,\Gamma^{(4)}_{\kp}(r+p+q,s,-q)\nonumber\\
&  & \hspace{0.2cm}+\,\dots\Bigg\}+(q\rightarrow -q).
\eeq
When truncated at $(L-2)$-loop order, this formula generalizes Eq.~(\ref{eq:dkIk}) and reads
\beq\label{eq:III}
\partial_\kappa{\cal I}_{L-2,\kappa}(q,p) & = & \Bigg\{-\left[\int_r \Gamma^{(4)}_{L-3,\kp}(r,q,p)\,\pk G_\kp(r)\,G_\kp(r+q+p)\,\Gamma^{(4)}_{L-3,\kp}(r,q,p)\right]_{L-2}\nonumber\\
&  & \hspace{0.2cm}+\,2\left[\int_s \pk G_\kp(s)\int_r \Gamma^{(4)}_{L-4,\kappa}(r,q,p)\,G_\kp(r)\,G_\kp(r+q+p)\right.\nonumber\\
& & \hspace{1.2cm}\left.\times\,\Gamma^{(4)}_{L-4,\kp}(-r,-s,-p)\,G_\kp(r+s+p)\,\Gamma^{(4)}_{L-4,\kp}(r+p+q,s,-q)\right]_{L-2}\nonumber\\
&  & \hspace{0.2cm}+\,\dots\Bigg\}+(q\rightarrow -q)
\eeq
with the dots representing higher order four-skeleton diagrams (up to $(L-2)$ loops) with vertices involving $\Gamma^{(4)}_{L-3,\kappa}$, \dots, $\Gamma^{(4)}_{0,\kappa}$. For instance, at order $L=3$, we would find the same terms as in Eq.~(\ref{eq:dkIk}) but with $\Gamma^{(4)}_{L,\kp}$ replaced by $\Gamma^{(4)}_{L+1,\kp}$ and $[\_]_{2}$ replaced by $[\_]_{3}$, and new three-loop four-skeleton diagrams with vertices given by $\Gamma^{(4)}_{0,\kp}$.

The analysis of this subsection has shown that the flow of ${\cal I}_\kappa$ involves four-skeleton diagrams, whose expressions are entirely determined by the propagator $G_\kappa$, as well as a set of functions $\Gamma^{(4)}_{L,\kp}$ that sit at the vertices of these four-skeleton diagrams. In order to complete the analysis, in the next subsection we   write the flow equations for these functions $\Gamma^{(4)}_{L,\kp}$.  


\subsection{Flow equations for the $\Gamma^{(4)}_{L,\kp}$'s}
The flow equations for the $\Gamma^{(4)}_{L,\kp}$'s are easily deduced from their diagrammatic expressions. For instance, from $\Gamma^{(4)}_{L=0,\kp}=\lambda_{\rm b}$, we find trivially
\beq\label{eq:trivial_flow}
\partial_\kp\Gamma^{(4)}_{L=0,\kp}=0\,.
\eeq
From Eq.~(\ref{Gamma4rqp}), we find
\beq\label{eq:flowGamma1}
\pk\Gamma^{(4)}_{L=1,\kp}(k,q,p) & = & -\left[(\Gamma^{(4)}_{L=0,\kp})^2\int_s \pk G_\kp(s)\,G_\kp(s+q+p)\right]_{1}+(k,q,p)_{\rm cyclic}\,,
\eeq
where $(k,q,p)_{\rm cyclic}$ denotes the contributions that are deduced from the first term in the right-hand side of Eq.~(\ref{eq:flowGamma1}) by cyclic permutations of $(k,q,p)$. To obtain the flow of $\Gamma^{(4)}_{L=2,\kappa}$, we note that the two-loop diagrams contributing to $\Gamma^{(4)}_{L=2,\kappa}$ have the same topologies as those that appear in the kernel ${\cal I}_{L=2,\kappa}$ that we treated in the previous subsection, with the exception that the three channels should be present in $\Gamma^{(4)}_{L=2,\kappa}$. Extending the equation for $\partial_\kappa {\cal I}_{L=2,\kappa}$ to an arbitrary configuration of momenta and including the missing channel,\footnote{Of course, there are other ways to obtain this equation, for instance using the BS equation truncated to $L$-loop order. The possibility to obtain $\Gamma^{(4)}_{L,\kappa}$ from ${\cal I}_{L,\kappa}$ (and therefore $\Phi_{L+2}$) generalizes to higher orders, as discussed in \ref{sec:relation} where we present a simple relation between $\Gamma^{(4)}_L$, ${\cal I}_L$,  and the fully two-particle-irreducible four-point function $\bar {\cal I}_L$.} one arrives at 
\beq\label{eq:b2}
\pk\Gamma^{(4)}_{L=2,\kp}(k,q,p)  & = & \left\{-\left[\int_r \Gamma^{(4)}_{L=1,\kp}(r,q,p)\,\pk G_\kp(r)\,G_\kp(r+q+p)\,\Gamma^{(4)}_{L=1,\kp}(r,q,p)\right]_{2}\right.\\
& & \hspace{0.2cm}\left.+\,2\left[(\Gamma^{(4)}_{L=0,\kappa})^3\int_s \pk G_\kp(s)\int_r G_\kp(r)\,G_\kp(r+q+p)\,G_\kp(r+s+p)\right]_{2}\right\}+(k,q,p)_{\rm cyclic}\,.\nonumber
\eeq
Equation (\ref{eq:b2}) bears a strong similarity with Eq.~(\ref{eq:dkIk}). And indeed, as we show in \ref{app:great}, the analysis that we performed for $\partial_\kp {\cal I}_\kp$ in terms of four-skeletons    extends straightforwardly to $\partial_\kappa\Gamma^{(4)}_\kp$. One can then write for the exact $\partial_\kappa\Gamma^{(4)}_\kp$ an equation analogous to  Eq.~(\ref{eq:dec_great}), namely  
\beq\label{eq:Ggreat2}
\pk\Gamma^{(4)}_{\kp}(k,q,p)  & = & \Bigg\{-\int_r \Gamma^{(4)}_{\kp}(r,q,p)\,\pk G_\kp(r)\,G_\kp(r+q+p)\,\Gamma^{(4)}_{\kp}(r,q,p)\nonumber\\
&  & \hspace{0.2cm}+\,2\int_s \pk G_\kp(s)\int_r \Gamma^{(4)}_{,\kappa}(r,q,p)\,G_\kp(r)\,G_\kp(r+q+p)\nonumber\\
& & \hspace{1.2cm}\times\,\Gamma^{(4)}_{\kp}(-r,-s,-p)\,G_\kp(r+s+p)\,\Gamma^{(4)}_{\kp}(r+p+q,s,-q)\nonumber\\
&  & \hspace{0.2cm}+\,\dots\Bigg\}+(k,q,p)_{\rm cyclic}\,.
\eeq
When truncated at $L$-loop order, this equation reads
\beq\label{eq:Ggreat}
\pk\Gamma^{(4)}_{L,\kp}(k,q,p)  & = & \Bigg\{-\left[\int_r \Gamma^{(4)}_{L-1,\kp}(r,q,p)\,\pk G_\kp(r)\,G_\kp(r+q+p)\,\Gamma^{(4)}_{L-1,\kp}(r,q,p)\right]_{L}\nonumber\\
&  & \hspace{0.2cm}+\,2\left[\int_s \pk G_\kp(s)\int_r \Gamma^{(4)}_{L-2,\kappa}(r,q,p)\,G_\kp(r)\,G_\kp(r+q+p)\right.\nonumber\\
& & \hspace{1.2cm}\left.\times\,\Gamma^{(4)}_{L-2,\kp}(-r,-s,-p)\,G_\kp(r+s+p)\,\Gamma^{(4)}_{L-2,\kp}(r+p+q,s,-q)\right]_{L}\nonumber\\
&  & \hspace{0.2cm}+\,\dots\Bigg\}+(k,q,p)_{\rm cyclic}\,,
\eeq
with the dots representing higher order four-skeleton diagrams (up to $L$ loops) with vertices involving $\Gamma^{(4)}_{L-3,\kappa}$, \dots, $\Gamma^{(4)}_{0,\kappa}$.


\subsection{Summary and initial conditions}\label{sec:summ}

What has been achieved in this section is a reformulation of the $L$-loop $\Phi$-derivable approximation in terms of a set of coupled  flow equations for $\Gamma^{(2)}_{L-1,\kp}\equiv p^2+m^2_{\rm b}+\Sigma_{L-1,\kp}$ and $\Gamma^{(4)}_{\Phi_L,\kp}$, that involve, in addition, flow equations for ``auxiliary'' functions $\Gamma^{(4)}_{L',\kp}$ that enter the determination of the flow of the irreducible kernel, $\del_\kappa{\cal I}_\kappa$.\footnote{Although it does not play a role in the determination of $G_\kappa$, we could add in this list the  flow equation for $f_\kappa$ in terms of $G_\kp$ (see Eq.~(\ref{eq:free})).}  The flow equations for $\Gamma^{(2)}_{L-1,\kp}$ and $\Gamma^{(4)}_{\Phi_L,\kp}$ read
\beq
\pk \Gamma^{(2)}_{L-1,\kappa}(p) & = & -\frac{1}{2}\int_q \,\pk R_\kappa(q)\,G^2_\kp(q)\,\Gamma^{(4)}_{\Phi_L,\kp}(q,p),\label{eq:g2k}
\eeq
and 
\beq\label{eq:22}
\partial_\kappa\Gamma^{(4)}_{\Phi_L,\kappa}(p,q) & = & \partial_\kappa{\cal I}_{L-2,\kappa}(p,q)\label{eq:s}\nonumber\\
& - & \frac{1}{2}\int_r\Gamma^{(4)}_{\Phi_L,\kappa}(p,r)\,\partial_\kappa G_\kappa^2(r)\,\Gamma^{(4)}_{\Phi_L,\kappa}(r,q)\nonumber\\
& - & \frac{1}{2}\int_r\partial_\kappa{\cal I}_{L-2,\kappa}(p,r)\,G_\kappa^2(r)\,\Gamma^{(4)}_{\Phi_L,\kappa}(r,q)\nonumber\\
& - & \frac{1}{2}\int_r\Gamma^{(4)}_{\Phi_L,\kappa}(p,r)\,G_\kappa^2(r)\,\partial_\kappa {\cal I}_{L-2,\kappa}(r,q)\nonumber\\
& + & \frac{1}{4}\int_r\int_s \Gamma^{(4)}_{\Phi_L,\kappa}(p,r)\,G_\kappa^2(r)\,\partial_\kappa {\cal I}_{L-2,\kappa}(r,s)\,G_\kappa^2(s)\,\Gamma^{(4)}_{\Phi_L,\kappa}(s,q)\,,\label{eq:g}
\eeq
with $G_\kappa(r)=\big[\Gamma^{(2)}_{L-1}(r)+R_\kappa(r)\big]^{-1}$.
In this equation, $\partial_\kappa{\cal I}_{L-2,\kappa}(p,q)$ is given in terms of the $\Gamma^{(4)}_{L',\kappa=0}$ (with $0\le L'\le L-3$) by Eq.~(\ref{eq:III}), with the flow of $\Gamma^{(4)}_{L',\kappa=0}$ being given in Eq.~(\ref{eq:Ggreat}).

In order to facilitate foregoing discussions, we shall rewrite these equations in a more compact form. Consider first the equation for the two-point function
\beq\label{flow2ptfct}
\pk\Gamma^{(2)}_{L-1,\kappa}(p) = \left.{\cal F}^{(2)}_{L-1,\kappa}\big[\Gamma^{(2)}_{L-1,\kappa},\Gamma^{(4)}_{\Phi_L,\kappa}\big](p)\right|_{\Lambda_{\rm uv}} .    
\eeq
The notation specifies that the flow of the two-point function, denoted ${\cal F}^{(2)}_{L-1,\kappa}$, depends functionally on the two-point function itself at the same loop order $(L-1)$, as well as on the four-point function $\Gamma^{(4)}_{\Phi_L,\kappa}$, solution of the BS equation (\ref{eq:flowGamma4gen}) with kernel ${\cal I}_{L=2,\kappa}(q,p)$. In addition it depends also on the momentum variable of the two-point function, as well as on the ultraviolet cutoff $\Lambda_{\rm uv}$. We write the flow equation  for $\Gamma^{(4)}_{\Phi_L,\kappa}$ in a similar fashion:
\beq\label{flow4Phi}
\pk\Gamma^{(4)}_{\Phi_L,\kappa}(p,q) = \left.{\cal F}^{(4)}_{\Phi_L,\kappa}\big[\Gamma^{(2)}_{L-1,\kappa},\Gamma^{(4)}_{\Phi_L,\kappa},\Gamma^{(4)}_{L-3,\kappa},\dots,\Gamma^{(4)}_{L=0,\kappa}\big](p,q)\right|_{\Lambda_{\rm uv}}
\eeq
where the dependence of the flow ${\cal F}^{(4)}_{\Phi_L,\kappa}$ on the $\Gamma^{(4)}_{L',\kappa}$  is through its dependence on $\partial_\kappa{\cal I}_{L-2,\kappa}$. The functions $\Gamma^{(4)}_{L,\kappa}$ satisfy the flow equations
\beq\label{flowGamma4L}
\pk \Gamma^{(4)}_{L',\kappa}(p_i)=\left. {\cal F}^{(4)}_{L',\kappa}\big[\Gamma^{(2)}_{L-1,\kappa},\Gamma^{(4)}_{L-1,\kappa},\dots,\Gamma^{(4)}_{L=0,\kappa}\big](p_i)\right|_{\Lambda_{\rm uv}}\quad (0\leq L'\leq L-3).
\eeq
Generically, the solution of the flow equations (\ref{flow2ptfct}), (\ref{flow4Phi}) and (\ref{flowGamma4L}) can be written as 
\beq
\Gamma_\kappa^{(n)}(p_i)=\Gamma_\Lambda^{(n)}(p_i)+ \int_\Lambda^\kappa \rmd\kappa' \left.{\cal F}^{(n)}_{\kappa'}(p_i)\right|_{\Lambda_{\rm uv}},\label{eq:71}
\eeq
which relates the value of the $n$-point function $\Gamma_\kappa^{(n)}(p_i)$  at a given scale $\kappa$ to that at the initial scale $\Lambda$. We have shown in the previous section that if this initial scale is chosen large enough $(\Lambda\gg\Lambda_{\rm uv}$), the initial values of the $n$-point functions, $\Gamma_\Lambda^{(n)}(p_i)$, are determined by the bare parameters, which may be adjusted so as to reach the desired values at $\kappa=0$ (see Eq.~(\ref{eq:init})). The solution  depends on the ultraviolet cutoff that is necessary at this point in order to render finite the integrals that enter the flows  ${\cal F}^{(n)}_{\kappa}$. The next three sections, devoted to renormalization,  will present three alternative ways to eliminate this dependence on $\Lambda_{\rm uv}$.

\section{Renormalized $\Phi$-derivable approximations from the flow equations}\label{sec:42}
Up until now, we have been dealing with $\Phi$-derivable approximations within the  theory with cutoff. We have seen that such approximations could be reformulated in terms of flow equations, and that, for an appropriate choice of the initial conditions, the solution of these flow equations reproduces the results obtained by solving directly the gap equation and the BS equation. The initial conditions involve explicitly the bare parameters, i.e. the parameters of the lagrangian, as indicated in Eq.~(\ref{eq:init}), and the solution depends on the ultraviolet cutoff $\Lambda_{\rm uv}$. We now address the question of whether and how we can remove the dependence on both the ultraviolet cutoff and the bare parameters by renormalization. In order to so so, it is useful to recall how this is achieved within the standard formulation of 1PI flow equations. 

\subsection{Renormalization within the functional RG}\label{sec:FRGren}
We then return to the functional renormalization group (see Sect.~\ref{frgbasics}) and  recall some general features of the usual flow equations for the 1PI $n$-point functions $\Gamma_\kappa^{(n)}$. These equations are obtained by taking successive functional derivatives of Eq.~(\ref{NPRGeq}), with  examples given in Eq.~(\ref{gamma2champnonnul}) for the two-point function $\Gamma^{(2)}_\kappa(p)$,  and Eq.~(\ref{eq:flow4}) for the four-point function $\Gamma^{(4)}_\kappa(q,p)$. There are three important properties of these flow equations that need to be underlined. The first one is  that they make no reference to the bare parameters, but are expressed instead solely in terms of the $n$-point functions, which we may identify with the renormalized $n$-point functions (see Sect.~\ref{sec:formal} below). The second property is that the flows (i.e., the derivatives $\del_\kappa \Gamma^{(n)}_\kappa$) are all given by one-loop integrals which are cut off by $\partial_\kappa R_\kappa$. This means that the ultraviolet cutoff $\Lambda_{\rm uv}$ can be removed from the loops without altering the solution in any significant way. Finally, the $n$-point functions have simple behaviors for $\kappa=\Lambda\gg \Lambda_{\rm phys}$, where $\Lambda_{\rm phys}$ is the scale of relevant physical momenta (we are typically interested in $n$-point functions $\Gamma^{(n)} (p_i)$ with external momenta  $p_i\lesssim \Lambda_{\rm phys}$). In the regime $\Lambda\gg\Lambda_{\rm phys}$, the external momenta $p_i$ contribute to $\Gamma^{(n)}(p_i)$ via propagators that always feature a scale much larger than $\Lambda_{\rm phys}$, of the order of $\Lambda$ or higher.\footnote{This is because, the total momentum of a given propagator is either less than $\Lambda$ in which case the propagator features a large mass of the order of $\Lambda$ as given by the regulator $R_\Lambda(q<\Lambda)\sim\Lambda^2$, or it is larger than $\Lambda$ in which case it provides itself the large momentum scale.} In this regime, it makes then sense to approximate the $n$-point functions by their Taylor expansion in powers of the $p_i$. In the case of the renormalized $n$-point functions, which are our focus here, the corresponding coefficients are given by simple power counting:\footnote{Subleading corrections to these coefficients involve terms controlled by $m/\Lambda$ where $m$ denotes the mass. We stress also that the possibility of Taylor expanding in powers of $p_i$ applies both to renormalized and to unrenormalized $n$-point functions. The important difference in the case of the renormalized fluctuations is that $\Lambda$ is the only large scale and the coefficients of the Taylor expansion are controlled by power counting with respect to this scale. In the case of unrenormalized fluctuations, in contrast, the coefficients of the Taylor expansion are functions of both $\Lambda$ and $\Lambda_{\rm uv}$, and for $\Lambda_{\rm uv}\gg\Lambda\gg\Lambda_{\rm phys}$, one does not have a simple power counting rule.}
\beq\label{eq:init3}
\Gamma^{(2)}_{\kappa=\Lambda}(p)\sim m^2_\Lambda+Z_\Lambda\,p^2\,, \quad \Gamma^{(4)}_{\kappa=\Lambda}(p,q)\sim \lambda_\Lambda\,, \quad \Gamma^{(2n\geq 6)}_{\kappa=\Lambda}(p_i)\approx 0\,, \quad \Lambda\gg \Lambda_{\rm phys} \,,
\eeq
where $m^2_\Lambda\sim\Lambda^2$, $Z_\Lambda\sim\ln\Lambda$ and $\lambda_\Lambda\sim\ln\Lambda$, and we have neglected contributions that are suppressed by powers of $\Lambda_{\rm phys}/\Lambda$. 
 
 It is interesting to see how this general behavior of the $n$-point functions indicated in Eq.~(\ref{eq:init3}) emerges as a  self-consistent solution of the flow equations. Let us indeed assume that the leading behaviors of the $n$-point functions are given by 
\beq\label{scalingsol}
\Gamma^{(2)}_\kp(p)\sim m^2_\kappa+Z_\kappa\,p^2\,,\quad   \Gamma^{(4)}_\kappa\sim \lambda_\kappa,\quad  \Gamma^{(n>4)}_\kappa\sim \kappa^{4-n},\eeq
with $m^2_\kappa\sim\kappa^2$, $Z_\kappa\sim \ln \kappa$ and $\lambda_\kappa\sim\ln\kappa$, where the notation $f_\kappa \sim \ln\kappa$, or equivalently $\kappa\del_\kappa f_\kappa\sim \kappa^0$, is meant to indicate that the growth of $f_\kappa$ at large $\kappa$ is slower than any power of $\kappa$. As a specific illustration of the argument, let us consider the flow equation for the four-point function, Eq.~(\ref{eq:flow4}). The loop momentum $r$ in this equation  is bounded by $\kappa$, because  $\partial_\kappa R_\kappa(r)$ plays the role of an ultraviolet  cutoff at the scale $\kappa$ (e.g. $\kappa\partial_\kappa R_\kappa(r)\sim \kappa^2\theta(\kappa-r)$). Then, by plugging the expected behaviors (\ref{scalingsol}) into Eq.~(\ref{eq:flow4}), using the fact that the regulated propagators $G_\kappa(r)$ at large $\kappa$ behave as $\kappa^{-2}$ and taking into account the four dimensional phase space integration $\sim \kappa^4$, one easily verifies that the leading order behaviors of the integrals in Eq.~(\ref{eq:flow4}) are indeed $\sim\kappa^0$, in agreement with the initial assumption that $\Gamma^{(4)}_\kappa\sim \ln \kappa$. This argument can be easily extended to arbitrary $n$-point functions, thereby verifying Eqs.~(\ref{scalingsol}).  Thus, when $\kappa=\Lambda$ becomes large, only the two and four-point functions are relevant, and their values are determined by the parameters  $m_\Lambda^2$, and $\lambda_\Lambda$ in addition to  $Z_\Lambda$: the effective action $\Gamma_\Lambda[\phi]$ eventually takes the form of the classical action of Eq.~(\ref{eactON}), with suitably defined parameters.\footnote{The general structure outlined above depends crucially on the presence of a derivative of the regulator in all the loop integrals. Indeed, for regulators with a sufficiently rapid decay or even compact support for momenta $p^2 \lesssim \kappa^2$ all flows are infrared and ultraviolet finite irrespective of the -perturbative- renormalizability of the theory at hand. For example,  the current $\phi^4$-theory in $d=6$ dimensions is not renormalizable but features finite flows:  the non-renormalizability of the theory reflects itself in infinitely many correlation functions flowing with positive power laws as $\kappa\to \infty$.} 

 We may furthermore argue that when $\kappa\gg \Lambda_{\rm phys}$ (and $p,q\lesssim \Lambda_{\rm phys}$) the dependence of the four-point function  $\Gamma^{(4)}_\kappa(p,q)$ on the external momenta $p$ and $q$ is negligible.  Recall that $\kappa$ plays a role similar to the external momenta in the loop integrals. When $\kappa\lesssim \Lambda_{\rm phys}$, the effect of changing $\kappa$ is negligible, there is essentially no flow: the external momenta play the role of the infrared regulator, and hide the effect of the variation of $\kappa$: . When $\kappa\gg\Lambda_{\rm phys}$, the opposite situation prevails: the $n$-point functions become independent of momenta. In other words, the flow is important only when $\kappa$ is of the order of the external momenta (or the mass). In the regime where $\Lambda_{\rm phys}\ll\kappa$,  the momentum dependence of the $n$-point functions shows up as small power corrections in $p/\kappa$ which are obtained by expanding the loop integrals in powers of $p/\kappa$, the presence of the regulator eliminating potential infrared divergences.

One concludes from this discussion that if $\Lambda$ is chosen large enough, $\Lambda_{\rm phys}\ll \Lambda\ll \Lambda_{\rm uv}$, then the appropriate initial conditions on the flows of the various $n$-point functions may be taken as indicated in Eq.~(\ref{eq:init3}). The values of the parameters $m^2_\Lambda$, $\lambda_\Lambda$ and $Z_\Lambda$  can be chosen so that $m^2_\kappa$, $\lambda_\kappa$ and $Z_\kappa$ take specified  values at $\kappa=0$, which correspond to the renormalized parameters (see renormalization conditions (\ref{eq:cond}) below). Note that the bare parameters, $m_{\rm b}$ and $\lambda_{\rm b}$, do not enter at all in the discussion: they have been eliminated in favor of the parameters $m^2_\Lambda$, $\lambda_\Lambda$ and $Z_\Lambda$ that specify the initial conditions of the flows. Similarly,  the ultraviolet cutoff drops out because the flow equations are finite, and fluctuations are integrated over the  finite range of momenta $[0,\Lambda]$. The independence of the renormalized parameters on the specific value of $\Lambda$ is ensured by the flow equations.  These govern the variations of the initial parameters  $m^2_\Lambda$, $\lambda_\Lambda$ and $Z_\Lambda$ under a change of $\Lambda$ in a way compatible with power counting, as expressed in Eqs.~(\ref{eq:init3}).

In the rest of this section, we shall verify that these features of the 1PI flow equations, in particular the  finiteness of the  flow equations and the absence of reference to the bare parameters,  hold, with some adjustments, for the flow equations for  $\Phi$-derivable approximations, which are  summarized in Sect.~\ref{sec:summ}.
 
 \subsection{Elimination of bare parameters }\label{sec:FRGren} 
We first examine the dependence of the flow equations on the bare parameters. The flow equations (\ref{flow2ptfct})-(\ref{flowGamma4L}) that we have derived in Sects.~\ref{Phitruncations} and \ref{sec:flowI}
 are nearly independent of these parameters.  If fact this is so for  Eq.~(\ref{flow2ptfct}) written in terms of $\Gamma^{(2)}_{L-1,\kappa}$ rather than in terms of  $\Sigma_{L-1,\kappa}$. Since $G_\kappa(r)=\big[\Gamma^{(2)}_{L-1,\kappa}(r)+R_\kappa(r)\big]^{-1}$, the bare mass non longer appears, as it would  if one would relate the propagator to the self-energy,   $G_\kappa(r)=\big[r^2+m^2_{\rm b}+\Sigma_{L-1,\kappa}(r)+R_\kappa(r)\big]^{-1}$.

As for the $\Gamma^{(4)}_{L,\kp}$'s that appear in the right-hand side of Eqs.~(\ref{eq:Ggreat}) they are still to be seen as cut-off regulated Feynman diagrams with vertices given by $\lambda_{\rm b}$, as specified by the operator $[\_]_{[L]}$ (see after Eq.~(\ref{eq:kernel_3loopa})), which is indeed based on the  expansion (\ref{eq:bare_exp}) in powers of $\lambda_{\rm b}$. To eliminate all reference to the bare coupling, we  rewrite the strict $l$-loop  contribution $\lambda_{\rm b}^{l+1} (-)^l\Gamma^{(4,l)}$ in Eq.~(\ref{eq:bare_exp}) as $\Delta\Gamma^{(4)}_{\kp,l}=\Gamma^{(4)}_{L=l,\kp}-\Gamma^{(4)}_{L=l-1,\kp}$ for $l\geq 1$ and $\Delta\Gamma^{(4)}_{0,\kp}=\Gamma^{(4)}_{L=0,\kp}$. We then replace in the  right-hand side of Eqs.~(\ref{eq:Ggreat}) any occurence of $\Gamma^{(4)}_{L,\kp}$ by the following expansion
\beq\label{eq:exp_fin}
\Gamma^{(4)}_{L,\kp}=\underbrace{\Gamma^{(4)}_{L=0,\kp}}_{\Delta\Gamma^{(4)}_{L=0,\kp}}+\Big[\underbrace{\Gamma^{(4)}_{L=1,\kp}-\Gamma^{(4)}_{L=0,\kp}}_{\Delta\Gamma^{(4)}_{L=1,\kp}}\Big]+\dots+\Big[\underbrace{\Gamma^{(4)}_{L,\kp}-\Gamma^{(4)}_{L-1,\kp}}_{\Delta\Gamma^{(4)}_{L,\kp}}\Big]\,,
\eeq
where each consecutive term counts one extra loop, and we redefine the expansion operator $[\_]_{L}$ with respect to this formal loop expansion. For greater clarity, we shall denote by $\{\_\}_{L}$ the expansion operator $[\_]_{L}$ in formulas where this substitution is made. Of course, as far as the bare $n$-point functions are concerned, using either $[\_]_{L}$ or $\{\_\}_{L}$ makes no difference. However, by using $\{\_\}_{L}$ in equations such as Eqs.~(\ref{eq:III}) and (\ref{eq:Ggreat}), we obtain equations  that make no reference to the bare coupling, and which therefore describe not only the bare $n$-point functions, but also the renormalized ones, the choice between one type of solution and the other being essentially determined by the initial conditions, as we discuss below.

\subsection{Finiteness of the flow equations}
Now that we have eliminated any reference to the bare parameters from the flow equations, let us show that they are finite. Before we proceed to this analysis a few words of caution are required however. 

For one thing, the present model features an ultraviolet Landau pole $\Lambda_{\rm L}$. Strictly speaking, this prevents us from taking the limit $\Lambda_{\rm uv}\to\infty$. However, at sufficiently weak coupling, the scale $\Lambda_L$ is so large that one can consider values of the cut off that are large with respect to the physical scales, without hitting the Landau pole (see the discussion after Eq.~(\ref{Landau}) in \ref{sec:twoloop}). It makes then sense to ask whether the flow equations are essentially insensitive to the cut-off in this range of cut-off values. This is what will be meant in what follows by ``checking the finiteness of the flow equations'', although we shall still use the short-hand notation $\Lambda_{\rm uv}\to\infty$ in place of $\Lambda_L\gg\Lambda_{\rm uv}\gg\Lambda_{\rm phys}$. 

The cut-off insensitivity of the flow equations will be granted in part by the presence of $\partial_\kappa R_\kappa$ which regulates some of the loop integrals in the UV. Some other loops, however, will not feature this natural regulator and proving their insensitivity to the cut-off will require a finer analysis based on power counting. We shall assume that, in estimating superficial degrees of divergence,  $G_\kappa(p)$ and $\Gamma^{(4)}_{\Phi_L,\kappa}(p,q)$ count respectively as $-2$ and $0$. This natural assumption follows from the application of Weinberg's theorem on the corresponding perturbative contributions (see \ref{sec:dI}). Such perturbative power counting needs to be taken with a small pinch of salt, however, as the resummations entailed by the $\Phi$-derivable approximations can potentially alter the asymptotic behavior of the $n$-point functions. In the considered cut-off range, we expect these modifications to be marginal, such that a superficial degree of divergence that is found to be negative based on the perturbative counting ($\delta\leq -1$), will remain so once resummation effects are considered. In other words, the perturbative power counting is sufficient to conclude on the finiteness of the flow equations.\footnote{We cannot exclude a situation where the resummation modifies the asymptotic behavior of the propagator in such a way that certain loops that are marginally divergent with a perturbative counting ($\delta=0$) become marginally convergent with the non-perturbative counting. In fact, this assumption has been implicitly made in \cite{Carrington:2016zlc} when checking the finiteness of certain loop integrals. Here, we shall not make this assumption because, first, it is not clear whether the resummation tends to improve the convergence (this could in fact depend on the considered truncation), and, second, because the obtained convergence is possibly quite slow and therefore does not necessarily ensure a strong cut off insensitivity before hitting the Landau pole. In fact, as stated above, such considerations are not necessary here because as we shall see, the perturbative counting is sufficient.}

With all these precautions taken, let us now analyze the finiteness of the flow equations. The flow of the two-point function is given by  Eq.~(\ref{eq:g2k}), and is identical to that obtained in the 1PI approach:  it contains indeed a derivative of the regulator in its right-hand side, so that the flow is finite if $\Gamma^{(4)}_\kappa(q,p)$ is finite. Thus, in this equation, we can ignore the cutoff $\Lambda_{\rm uv}$ that controls the loop momentum without affecting the solution in any significant way.

The situation is a priori different for the 
 flow equation for $\Gamma^{(4)}_\kappa(q,p)$, given by Eq.~\eq{eq:flowGamma4gen}. The loop integrals in the right-hand side of this equation are not cut off by  a factor $\pk R_\kp(q)$. Furthermore it contains convolutions of $n$-point functions whose external momenta are not limited to the range $p_i\lesssim \Lambda_{\rm phys}$. Thus, even though we are a priori interested in $\Gamma^{(4)}_\kappa(p,q)$ for $p$ and $q$  in the physical range,  $p,q\lesssim \Lambda_{\rm phys}$, in order to calculate the integrals in the right-hand side of Eq.~\eq{eq:flowGamma4gen}, we need the function $\Gamma^{(4)}_\kappa(p,q)$ with at least one of its arguments allowed to run freely to infinite values. The same remark applies to the irreducible kernel ${\cal I}_\kappa(p,q)$, or to the self-energy $\Sigma_\kappa(r)$ that enters the propagators in Eq.~\eq{eq:flowGamma4gen}. We need therefore to examine more closely whether Eq.~\eq{eq:flowGamma4gen} remains finite as $\Lambda_{\rm uv}\to\infty$, by using power counting.
 
Let us then consider the first integral in the right-hand side of Eq.~(\ref{eq:flowGamma4gen}), which we rewrite here for convenience
\beq\label{eq:int1}
\int_r\Gamma^{(4)}_{\Phi_L,\kappa}(p,r)\,G_\kappa(r)\,\partial_\kappa G_\kappa(r)\,\Gamma^{(4)}_{\Phi_L,\kappa}(r,q)\,.
\eeq 
To analyze the convergence of this integral, we need to determine how much $\partial_\kappa G_\kappa$ contributes to the power counting. This is easily done by combining the relation $G_\kappa^{-1}(r)=\Gamma^{(2)}_{L-1,\kappa}(r)+R_\kappa(r)$ and  Eq.~(\ref{eq:g2k}), to obtain 
\beq\label{eq:dGa2}
\partial_\kappa G_\kappa(r)=-\partial_\kappa R_\kappa(r)\,G_\kappa^2(r)+\frac{1}{2}\int_s \partial_\kappa R_\kappa(s)\,G^2_\kappa(s)\,\Gamma^{(4)}_{\Phi_L,\kappa}(s,r)\,G^2_\kappa(r)\,.
\eeq
Since $\partial_\kappa R_\kappa(r)$ is strongly suppressed at large $r$, the dominant asymptotic behavior of $\partial_\kappa G_\kappa (r)$ is given by the second term in Eq.~(\ref{eq:dGa2}), which is a finite integral. It follows that $\partial_\kappa G_\kappa(r)$ contributes as $G^2_\kappa(r)$, i.e. as $-4$ to the power counting. The superficial degree of divergence of the integral (\ref{eq:int1}) is then $4+2\times 0-2-4=-2$, where we counted $4$ for the integration measure, $2\times 0$ for the two four-point functions, $-2$ for the propagator, and $-4$ for the flow of the propagator. The integral is then finite. As for the other integrals in Eq.~\eq{eq:flowGamma4gen}, they are finite thanks to the fact that $\partial_\kappa {\cal I}_\kappa(p,q)$ contributes as $-2$ to the power counting rules. This property is intimately related to the s-channel two-particle irreducibility of ${\cal I}_\kappa(p,q)$, as we discuss in \ref{sec:dI}. The two one-loop integrals involving $\partial_\kappa {\cal I}_\kappa$ in Eq.~\eq{eq:flowGamma4gen} have a superficial degree of divergence equal to $4+0-2\times 2-2=-2$, where we counted $4$ for the integration measure, $0$ for the four-point function, $2\times (-2)$ for the two propagators and $-2$ for $\partial_\kappa {\cal I}_\kappa$. The same goes for the two-loop integral in Eq.~\eq{eq:flowGamma4gen} since both its sub-loop integrals and the two-loop integral itself have superficial degrees of divergence equal to $-2$. One concludes therefore that all the integrals involved in Eq.~\eq{eq:flowGamma4gen} are finite by power counting.   

Equations (\ref{eq:dec_great}) and (\ref{eq:Ggreat}) are easier to handle. Because they involve diagrams with no explicit four-point subdiagrams (assuming the $\Gamma_L^{(4)}$'s to be finite), the only possible divergence is an overall divergence. However, it is easily seen that the superficial degree of divergence of any integral appearing in $\partial_\kappa {\cal I}_\kp$ is $\delta=-2$. This is because the diagrams of $\partial_\kappa {\cal I}_\kappa$ are nothing but those of ${\cal I}_\kappa$ with one of the propagators $G_\kappa$ replaced by $\partial_\kappa G_\kappa$. This implies that $\delta_{\partial{\cal I}}=\delta_I-4+2=0-2=-2$, where we used that $\delta_I=0$ and that $\delta_{\partial_\kappa G_\kappa}=-4$. The same argument applies to $\partial_\kappa \Gamma^{(4)}_{L,\kappa}$.

One conclude that all the integrals that are involved in the flow equations for $\Phi$-derivable approximations are finite. We may therefore safely eliminate the ultraviolet cutoff. 

\subsection{Initial conditions}
Starting from the $L$-loop $\Phi$-derivable approximation in the bare theory, we have derived a system of coupled finite flow equations for $\Gamma^{(2)}_{L-1,\kappa}$, $\Gamma^{(4)}_{\Phi_L,\kp}$ and the $\Gamma^{(4)}_{L',\kp}$ (for $0\leq L'\leq L-3$), that make no reference to the bare parameters. These equations can be written in the same concise form as in Sect.~\ref{sec:summ}:
\beq\label{flow2ptfctb}
\pk\Gamma^{(2)}_{L-1,\kappa}(p) = {\cal F}^{(2)}_{L-1,\kappa}\big[\Gamma^{(2)}_{L-1,\kappa},\Gamma^{(4)}_{\Phi_L,\kappa}\big](p).    
\eeq
\beq\label{flow4Phib}
\pk\Gamma^{(4)}_{\Phi_L,\kappa}(p,q)= {\cal F}^{(4)}_{\Phi_L,\kappa}\big[\Gamma^{(2)}_{L-1,\kappa},\Gamma^{(4)}_{\Phi_L,\kappa},\Gamma^{(4)}_{L-3,\kappa},\dots,\Gamma^{(4)}_{L=0,\kappa}\big](p,q)\eeq
\beq\label{flowGamma4Lb}
\pk \Gamma^{(4)}_{L',\kappa}(p_i)={\cal F}^{(4)}_{L',\kappa}\big[\Gamma^{(2)}_{L-1,\kappa},\Gamma^{(4)}_{L-1,\kappa},\dots,\Gamma^{(4)}_{L=0,\kappa}\big](p_i)\quad (0\leq L'\leq L-3).
\eeq
These equations have the same form as Eqs.~(\ref{flow2ptfct})-(\ref{flowGamma4L}). The main difference is that here, thanks to the use of the $\{\_\}_{L}$ expansion operator, the flows do not make any reference to the bare parameters. Moreover, since the flows are finite, and we restrict $\kappa$ to values smaller than $\Lambda\ll \Lambda_{\rm uv}$, the ultraviolet cuttoff $\Lambda_{\rm uv}$ does not play any major role and can be dropped.

One can write the solution of these equations generically as in Eq.~(\ref{eq:71})
\beq\label{flowLambda}
\Gamma_\kappa^{(n)}(p_i)=\Gamma_\Lambda^{(n)}(p_i)+ \int_\Lambda^\kappa \rmd\kappa' {\cal F}^{(n)}_{\kappa'}(p_i).
\eeq
Assuming the flow to be initialized at the scale $\Lambda\gg \Lambda_{\rm phys}$, the  initial conditions, that is the values of the $\Gamma_\Lambda^{(n)}(p_i)$'s, are essentially constant, with only a very weak dependence on the momenta as discussed  earlier (see Eqs.~(\ref{eq:init3})). They can be chosen in the form 
\beq\label{eq:init_ren2}
\Gamma^{(2)}_{L-1,\kappa=\Lambda}(p)\sim  m^2_\Lambda+Z_\Lambda \,p^2\,, \quad \Gamma^{(4)}_{\Phi_L,\kappa=\Lambda}(p,q)\sim \lambda_\Lambda\,, \quad \Gamma^{(4)}_{L',\kp=\Lambda}(p,q,r)=\lambda_{L',\Lambda}\,, \quad (0\leq L'\leq L-3)\,,
\eeq
for $\Lambda\gg \Lambda_{\rm phys}$. The need for different initial conditions for the $\Gamma^{(4)}_{L,\kp}$'s, which also differ from the initial condition for $\Gamma^{(4)}_{\Phi_L,\kp}$, reflects the fact that these are different approximations to the exact four-point function. In particular even though $\Gamma^{(4)}_{L,\kp}$ and $\Gamma^{(4)}_{\Phi_L,\kp}$ should coincide in the exact case (correspoding to $L\to\infty$), they have no reason to coincide for any finite $L$. The best we can do is to adjust the initial conditions such that these functions coincide at a given value of $\kappa$ and a given value of the external momenta, see below. Contact with the original formulation of $\Phi$-derivable approximations is made by fixing the values of the  parameters  $Z_\Lambda$, $m_\Lambda$ and $\lambda_\Lambda$  via the ``renormalization conditions''
\beq\label{eq:cond}
\Gamma^{(2)}_{L-1,\kappa=0}(p=0)=m^2\,, \quad \left.\frac{\rmd \Gamma^{(2)}_{L-1,\kappa=0}}{\rmd p^2}\right|_{p=0}=1\,, \quad \Gamma^{(4)}_{\Phi_L,\kappa=0}(p_i=0)=\lambda\,,
\eeq
whereas the $\lambda_{L,\Lambda}$'s are fixed via the ``consistency'' conditions\footnote{Note that these consistency conditions have nothing to do with those introduced in \cite{Carrington:2017lry}. Our choice follows the general strategy described in \cite{Berges:2005hc}.}
\beq\label{eq:cons}
\Gamma^{(4)}_{L',\kappa= 0}(p_i=0)=\Gamma^{(4)}_{\Phi_L,\kappa= 0}(p_i=0)\,, \quad (0\leq L'\leq L-3)\,.
\eeq
This last condition ensures that $\Gamma^{(4)}_{L,\kp}$ and $\Gamma^{(4)}_{\Phi_L,\kappa}$ coincide as $L\to \infty$, a condition that is not necessarily met if one fixes the finite parts of the $\lambda_{L,\Lambda}$'s in an arbitrary way.\footnote{In a sense, consistency conditions are implicitly used also in perturbative calculations where, for instance, the four-point function at any loop order is required to equal $\lambda$ at the renormalization point.}

A convenient feature that results from this choice of renormalization conditions, is that the renormalized solution for $\Gamma^{(4)}_{L,\kp}$ is a polynomial of order $L+1$ in the renormalized coupling $\lambda$, whose coefficients do not depend on $L$. That is
\beq\label{eq:polynomial}
\Gamma^{(4)}_{L,\kp}=\lambda-\lambda^2\tilde\Gamma^{(4,1)}_\kp+\dots+\lambda^{L+1}(-)^L\tilde\Gamma^{(4,L)}\,.
\eeq
This results is established in \ref{sec:rsexp} and will be  exploited in the next section. 
Since the initial conditions $\lambda_{L,\Lambda}$ are nothing but the asymptotic form of the $\Gamma^{(4)}_{L,\kp=\Lambda}$'s for $\Lambda\gg p_i$, it follows that the $\lambda_{L,\Lambda}$'s have a similar polynomial structure.\\

In summary, we have shown in this section that the flow equations for $\Phi$-derivable approximations, that are summarized  in Set.~\ref{sec:summ}, enjoy essentially the same properties as the usual flow equations for the 1PI $n$-point functions. In particular, they are finite, and can be made independent of the bare parameters. The renormalization process can then be carried out in very much the same way as for the 1PI flow equations. The strength of this approach resides in that it gives direct access to the renormalized $n$-point functions, without having to deal with the ultraviolet divergences and the counterterms that are omnipresent in the diagrammatic approach, the topic of the next section. 

The derivation of the flow equations in Sect.~\ref{Phitruncations} was initially motivated by the search for a truncation of the 1PI equations at the level of the four-point function. The obvious  four-point function that appears in this context is that which solves the Bethe-Salpeter equation. However, as was discussed at length in Sect.~\ref{sec:flowI}, other four-point functions naturally appear when calculating the flow of the irreducible kernel of the BS equation. As we saw in the present section, these functions $\Gamma^{(4)}_L$ play an important role in the renormalization, and they need to be properly initialized. We may be concerned by the fact that the number of parameters that need to be specified to fix the initial conditions seem to exceed the usual number of  available parameters in a renormalizable field theory, and furthermore this number grows with the loop order of the considered approximation. However, as we have seen, this is not the case, since the excess number of parameters is balanced by an equal number of consistency conditions. As we shall see in the next section,  the very same issue arises when we examine renormalization from the diagrammatic point of view.

\section{Renormalized $\Phi$-derivable approximations from the diagrammatic approach}\label{sec:2PIren}
The approach based on flow equations, which was shown in the previous section  to be so effective in the calculation of  renormalized $n$-point functions, proves also to be quite helpful in analyzing the renormalization of $\Phi$-derivable approximations from the diagrammatic point of view.  
In the previous section, we wrote the flow equations in the generic form 
\beq\label{flowdiagrams}
\Gamma_\kappa^{(n)}(p_i)=\Gamma_\Lambda^{(n)}(p_i)+ \int_\Lambda^\kappa \rmd\kappa' \left.{\cal F}^{(n)}_{\kappa'}(p_i)\right|_{\Lambda_{\rm uv}}.
\eeq
We argued that since the flows ${\cal F}^{(n)}$ are finite, and $\Lambda$ is also kept finite, we could ignore $\Lambda_{\rm uv}$. In this section, the same flow equations will be used, but we shall let   $\Lambda\to\infty$ (as in Sect.~\ref{sec:flowI}), keeping $\Lambda_{\rm uv}$ fixed. We shall then examine the divergences that arise when $\Lambda_{\rm uv}\to \infty$ and see how these can  be absorbed in counterterms present in the initial conditions $\Gamma_\Lambda^{(n)}(p_i)$. Before we see how this works in detail, it is useful to recall how the renormalization problem is formulated in the diagrammatic approach of $\Phi$-derivable approximations.

\subsection{Diagrammatic renormalization of $\Phi$-derivable approximations: setting-up the problem}\label{renormdiag}
 Following the standard techniques of perturbative renormalization, one rewrites the action (\ref{eactON}) in terms of a rescaled field, $\varphi_{\rm b}= Z^{1/2}\varphi$,
\beq
S[\varphi] =\int \rmd^{d}x\,\left\lbrace{ \frac{Z}{2}}\left(\del\varphi(x)\right)^2  + \frac{m^2+\delta m^2}{2} \, \varphi^2(x) + \frac{\lambda+\delta\lambda}{4!}\,\varphi^4(x) \right\rbrace,
\eeq 
where 
\beq
Zm^2_{\rm b}\equiv m^2+\delta m^2,\quad Z^2\lambda_{\rm b}\equiv\lambda+\delta\lambda ,\quad Z=1+\delta Z,
\eeq 
are conventionally split into  renormalized parameters $\lambda$, $m$, and counterterms $\delta\lambda$, $\delta m^2$ and $\delta Z$.  To the rescaling of the field corresponds a rescaling of the $n$-point functions: $\Gamma^{(n)}_{\rm b}=Z^{-n/2}\Gamma^{(n)}$. In particular, the propagator rescales as $G_{\rm b}= ZG$ and, to within an inessential additive constant term, the 2PI effective action (\ref{eq:2PIpot}) becomes
\beq
\Gamma[G]=\frac{1}{2} \int_p \log G^{-1}(p)+\frac{1}{2} \int_p \,ZG_0^{-1}(p)G(p) +\Phi[G]\,,
\eeq
with $ZG_0^{-1}(p)=Zp^2+m^2+\delta m^2$. The functional $\Phi[G]$ is made of the same 2PI diagrams as before but with the coupling $\lambda_{\rm b}$ replaced by $\lambda+\delta\lambda$.\footnote{The reason is quite simple: a given diagram of $\Phi[G]$ has no external leg, so the number $I$ of is internal lines is twice the number $V$ of its vertices. The rescaling of the propagator, $G_{\rm b}= Z G$, generates a factor $Z^I=Z^{2V}$ that can be combined with the prefactor  $\lambda_{\rm b}^V$ into $Z^{2V}\lambda_{\rm b}^V=(\lambda+\delta\lambda)^V$.} The corresponding gap equation is now
\beq\label{Dysonb}
G^{-1}=ZG_0^{-1}+\Sigma[G],
\eeq
where the self-energy functional $\Sigma[G]$ is related to  $\Phi[G]$ as before, see Eq.~(\ref{PhiPi}).\footnote{It is sometimes convenient to reabsorb the one-loop term $\int (\delta Zp^2+\delta m^2)G(p)/2$ in the definition of $\Phi[G]$, in which case the 2PI effective action and the gap equation can be written with $Z=1$ and $G_0^{-1}=p^2+m^2$. The relation between the self-energy functional and $\Phi[G]$ remains unchanged.} To take a definite example, at $4$-loop order in $\Phi$, the gap equation reads
\beq\label{eq:gap4}
G^{-1}(p) &=& Zp^2+m^2+\delta m^2+\frac{\lambda+\delta\lambda}{2}\int_q G(q)-\frac{(\lambda+\delta\lambda)^2}{6}\int_q\int_r G(q)G(r)G(r+q+p)\nn
&+& \frac{(\lambda+\delta\lambda)^3}{4}\int_q\int_r\int_k\, G(q)G(r) G(k)G(r+q+p)G(k+q+p)\,.
\eeq

If one were doing perturbation theory (e.g. calculate contributions to the inverse propagator from the equation above with, in the right-hand side, $G$ substituted by $G_0$) one would follow the standard route. The counterterms $\delta Z$, $\delta m^2$, $\delta \lambda$ contain the parts of the bare parameters that diverge as $\Lambda_{\rm uv}\to\infty$.\footnote{Again, the notation $\Lambda_{\rm uv}\to\infty$ is to be understood as $\Lambda_{\rm Landau}\gg\Lambda_{\rm uv}\gg\Lambda_{\rm phys}$.} They are  functions of $m$, $\lambda$ and the cutoff $\Lambda_{\rm uv}$, and are determined as formal series in  powers of $\lambda$, with their divergent part chosen so as to absorb the ultraviolet divergences order by order in the expansion in $\lambda$.
Their finite parts are fixed by the same renormalization conditions as in Eq.~(\ref{eq:cond}), namely\footnote{We consider only the massive case in this paper.}
\beq\label{renormcond}
\left. \frac{\rmd \Gamma^{(2)}(p)}{\rmd p^2}\right|_{p=0}=1, \qquad \Gamma^{(2)}(p=0)=m^2,
\qquad
\Gamma^{(4)}(0,0,0)=\lambda.
\eeq

This method, rooted in perturbation theory, can be extended to the $\Phi$-derivable approximations \cite{vanHees:2001ik,Blaizot:2003an,Berges:2005hc}. For the specific 4-loop example considered above, this implies solving the full non-linear gap equation, Eq.~(\ref{eq:gap4}), with the self-consistent propagator $G$ in its right-hand side. The main difficulty that one has to face in this approach is the proper identification and treatment of the divergences and subdivergences in the gap equation. In particular, the various occurrences of the coupling counterterm $\delta\lambda$, because they absorb distinct subdivergences, need to be treated on different footings. This is where the  flow equations bring a significant clarification by providing a precise map of how these divergences are  distributed among the relevant  $n$-point functions and how they can be absorbed in  counterterms. This is what we discuss in the next two subsections.

\subsection{Implicit renormalization from the flow}\label{sec:formal}
As was shown in Sects.~\ref{Phitruncations} and \ref{sec:summ}, the bare diagrammatic expansion is a solution to the flow equations (\ref{flow2ptfct}-\ref{flowGamma4L}) with initial conditions (\ref{eq:init}) taken at a scale $\Lambda$ well above the ultraviolet cutoff. Taking $\Lambda\to\infty$, one can write the solutions of these equations formally as follows
\beq
\Gamma^{(2)}_{L-1,\kappa=0}(p) & = & p^2+m^2_{\rm b}+\int_\infty^0 \rmd\kappa\,{\cal F}^{(2)}_{L-1,\kappa}\big[\Gamma^{(2)}_{L-1,\kappa},\Gamma^{(4)}_{\Phi_L,\kappa}\big](p)\Big|_{\Lambda_{\rm uv}}\,,\\
\Gamma^{(4)}_{\Phi_L,\kappa=0}(p,q) & = & \lambda_{\rm b}+\int_\infty^0 \rmd\kappa\,{\cal F}^{(4)}_{\Phi_L,\kappa}\big[\Gamma^{(2)}_{L-1,\kappa},\Gamma^{(4)}_{\Phi_L,\kappa},\Gamma^{(4)}_{L-3,\kappa},\dots,\Gamma^{(4)}_{L=0,\kappa}\big](p,q)\Big|_{\Lambda_{\rm uv}}\,,\\
\Gamma^{(4)}_{L',\kappa=0}(p_i) & = & \lambda_{\rm b}+\int_\infty^0 \rmd\kappa\,{\cal F}^{(4)}_{L',\kappa}\big[\Gamma^{(2)}_{L-1,\kappa},\Gamma^{(4)}_{L-1,\kappa},\dots,\Gamma^{(4)}_{L=0,\kappa}\big](p_i)\Big|_{\Lambda_{\rm uv}}\,, (0\leq L'\leq L-3)\,.
\eeq
In the above equations, the notation $|_{\Lambda_{\rm uv}}$ reminds one  that the cutoff is kept in the integrals entering the flows ${\cal F}^{(n)}$, and the $n$-point functions there should be viewed as the bare $n$-point functions. 

At this point it is easy to make contact with the renormalized theory. To do so, we first move to the rescaled theory, which is easily done since the flow equations are invariant under this operation, as is easily verified. Only the initial conditions are affected. One obtains then, for the renormalized functions,
\beq
\Gamma^{(2)}_{L-1,\kappa=0}(p) & = & Zp^2+m^2+\delta m^2+\int_\infty^0 \rmd\kappa\,{\cal F}^{(2)}_{L-1,\kappa}\big[\Gamma^{(2)}_{L-1,\kappa},\Gamma^{(4)}_{\Phi_L,\kappa}\big](p)\Big|_{\Lambda_{\rm uv}}\,,\label{eq:0111}\\
\Gamma^{(4)}_{\Phi_L,\kappa=0}(p,q) & = & \lambda+\delta\lambda+\int_\infty^0 \rmd\kappa\,{\cal F}^{(4)}_{\Phi_L,\kappa}\big[\Gamma^{(2)}_{L-1,\kappa},\Gamma^{(4)}_{\Phi_L,\kappa},\Gamma^{(4)}_{L-3,\kappa},\dots,\Gamma^{(4)}_{L=0,\kappa}\big](p,q)\Big|_{\Lambda_{\rm uv}}\,,\label{eq:0222}\\
\Gamma^{(4)}_{L',\kappa=0}(p_i) & = & \lambda+\delta\lambda+\int_\infty^0 \rmd\kappa\,{\cal F}^{(4)}_{L',\kappa}\big[\Gamma^{(2)}_{L-1,\kappa},\Gamma^{(4)}_{L-1,\kappa},\dots,\Gamma^{(4)}_{L=0,\kappa}\big](p_i)\Big|_{\Lambda_{\rm uv}}\!\!\,, (0\leq L'\leq L-3)\,.\label{eq:0333}
\eeq
A first important observation is that these equations  clearly separate the subdivergences and the overall divergences of a given $n$-point function. Indeed, since overall $\Lambda_{\rm uv}$-divergences are associated to the regime where all loop momenta of a given diagram contributing to $\Gamma^{(n)}$ grow large, they cannot depend on $\kappa$. It follows that the ${\cal F}^{(n)}|_{\Lambda_{\rm uv}}$'s in the above equations, which are nothing but the $\kappa$-derivatives of the corresponding $n$-point functions, are only sensitive to the subdivergences of these $n$-point functions, while the overall divergences occur due to the explicit unbounded $\kappa$-integrals in Eqs.~(\ref{eq:0111})-(\ref{eq:0333}). A second observation is that, because the explicit loops that enter the ${\cal F}^{(n)}_\kappa|_{\Lambda_{\rm uv}}$'s are convergent for $\Lambda_{\rm uv}\gg\kappa$, the subdivergences never appear explicitly but only  as hidden divergences (including overall divergences) of the various $n$-point functions that enter as arguments of the ${\cal F}^{(n)}_\kappa|_{\Lambda_{\rm uv}}$'s. It results from these observations that overall divergences encompass actually all the possible divergences that can appear in the $n$-point functions (a well known feature, also in perturbation theory). It is a strength of the flow formulation to remove all divergences at once by removing only overall divergences, since the counterterms never appear explicitly in ${\cal F}^{(n)}_\kappa|_{\Lambda_{\rm uv}}$, but only as tree-level contributions in Eqs.~(\ref{eq:0111})-(\ref{eq:0333}) via the initial conditions. Thus, by fixing  these counterterms in order to satisfy renormalization conditions at $\smash{\kappa=0}$, one handles simultaneously  both the overall divergences and the subdivergences present in the $n$-point functions. This consitutes in fact an essential ingredient towards a formal proof of renormalization in the diagrammatic formulation.

The previous argument needs to be slightly amended to take into account the fact that $\Gamma^{(4)}_{\Phi_L}$ and the $\Gamma^{(4)}_L$'s are different approximations to the four-point function, and there is no reason for their overall divergences  to be absorbed by the same counterterm $\delta\lambda$. Such an observation was already made in the previous section (see Eq.~(\ref{eq:init_ren2})). To account for this possibility, we modify the initial conditions and set
\beq
\Gamma^{(2)}_{L-1,\kappa=0}(p) & = & Zp^2+m^2+\delta m^2+\int_\infty^0 \rmd\kappa\,{\cal F}^{(2)}_{L-1,\kappa}\big[\Gamma^{(2)}_{L-1,\kappa},\Gamma^{(4)}_{\Phi_L,\kappa}\big](p)\Big|_{\Lambda_{\rm uv}}\,,\label{eq:1111}\\
\Gamma^{(4)}_{\Phi_L,\kappa=0}(p,q) & = & \lambda+\delta\tilde\lambda+\int_\infty^0 \rmd\kappa\,{\cal F}^{(4)}_{\Phi_L,\kappa}\big[\Gamma^{(2)}_{L-1,\kappa},\Gamma^{(4)}_{\Phi_L,\kappa},\Gamma^{(4)}_{L-3,\kappa},\dots,\Gamma^{(4)}_{L=0,\kappa}\big](p,q)\Big|_{\Lambda_{\rm uv}}\,,\label{eq:2222}\\
\Gamma^{(4)}_{L',\kappa=0}(p_i) & = & \lambda+\delta\lambda_{L'}+\int_\infty^0 \rmd\kappa\,{\cal F}^{(4)}_{L',\kappa}\big[\Gamma^{(2)}_{L-1,\kappa},\Gamma^{(4)}_{L-1,\kappa},\dots,\Gamma^{(4)}_{L=0,\kappa}\big](p_i)\Big|_{\Lambda_{\rm uv}}\!\!\,, (0\leq L'\leq L-3)\,.\nonumber\\\label{eq:3333}
\eeq
As shown by these equations, the counterterms determine the $n$-point functions at the large initial scale $\Lambda\to\infty$. Upon imposing the same renormalization conditions as in the previous section (see Eqs.~(\ref{renormcond})), we extract the following formal expressions:
\beq
\delta m^2 & = & -\int_\infty^0 \rmd\kappa\,{\cal F}^{(2)}_{L-1,\kappa}\big[\Gamma^{(2)}_{L-1,\kappa},\Gamma^{(4)}_{\Phi_L,\kappa}\big](0)\Big|_{\Lambda_{\rm uv}}\,,\label{eq:ct1}\\
\delta Z & = & -\int_\infty^0 \rmd\kappa\,\left.\frac{\rmd}{\rmd p^2}{\cal F}^{(2)}_{L-1,\kappa}\big[\Gamma^{(2)}_{L-1,\kappa},\Gamma^{(4)}_{\Phi_L,\kappa}\big](p)\Big|_{\Lambda_{\rm uv}}\right|_{p=0}\,,\label{eq:ct2}\\
\delta\tilde\lambda & = & -\int_\infty^0 \rmd\kappa\,{\cal F}^{(4)}_{\Phi_L,\kappa}\big[\Gamma^{(2)}_{L-1,\kappa},\Gamma^{(4)}_{\Phi_L,\kappa},\Gamma^{(4)}_{L-3,\kappa},\dots,\Gamma^{(4)}_{L=0,\kappa}\big](0,0)\Big|_{\Lambda_{\rm uv}}\,,\label{eq:ct3}\\
\delta\lambda_{L'} & = & -\int_\infty^0 \rmd\kappa\,{\cal F}^{(4)}_{L',\kappa}\big[\Gamma^{(2)}_{L-1,\kappa},\Gamma^{(4)}_{L-1,\kappa},\dots,\Gamma^{(4)}_{L=0,\kappa}\big](p_i=0)\Big|_{\Lambda_{\rm uv}}\!\!\,, (0\leq L'\leq L-3)\,.\label{eq:ct4}
\eeq
 While these expressions may  be of limited practical use, they summarize the general structure of the counterterms that are needed to  cope with all the divergences of the $L$-loop $\Phi$-derivable approximation. They exhibit the basic $n$-point functions on which to impose the renormalization conditions necessary to the complete determination of these counterterms, namely $\Gamma^{(2)}_{L-1}$,  $\Gamma^{(4)}_{\Phi_L}$ and the $\Gamma^{(4)}_{L'}$'s. 
In the next subsection, we shall turn these formal considerations into a more explicit construction by integrating exactly the flow equations (\ref{eq:1111})-(\ref{eq:3333}) in terms of the diagrams. This will also allow us to derive diagrammatic expressions for the counterterms (\ref{eq:ct1})-(\ref{eq:ct4}), thus providing an explicit and synthetic procedure for the renormalization of $\Phi$-derivable approximations in their diagrammatic formulation.

\subsection{Diagrammatic form of the renormalized solution}
We begin with an observation concerning the general structure of the solution, and in particular on the way it depends on the parameters of the initial conditions. Let us consider the original 2PI effective action, Eq.~(\ref{eq:2PIpot}
),  truncated at order $L$-loop,  and write it as follows 
\beq\label{eq:mod2PI}
\Gamma_L[G]=\frac{1}{2} \int_p \log G^{-1}(p)+\frac{1}{2} \int_p \,(z p^2+y)\,G(p) +\Phi_L[G]\,,
\eeq
with
\beq\label{eq:modPhi}
\Phi_L[G]=u\,\Phi^{(2)}[G]-v^2\,\Phi^{(3)}[G]+\dots+v^{L-1} (-)^L\Phi^{(L)}\,.
\eeq
Here, $u$, $v$, $y$ and $z$ are arbitrary constants and $\Phi^{(l)}[G]$ is built with the same cut  off integrals as in Eq.~(\ref{eq:Phi_exp}). It is then easily verified, by following the same steps as in Sects.~\ref{Phitruncations} and \ref{sec:flowI}, that $\Gamma^{(2)}_{L-1,\kp}\equiv zp^2+y+\Sigma_{L-1,\kappa}$ and $\Gamma^{(4)}_{\Phi_L,\kappa}$, defined from the LW functional $\Phi_L[G_\kappa]$ just given,  obey the same flow equations as those derived in  Sects.~\ref{Phitruncations} and \ref{sec:flowI}. The $\Gamma^{(4)}_{L,\kp}$'s  now involve $v$ as coupling  constant rather than $\lambda_{\rm b}$, and  correspondingly (see Eq.~(\ref{eq:bare_exp}))
\beq\label{eq:bare_exp_v}
\Gamma^{(4)}_{L,\kp}=v -v^2\Gamma^{(4,1)}_\kp +\cdots +v^{L+1} (-)^L\Gamma^{(4,L)}_\kp.
\eeq
This result is in fact trivial: first, $y$ and $z$ disappear entirely from the flow equations once these are written in terms of $\Gamma^{(2)}_{L-1,\kp}$ rather than  $\Sigma_{L-1,\kappa}$ (see Eq.~(\ref{eq:g2k})); second, the term involving the coefficient $u$ in Eq.~(\ref{eq:modPhi}) drops entirely from the flow equation since it gives a constant contribution to ${\cal I}_{L-2,\kappa}$ which disappears in $\del_\kappa{\cal I}_{L-2,\kappa}$; finally, $v$ is a mere renaming of the bare coupling. 

Clearly, the solution to the flow equations that corresponds to Eqs.~(\ref{eq:modPhi}) and (\ref{eq:bare_exp_v}) obeys the initial conditions
\beq
\Gamma^{(2)}_{L-1,\kappa=\infty}(p)=zp^2+y\,, \quad \Gamma^{(4)}_{\Phi_L,\kappa=\infty}=u\,, \quad \Gamma^{(4)}_{L,\kappa=\infty}(p)=v\,.
\eeq
This type of initial conditions  accommodates the solution (\ref{eq:0111})-(\ref{eq:0333}), upon the choice $z=Z$, $y=m^2+\delta m^2$, $u=v=\lambda+\delta\lambda$, which corresponds to a simple rescaling of the $n$-point functions. But it  cannot correspond to the renormalized solution (\ref{eq:1111})-(\ref{eq:3333}) because, although $u$ can be chosen independently of $v$, as in Eq.~(\ref{eq:2222}), all the $\Gamma^{(4)}_{L,\kp}$'s are initialized at the same value, independent of $L$.

We know, however, that the different $\delta\lambda_L$'s in Eq.~(\ref{eq:3333}) are finite order truncations of a unique series $\delta\lambda$ in powers of $\lambda$:
\beq\label{eq:pol}
\delta\lambda=\sum_{l\geq 1} \lambda^{l+1}\delta\lambda^{(l)},\qquad \delta\lambda_L=\sum_{1\le l\le L} \lambda^{l+1}\delta\lambda^{(l)}\,.  
\eeq 
This is because the counterterms are essentially nothing but the initial conditions for the renormalized solution at the scale $\Lambda\gg\Lambda_{\rm uv}$. We have seen in the previous section that the renormalized $\Gamma^{(4)}_{L,\kp}$'s are polynomials in $\lambda$ with $L$-independent coefficients (see Eq.~(\ref{eq:polynomial})). It follows that the same property holds for the $\delta\lambda_L$'s, which is precisely what Eq.~(\ref{eq:pol}) states.

Let us then consider the extension of $\Phi[G]$ in Eq.~(\ref{eq:modPhi}) to arbitrary order in $v$, and  define $\Gamma^{(2)}_{\kp}\equiv zp^2+y+\Sigma_{\kappa}$ and $\Gamma^{(4)}_\kp$ from the (exact) gap and BS equations. It is easily seen that these functions satisfy the flow equations (\ref{eq:g2k}) and (\ref{eq:g}), with $\partial_\kappa {\cal I}_\kappa$ given by the exact relation (\ref{eq:dec_great}), while 
\beq\label{eq:bare_exp_w}
\Gamma^{(4)}_{\kp}=v -v^2\Gamma^{(4,1)}_\kp +\cdots +v^{L+1} (-)^L\Gamma^{(4,L)}_\kp+\cdots  
\eeq
also obeys the exact equation (\ref{eq:Ggreat2}). At this point we substitute $v\to \lambda+\delta\lambda$ and expand the exact equations (\ref{eq:dec_great}) and (\ref{eq:Ggreat2}) in a loop expansion in power of $\lambda$, using wherever needed the expansion operator $\{\_\}_{L}$ defined after Eq.~(\ref{eq:exp_fin}), as well as the expansion (\ref{eq:pol}) of $\delta\lambda$. By doing so, one easily verifies that $\Gamma^{(2)}_{L-1,\kp}$ and $\Gamma_{\Phi_L,\kp}$ as derived from
\beq\label{eq:modPhi2}
\Phi_L[G] & = & u\,\Phi^{(2)}[G]+\Big[-(\lambda+\delta\lambda)^2\,\Phi^{(3)}[G]+\dots+(\lambda+\delta\lambda)^{L-1} (-)^L\Phi^{(L)}\Big]_{L}\nonumber\\
& = & u\,\Phi^{(2)}[G]-\big[(\lambda+\delta\lambda)^2\big]_{L-3}\,\Phi^{(3)}[G]+\dots+\big[(\lambda+\delta\lambda)^{L-1}\big]_{L=0} (-)^L\Phi^{(L)}\,,
\eeq
as well as $\Gamma^{(4)}_{L,\kp}$ defined as
\beq
\Gamma^{(4)}_{L,\kp} & = & \Big[\lambda+\delta\lambda -(\lambda+\delta\lambda)^2\Gamma^{(4,1)}_\kp +\cdots +(\lambda+\delta\lambda)^{L+1} (-)^L\Gamma^{(4,L)}_\kp\Big]_{L}\nonumber\\
& = & \big[\lambda+\delta\lambda\big]_L -\big[(\lambda+\delta\lambda)^2\big]_{L-1}\Gamma^{(4,1)}_\kp +\cdots +\big[(\lambda+\delta\lambda)^{L+1}\big]_{L=0} (-)^L\Gamma^{(4,L)}_\kp
\eeq
where $[\_]_{L}$ refers now the loop expansion in powers of $\lambda$, obey the same flow equations as those in Sect.~\ref{sec:FRGren}.

The solution just obtained satisfies now the initial conditions in Eqs.~(\ref{eq:0111})-(\ref{eq:0333}). It corresponds to the following expression for the LW functional
\beq\label{eq:newPhi}
\Phi[G] & = & (\lambda+\delta\tilde\lambda)\,\Phi^{(2)}[G]+\sum_{3\leq l\leq L}\big[(\lambda+\delta\lambda)^{l-1}\big]_{L-l}\,(-)^l\,\Phi^{(l)}[G]\,,
\eeq
where  the counterterms $\delta\tilde\lambda$ and $\delta\lambda_L$ are explicitly indicated. These  are determined from the renormalization conditions imposed respectively on $\Gamma^{(4)}_{\Phi_L}$, as given by the BS equation, and $\Gamma^{(4)}_L$, as given by
\beq\label{eq:63}
\Gamma^{(4)}_{L}=\sum_{0\leq l\leq L}\big[(\lambda+\delta\lambda)^{l+1}\big]_{L-l}\,(-)^l\,\Gamma^{(4,l)}\,.
\eeq

\subsection{Diagrammatic form of the counterterms}
The preceding analysis has taught us how the counterterms $\delta\lambda_L$, $\delta\tilde\lambda$, $\delta m^2$ and $\delta Z$ should be implemented in the diagrammatic formulation of $\Phi$-derivable approximations.  In this subsection we turn to their explicit determination.

\subsubsection{Counterterm $\delta\lambda_L$ and renormalization of the $\Gamma^{(4)}_L$'s}
The renormalized expression (\ref{eq:63}) is similar to that obtained in perturbation theory, the only difference being that the self-consistent propagator $G$ is involved in the calculation of the integrals $\Gamma^{(4,l)}$ instead of the perturbative one, $G_0$. So the renormalization works along the same lines as in perturbation theory, which we shall follow  to determine  the first contributions to $\delta\lambda$. 

Using the expansion (\ref{eq:pol}) of $\delta \lambda$, one can recast Eq.~(\ref{eq:63}) in the form
\beq\label{eq:ren_exp_2}
\Gamma^{(4)}_{L}=\sum_{0\leq l\leq L} \lambda^{l+1}(-)^l\,\tilde\Gamma^{(4,l)},
\eeq
with
\beq
\tilde\Gamma^{(4,0)} & = & 1\,,\\
\tilde\Gamma^{(4,1)} & = & \Gamma^{(4,1)}-\delta\lambda^{(1)}\,,\label{eq:g41}\\
\tilde\Gamma^{(4,2)} & = & \Gamma^{(4,2)}+\delta\lambda^{(2)}-2\delta\lambda^{(1)}\Gamma^{(4,1)}\,,\label{eq:g42}\\
& \dots & \nonumber
\eeq
Higher orders feature the same triangular structure. For a given $\tilde\Gamma^{(4,l)}$, $\delta\lambda^{(l)}$ appears as a tree-level contribution whereas the $\delta\lambda^{(l'<l)}$ multiply lower instances of $\tilde\Gamma^{(4,l)}$. The role of these $\delta\lambda^{(l'<l)}$ is to eliminate subdivergences within $\tilde\Gamma^{(4,l)}$, whereas the remaining overall divergence is absorbed into $\delta\lambda^{(l)}$. The finite parts of the $\delta\lambda^{(l)}$ are fixed by the renormalizations conditions. Here we choose $\tilde\Gamma^{(4)}_L(p_i=0)=\lambda$ for any $L$ (just as one would do in the perturbative loop expansion), which corresponds to $\tilde\Gamma^{(4,l>0)}(p_i)=0$.

For the first non-trivial orders, we have
\beq
\Gamma^{(4,1)}(k,q,p) & = & \frac{1}{2} J(p+q)+(k,q,p)_{\rm cyclic}\,,\label{eq:111}\\
\Gamma^{(4,2)}(k,q,p) & = & \frac{1}{4} J(p+q)^2+\frac{1}{2}\int_r \big[J(p+r)+J(k-r)\big] G(r)G(r+p+q)+(k,q,p)_{\rm cyclic}\,,\label{eq:222}
\eeq
where the one-loop integral $J_\kappa(p)$ is defined in Eq.~(\ref{Jdefp}).
Since $\Gamma^{(4,1)}$ is made of one-loop contributions, there is only an overall divergence in Eq.~(\ref{eq:g41}) which is absorbed in $\delta\lambda^{(1)}$. The chosen renormalization condition fixes  
\beq\label{deltalambda1}
\delta\lambda^{(1)}=\Gamma^{(4,1)}(p_i=0)\,,
\eeq
with $\Gamma^{(4,1)}(p_i=0)=(3/2)J_{\kp=0}(0)$. In Eq.~(\ref{eq:g42}), the term proportional to $\delta\lambda^{(1)}$ eliminates the subdivergence present in $\Gamma^{(4,2)}$. To see that, we set $\bar J(p)=J(p)-J(0)$, and get
\beq
\Gamma^{(4,2)}(k,q,p) & = & \frac{1}{4} \left[\bar J(p+q)^2+6\bar J(p+q) J(0)+5J(0)^2\right]+(k,q,p)_{\rm cyclic}\nonumber\\
& + & \frac{1}{2}\int_r \big[\bar J(p+r)+\bar J(k-r)\big] G(r)G(r+p+q)+(k,q,p)_{\rm cyclic}\,,
\eeq
where the subdivergence  is located in the second term of the first line. By using the expression of $\Gamma^{(4,1)}$ in Eq.~(\ref{eq:111}), and the just determined $\delta\lambda^{(1)}$, we see that the last term in the right-hand side of Eq.~(\ref{eq:222}) cancels this subdivergence. We are then left with an overall divergence that can be absorbed in $\delta\lambda^{(2)}$. The renormalization conditions fixes  
\beq
\delta\lambda^{(2)}=-\Gamma^{(4,2)}(p_i=0)+2\big(\Gamma^{(4,1)}(p_i=0)\big)^2\,.
\eeq 
The procedure can be continued to higher loop orders.

\subsubsection{Counterterm $\delta\tilde\lambda$ and renormalization of $\Gamma^{(4)}_{\Phi_L}$}
The counterterm $\delta\tilde\lambda$ is determined from the renormalization condition $\Gamma^{(4)}_{\Phi_L}(p_i)=\lambda$ which reads
\beq\label{rc}
\lambda={\cal I}_{L-2}(0,0)-\frac{1}{2}\int_r \Gamma^{(4)}_{\Phi_L}(0,r)G^2(r){\cal I}_{L-2}(r,0)\,.
\eeq
This equation should be seen as an equation for $\delta\tilde\lambda$ that appears as a tree-level contribution to ${\cal I}_{L-2}$. To extract $\delta\tilde\lambda$, we use Eq.~(\ref{eq:newPhi})  and write
\beq\label{eq:dec}
{\cal I}_{L=2}=\lambda+\delta\tilde\lambda+\tilde {\cal I}_{L-2}\,,
\eeq
where
\beq
\tilde {\cal I}_{L-2}(q,p)\equiv\sum_{3\leq l\leq L}\big[(\lambda+\delta\lambda)^{l-1}\big]_{L-l}\,(-)^l\,\frac{\delta^2\Phi^{(l)}[G]}{\delta G(q)\delta G(p)}\,.
\eeq
We can now solve for $\delta\tilde\lambda$ in Eq.~(\ref{rc}) and find
\beq
\delta\tilde\lambda=-\frac{\tilde{\cal I}_{L-2}(0,0)-\frac{1}{2}\int_r \Gamma^{(4)}_{\Phi_L}(0,r)G^2(r)(\lambda+\tilde {\cal I}_{L-2}(r,0))}{1-\frac{1}{2}\int_r \Gamma^{(4)}_{\Phi_L}(0,r)G^2(r)}\,.
\eeq
From the general discussion in Sect.~\ref{sec:formal} we know that the  counterterms $\delta\lambda$ and $\delta\tilde\lambda$ that we have just determined allow us to renormalize $\Gamma^{(4)}_{\Phi_L}$. Note that this result was obtained without referring  explicitly to diagrams. It is however interesting to see how the renormalization of $\Gamma^{(4)}_{\Phi_L}$ works by inspecting the resummation  of the relevant diagrams.

Clearly, $\Gamma^{(4)}_{\Phi_L}$ can be viewed as an infinite sum of ladder diagrams with rungs given by ${\cal I}_{L-2}$. These diagrams  are obtained by iterating the BS equation. It is then convenient to classify the four-point divergences in $\Gamma^{(4)}_{\Phi_L}$ into two groups: the four-point subdivergences of a given rung ${\cal I}_{L-2}$, and the rest. Due to the two-particle irreducibility of the rungs, it is easily checked that these other divergences involve an integer number of rungs, including the case of a single rung. Let us now discuss these two classes of divergences separately.

\begin{figure}[h]
\begin{center}
\includegraphics[angle=0,width=10cm]{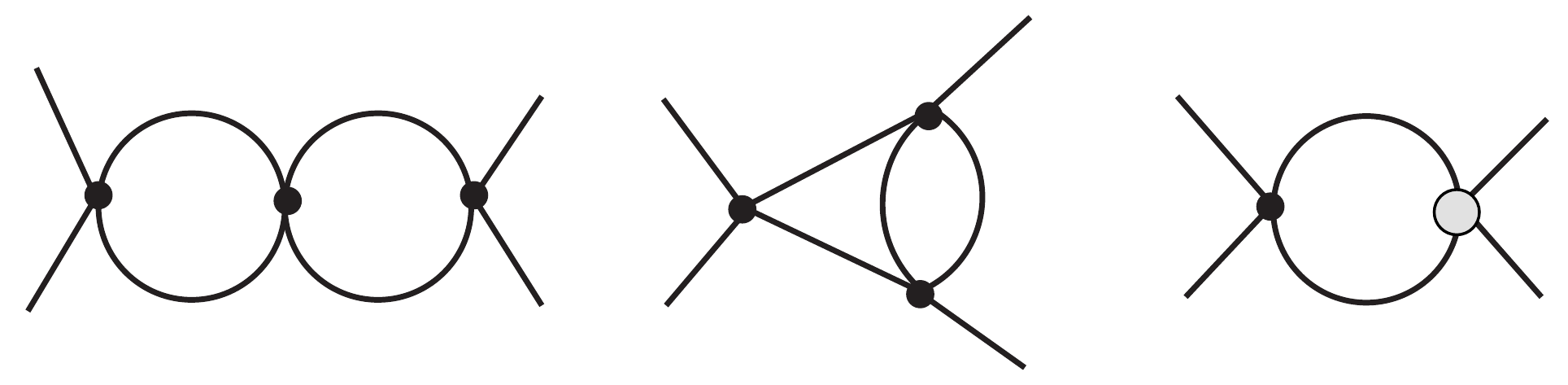}
\caption{The two left diagrams are the two-loop 2PI diagrams that contribute to the kernel ${\cal I}(q,p)$ . The grey blob of the right most diagram is the  counterterm $\delta\lambda^{(1)}$ that renormalizes  $\Gamma^{(4)}_{L=1}$.}\label{fig:diagramscalI}
\end{center}
\end{figure}

We have seen that the subdivergences of ${\cal I}_{L-2}$ are nothing but the divergences of $\Gamma^{(4)}_{L'}$ (for $0\leq L'\leq L-3$). Let us illustrate this in the case of ${\cal I}_2$ which reads
\begin{eqnarray}\label{calI2qp}
{\cal I}_2(q,p) & = & \lambda+\delta\tilde\lambda-(\lambda^2+2\lambda^3\,\delta\lambda^{(1)})\,{\cal I}^{(1)}(q,p)+\lambda^3\,{\cal I}^{(2)}(q,p)\,,
\eeq
with
\beq
{\cal I}^{(1)}(q,p) & = & \frac{1}{2}\left[J(p+q) +J(p-q)\right]\,,\\
{\cal I}^{(2)}(q,p) & = & \frac{1}{4}\left[J^2(p+q) +J^2(p-q)\right]+\int_r J(p+r) G(r)\left[G(r+p+q)+G(r+p-q) \right].
\label{eq:Ika3}
\end{eqnarray}
Diagrams that contribute to ${\cal I}^{(2)}(q,p)$ are displayed in Fig.~\ref{fig:diagramscalI}.
We observe that ${\cal I}^{(1)}(q,p)$ and ${\cal I}^{(2)}(q,p)$ involve exactly the same topologies as $\Gamma^{(4,1)}$ and $\Gamma^{(4,2)}$, the only difference being that a particular configuration of the external momenta has been chosen, and that only two out of the three channels are present in ${\cal I}^{(1)}(q,p)$ and ${\cal I}^{(2)}(q,p)$. Since the analysis of the subdivergences of $\Gamma^{(4,2)}$ was made separately for each channel, it is clear that a similar analysis applies for ${\cal I}^{(2)}(q,p)$. Thus, one verifies that the counterterm $\delta\lambda^{(1)}$ (which renormalizes $\Gamma^{(4,1)}$, see Eq.~(\ref{deltalambda1})), when multiplied by ${\cal I}^{(1)}$ in Eq.~(\ref{calI2qp}),  absorbs  precisely the subdivergence of ${\cal I}_2$, as expected. The same occurs at higher orders.

We now turn to the remaining divergences in $\Gamma^{(4)}_{\Phi_L}$, namely those associated to sets of consecutive rungs. These divergences have the special property of being simultaneously subdivergences  and overall divergences of $\Gamma^{(4)}_{\Phi_L}$. This can be seen already from the flow equations. According to Eq.~(\ref{eq:22}), the subdivergences of $\Gamma^{(4)}_{\Phi_L}$ are the divergences of $\partial_\kappa {\cal I}_{L-2}$ (and hence the subdivergences of ${\cal I}_{L-2}$) and the divergences of $\Gamma^{(4)}_{\Phi_L}$ itself, including overall divergences. Thus, once the subdivergences of ${\cal I}_{L-2}$ have been eliminated, the divergences that remain in $\Gamma^{(4)}_{\Phi_L}$ play both the role of subdivergences and overall divergences. To see how this emerges diagrammatically, take for instance a subdivergence coming from a set of $r'$ consecutive rungs in a ladder made of $r>r'$ rungs. It is clear that this ladder of $r'$ rungs also contributes to $\Gamma^{(4)}_{\Phi_L}$ as a whole diagram. It contributes therefore to an overall divergence.

To show that these divergences can be absorbed in the tree-level contribution $\lambda+\delta\tilde\lambda$ to $\Gamma^{(4)}_{\Phi_L}$, we rescale momentarily the rungs ${\cal I}_{L-2}$ by a parameter $\eta$ (that we shall eventually set to $1$) whose role is to count the number of rungs. We then define the bare rung expansion of $\Gamma^{(4)}_{\Phi_L}$ up tor $R$ rungs as
\beq
\big[\Gamma^{(4)}_{\Phi_L}\big]_{[R]}=\sum_{1\leq r\leq R} \eta^r\,\Gamma^{(4,r)}_{\Phi_L}\,,
\eeq
where $\Gamma^{(4,r)}_{\Phi_L}$ is the contribution of exactly $r$ rungs, viz. $(-1/2)^{r-1}{\cal I}_{L-2}(G^2{\cal I}_{L-2})^{r-1}$. We next define the renormalized rung expansion up to $R$ rungs as follows. We split the one-rung contribution  as in Eq.~(\ref{eq:dec}), and write formally the counterterm $\delta\tilde\lambda$ as an infinite series in $\eta$:
\beq\label{eq:ltilde}
\delta\tilde\lambda=\sum_{r\geq 1} \eta^{r-1}\,\delta\tilde\lambda^{(r)}\,.
\eeq
The renormalized rung expansion  is obtained by expanding $\Gamma^{(4)}_{\Phi_L}$ in powers of  $\eta$, taking into account those $\eta$'s that come from the expansion of the counterterm $\delta\tilde\lambda$, that is
\beq
\big[\Gamma^{(4)}_{\Phi_L}\big]_{[R]}=\sum_{1\leq r\leq R} \eta^r\,\tilde\Gamma^{(4,r)}_{\Phi_L},
\eeq
with
\beq
\tilde\Gamma^{(4,1)}_{\Phi_L} & = & \lambda+\delta\tilde\lambda^{(1)}-\frac{1}{2}\tilde {\cal I}_{L-2}\,,\label{eq:r1}\\
\tilde\Gamma^{(4,2)}_{\Phi_L} & = & \delta\tilde\lambda^{(2)}+\frac{1}{2}(\lambda+\delta\tilde\lambda^{(1)})G^2\tilde{\cal I}_{L-2}+\frac{1}{4}\tilde{\cal I}_{L-2}G^2\tilde{\cal I}_{L-2}\,\label{eq:r2}\\
& \dots & \nonumber
\eeq
The counterterm contribution $\delta\tilde\lambda^{(1)}$ renormalizes the first rung (which has only an overall divergence since its subdivergences have been removed by the renormalization of the $\Gamma^{(4)}_L$'s). This same counterterm allows to remove a subdivergence in the two-rung contribution, as it is clear from (\ref{eq:r2}). There only remains an overall divergence that can be absorbed in $\delta\tilde\lambda^{(2)}$. The same logic extends in higher rung contributions. 

\begin{figure}[h]
\begin{center}
\includegraphics[angle=0,width=10cm]{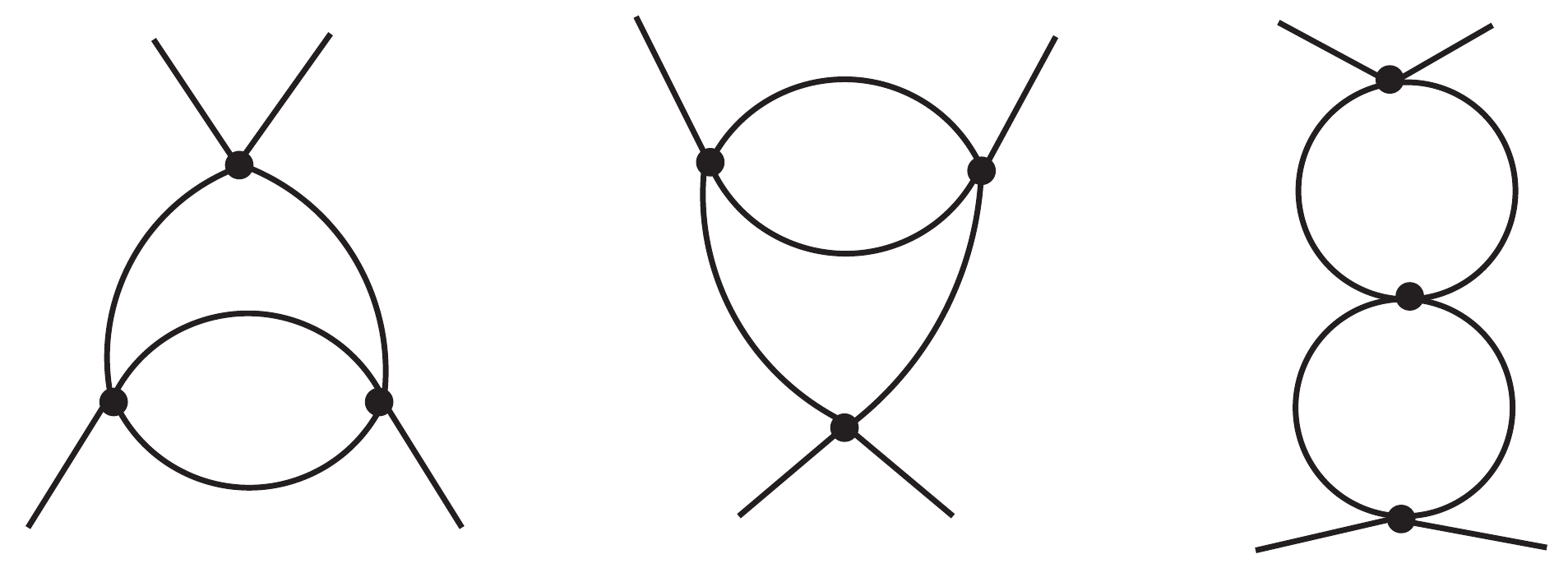}
\caption{The contributions to $\Gamma^{(4,2)}$ that involve iterations of ${\cal I}_1$ in the BS equation up to order $\lambda^3$.  }\label{fig:renormGammaL}
\end{center}
\end{figure}

The counterterms $\delta \lambda$ and $\delta\tilde\lambda$ have  different structures and renormalize their respective four-point functions in  different ways. Thus, as all the ladders are resummed in $\Gamma^{(4)}_{\Phi_L}$, the same counterterm (\ref{eq:ltilde}) ends up renormalizing the subdivergences and the overall divergences of the rung expansion, as we have just discussed. In contrast, in the case of a given $\Gamma^{(4)}_L$, different expansions of the same counterterm $\delta\lambda$ renormalizes the overall divergences and the subdivergences. We should note however that all these countertems are not completely independent since, as discussed at the end of  Sect.~\ref{sec:2PI}, at a given loop order the diagrams that contribute to $\Gamma^{(4)}_L$ can be separated into contributions to ${\cal I}$ and diagrams resulting from the iteration of the BS equation (see e.g. Fig.~\ref{fig:renormGammaL}). Thus, when restricted at a given loop order, the counterterm $\delta\tilde \lambda_L$ obviously coincides with $\delta \lambda_L$, the difference between the two quantities being of order $\lambda^{L+2}$.

 Returning now to the general form of the BS equation, Eq.~(\ref{BS1}), we use the equation for the counterterm $\delta\tilde\lambda$ in the form of Eq.~(\ref{rc}) and subtract it from the BS equation. One obtains
\beq\label{eq:ttt}
\Gamma^{(4)}_{\Phi_L}(q,p)-\lambda & = & \tilde {\cal I}_{L-2}(q,p)-\tilde {\cal I}_{L-2}(0,0)-\frac{1}{2}\int_r  \Gamma^{(4)}_{\Phi_L}(q,r)G^2(r){\cal I}_{L-2}(r,p)+\frac{1}{2}\int_r  \Gamma^{(4)}_{\Phi_L}(0,r)G^2(r){\cal I}_{L-2}(r,0)\nonumber\\
& = & \tilde {\cal I}_{L-2}(q,p)-\tilde {\cal I}_{L-2}(0,0)-\frac{1}{2}\int_r  \Gamma^{(4)}_{\Phi_L}(q,r)G^2(r)[\tilde {\cal I}_{L-2}(r,p)-\tilde {\cal I}_{L-2}(r,0)]\nonumber\\
& & \hspace{2.5cm}+\,\frac{1}{2}\int_r  \Gamma^{(4)}_{\Phi_L}(0,r)G^2(r)[\tilde{\cal I}_{L-2}(r,0)-\tilde{\cal I}_{L-2}(r,q)]\,.
\eeq
In going from the first to the second line, we have added and subtracted the same quantity but written in two equivalent ways (that follows from the ladder structure of the diagrams resummed by the BS equation, as well as the symmetry under the exchanges of the arguments):
\beq
\frac{1}{2}\int_r  \Gamma^{(4)}_{\Phi_L}(q,r)G^2(r){\cal I}_{L-2}(r,0)=\frac{1}{2}\int_r  {\cal I}_{L-2}(q,r)G^2(r)\Gamma^{(4)}_{\Phi_L}(r,0).
\eeq
The benefit of the second equality in Eq.~(\ref{eq:ttt}) is that it provides an explicitly finite equation for $\Gamma^{(4)}_{\Phi_L}(q,p)$. Indeed $\tilde {\cal I}_{L-2}(r,q)$ has no subdivergences thanks to the renormalization of the $\Gamma^{(4)}_L$'s, and the remaining overall divergence cancels in the difference $\tilde {\cal I}_{L-2}(r,q)-\tilde {\cal I}_{L-2}(r,0)$. Moreover, as we argue in \ref{sec:dI}, due to the two-particule irreductibility of $\tilde{\cal I}_{L-2}(r,q)$, the leading asymptotic behavior at large $r$ does not depend on $q$. It therefore  cancels in the difference $\tilde {\cal I}_{L-2}(r,q)-\tilde {\cal I}_{L-2}(r,0)$, ensuring the convergence of the integrals in the second equality of Eq.~(\ref{eq:ttt}).

\subsubsection{Counterterms $\delta Z$ and $\delta m^2$ and renormalization of $G$}
We come now to the final steps involved in the renormalization of the gap equation. 
From Eq.~(\ref{eq:newPhi}), one derives the gap equation in the form
\beq
G^{-1}(p)=Zp^2+m^2+\delta m^2+(\lambda+\delta\tilde\lambda)\frac{\delta\Phi^{(2)}}{\delta G(p)}+\sum_{3\leq l\leq L}\big[(\lambda+\delta\lambda)^{l-1}\big]_{L-l}\,(-)^l\,\frac{\delta\Phi^{(l)}[G]}{\delta G(p)}\,.
\eeq
Imposing the renormalization conditions, we find
\beq
\delta m^2=-(\lambda+\delta\tilde\lambda)\frac{\delta\Phi^{(2)}}{\delta G(0)}-\sum_{3\leq l\leq L}\big[(\lambda+\delta\lambda)^{l-1}\big]_{L-l}\,(-)^l\,\frac{\delta\Phi^{(l)}[G]}{\delta G(0)},
\eeq
and
\beq
\delta Z=-\sum_{3\leq l\leq L}\big[(\lambda+\delta\lambda)^{l-1}\big]_{L-l}\,(-)^l\,\left.\frac{\rmd}{\rmd p^2}\frac{\delta\Phi^{(l)}[G]}{\delta G(p)}\right|_{p=0},
\eeq
where we have used that $\delta\Phi^{(2)}/\delta G(p)$ does not depend on $p$. Again, we know from the general considerations of Sect.~\ref{sec:formal} that, together with the previously determined $\delta\lambda$ and $\delta\tilde\lambda$, the counterterms $\delta Z$ and $\delta m^2$ renormalize the gap equation. We can use their expressions above to rewrite the gap equation as
\beq
G^{-1}(p)=p^2+m^2+\sum_{l=3}^L\big[(\lambda+\delta\lambda)^{l-1}\big]_{L-l}\,(-)^l\,\left[\frac{\delta\Phi^{(l)}[G]}{\delta G(p)}-\frac{\delta\Phi^{(l)}[G]}{\delta G(0)}-p^2\left.\frac{\rmd}{\rmd p^2}\frac{\delta\Phi^{(l)}[G]}{\delta G(p)}\right|_{p=0}\right].
\eeq
This equation still depends on the counterterm $\delta\lambda$ needed to absorb the divergences associated to the $\Gamma^{(4)}_L$'s (see Eq.~(\ref{eq:63})).  Note, however, that the value of $\delta\tilde\lambda$ is not used at all. This is because the term in which it would appear in the gap equation is momentum independent, and it is cancelled by an identical contribution to $\delta m^2$ that has the opposite sign. This is why, for instance,  the renormalization of the gap equation in the vacuum can be done independently from that of the BS equation \cite{VanHees:2001pf,Marko:2012wc}. At finite temperature, this would remain true in a thermal scheme where renormalization conditions  are imposed at finite temperature. However, in this case the renormalized mass becomes temperature dependent and the analysis of this temperature dependence relates to the BS equation and its renormalization \cite{Reinosa:2011ut}. The BS is also mandatory when one does not use a thermal scheme (see the discussion in \ref{sec:twoloop}).

Similar considerations can be used to renormalize the free-energy density, to within an additive global counterterm.

\subsection{Renormalization recipe}
Let us summarize the above findings by providing a synthetic renormalization procedure for $\Phi$-derivable approximations. Consider the $L$-loop $\Phi$-derivable approximation. In a first step, one replaces the bare $L$-loop truncation by
\beq
\Gamma[G] & = & \frac{1}{2} \int_p \log G^{-1}(p)+\frac{1}{2} \int_p \,(Zp^2+m^2+\delta m^2)\,G(p)\nonumber\\
& + & (\lambda+\delta\tilde\lambda)\Phi^{(2)}+\sum_{3\leq l\leq L}\big[(\lambda+\delta\lambda)^{l-1}\big]_{L-l}\,(-)^l\,\Phi^{(l)}[G]\,.\label{eq:G22}
\eeq
Second, the counterterms $\delta Z$, $\delta m^2$, $\delta\tilde\lambda$ and $\delta\lambda$ are given by
\beq
\delta Z & = & -\sum_{3\leq l\leq L}\big[(\lambda+\delta\lambda)^{l-1}\big]_{L-l}\,(-)^l\,\left.\frac{\rmd}{\rmd p^2}\frac{\delta\Phi^{(l)}[G]}{\delta G(p)}\right|_{p=0}\,,\\
\delta m^2 &= & -(\lambda+\delta\tilde\lambda)\frac{\delta\Phi^{(2)}}{\delta G(0)}-\sum_{3\leq l\leq L}\big[(\lambda+\delta\lambda)^{l-1}\big]_{L-l}\,(-)^l\,\frac{\delta\Phi^{(l)}[G]}{\delta G(0)}\,,\\
\delta\tilde\lambda & = & -\frac{\tilde{\cal I}_{L-2}(0,0)-\frac{1}{2}\int_r \Gamma^{(4)}_{\Phi_L}(0,r)G^2(r)(\lambda+\tilde {\cal I}_{L-2}(r,0))}{1-\frac{1}{2}\int_r \Gamma^{(4)}_{\Phi_L}(0,r)G^2(r)}\,.\\
\delta\lambda & = & \lambda^2\Gamma^{(4,2)}(p_i=0)+\lambda^3\Big[-\Gamma^{(4,2)}(p_i=0)+2\big(\Gamma^{(4,1)}(p_i=0)\big)^2\Big]+\dots
\eeq
where the higher terms in $\delta\lambda$ can be determined systematically by renormalizing the skeleton expansion of the four-point function, as explained above, and where
\beq\label{eq:147}
\tilde {\cal I}_{L-2}(q,p)\equiv\sum_{3\leq l\leq L}\big[(\lambda+\delta\lambda)^{l-1}\big]_{L-l}\,(-)^l\,\frac{\delta^2\Phi^{(l)}[G]}{\delta G(q)\delta G(p)}\,.
\eeq
These counterterms renormalize not only the gap and Bethe-Salpeter equations that one derives from (\ref{eq:G22}), but also the free-energy density up to an overall shift.

The equations (\ref{eq:G22})-(\ref{eq:147}) provide a concise summary of the renormalization of $\Phi$-derivable approximations, to all orders, in their diagrammatic formulation.

\section{General remarks and practical implementations}\label{sec:practical}
We have seen in the previous section that  the flow equations (\ref{flow2ptfctb})-(\ref{flowGamma4Lb}) allow us to understand the renormalization of $\Phi$-derivable approximations in their original diagrammatic formulation. This is because  the diagrammatic formulation can be based on the same flow equations  that were used in the previous section, but keeping the ultraviolet cutoff and choosing the initial scale  $\Lambda=\infty$, as indicated at the beginning of this section, see Eq.~(\ref{flowdiagrams}). 
 
As described above, this exact rewriting of the diagrammatic formulation provides a clear map of how divergences are distributed among the relevant $n$-point functions and shows that their cancellations can be achieved by removing solely  overall divergences. This is done by adjusting the counterterms in the initial conditions in Eqs.~(\ref{eq:1111}-\ref{eq:3333}) such that the renormalization and consistency conditions are met at $\kappa=0$. By imposing these conditions on the relevant $n$-point functions written in terms of the diagrams (and not as a formal solution to the flow equations), we can obtain the explicit expressions for the counterterms. 

In fact, for a given set of renormalization and consistency conditions, the renormalized solution defined by Eq.~(\ref{flowLambda}) and the renormalized solution defined by Eq.~(\ref{flowdiagrams}), with the appropriate modifications referred to above, are one and the same renormalized solution. To make this clearer, let us rewrite Eq.~(\ref{flowdiagrams}) here:
\beq\label{flowdiagrams2b}
\Gamma_\kappa^{(n)}(p_i)=\Gamma_\Lambda^{(n)}(p_i)+ \int_\Lambda^\kappa \rmd\kappa' \left.{\cal F}^{(n)}_{\kappa'}(p_i)\right|_{\Lambda_{\rm uv}}.
\eeq
On may read this equation backwards,  that is as an equation specifying the $n$-point functions at scale $\Lambda$ in terms of their values at scale $\kappa=0$, i.e., in terms of their renormalized values. What was done in Sect.~\ref{sec:42} was to choose an initial scale large compared to the physical momenta,  $\Lambda\gg \Lambda_{\rm phys}$, but small compared to the ultraviolet cutoff. In fact, as long as $\Lambda$ is kept finite, the ultraviolet cutoff is not needed, and can just be dropped. One gets then
\beq\label{flowdiagrams2c}
\Gamma_\Lambda^{(n)}(p_i)=\Gamma_{\kappa=0}^{(n)}(p_i)+ \int^\Lambda_0 \rmd\kappa' {\cal F}^{(n)}_{\kappa'}(p_i).
\eeq
The flow equation ensures that $\Gamma_{\kappa=0}^{(n)}(p_i)$,  the renormalized $n$-point functions at scale $\kappa=0$, remain constant as we change $\Lambda$: the change of $\Gamma_{\kappa=0}^{(n)}(p_i)$ is  compensated by the change of the integration boundary.  
What was done in the previous section was to rewrite the same solution with a different treatment of the initial conditions. Because we let $\Lambda\to\infty$, we need to keep $\Lambda_{\rm uv}$ finite to avoid divergences.  One gets then 
\beq\label{flowdiagrams2d}
\Gamma_{\Lambda\to\infty}^{(n)}(p_i)=\Gamma_{\kappa=0}^{(n)}(p_i)+ \int^\infty_0 \rmd\kappa' \left.{\cal F}^{(n)}_{\kappa'}(p_i)\right|_{\Lambda_{\rm uv}}.
\eeq
Divergences appear as one lets $\Lambda_{\rm uv}\to \infty$. What we have seen in the previous section is that these divergences can be handled by counterterms in the initial values $\Gamma_{\Lambda\to\infty}^{(n)}(p_i)$. These counterterms eliminate global divergences, subdivergences being effectively  taken care of by the coupled flow equations. The initial conditions $\Gamma_{\Lambda\to\infty}^{(n)}(p_i)$ depend on $\Lambda_{\rm uv}$ in a similar way as $\Gamma_\Lambda^{(n)}(p_i)$ depend on $\Lambda$, this being dominated by power counting. 
 In a way, we could view the flow approach based on Eq.~(\ref{flowdiagrams2c}) as a (powerful) regularisation, which avoids the needs to determine the counterterms of the more conventional  regularizations of diagrammatic approaches, such as summarized in Eq.~(\ref{flowdiagrams2d}). Because we are dealing with the same solution, it is clear that the functions that need to be initialized in this ``flow regularization'' are nothing but those functions that need to be renormalized in the cutoff regularization.

Our analysis has been based on a specific treatment of the flow of the irreducible kernel which, we believe, is that which best reveals the general structure of the approximations that we are considering. However, in practical applications, some hybrid alternative could be advantageous. For instance, since we have an explicit form for the renormalized $\Gamma^{(4)}_{L,\kp}$'s, Eq.~(\ref{eq:63}), we can use it directly in Eq.~(\ref{eq:III}) with the expansion operator $[\_]_L$ replaced by $\{\_\}_L$ or reinterpreted as referring to the loop expansion in powers of $\lambda$. One is then left with the flow equations for the two-point function $\Gamma^{(2)}_{L-1,\kp}$ and for the four-point function $\Gamma^{(4)}_{\Phi_{L,\kappa}}$. The practical advantage of such an hybrid scheme is that computing directly the renormalized $\Gamma^{(4)}_{L,\kp}$'s  could be more economical than solving the tower of their flow equations. Yet another possibility is to couple the flow equation for $\Gamma^{(2)}_{L-1,\kappa}$ to the renormalized BS equation for $\Gamma^{(4)}_{\Phi_L,\kp}$, obtained from a straightforward extension of Eq.~(\ref{eq:ttt}) in the presence of $R_\kappa$.

\section{Improving on $\Phi$-derivable approximations using their fRG reformulation}\label{sec:beyond}

In the previous sections, we have shown that the reformulation of $\Phi$-derivable approximations in terms of flow equations provides much insight into their renormalization. The resulting flow equations form a finite set of equations, realizing a specific truncation of the fRG equations, and we have indicated several strategies that can be implemented to solve these equations in practice. 
In this section, we exploit the flexibility of this reformulation and show how it can be used to construct approximations that extend $\Phi$-derivable approximations beyond their standard diagrammatic derivation. As an example, we derive  approximations that are crossing-symmetric.  We also provide possible generalizations of the present truncation of the fRG equations. 

\subsection{Crossing symmetric approximations}
Let us first recall that the standard 1PI flow equations  are explicitly crossing symmetric, and remain so even in the presence of approximations. This is perhaps not completely obvious at the level of Eq.~(\ref{eq:flow4}) because this equation is restricted to a particular momentum configuration. However, one may derive a more general version of the flow equation, valid for an arbitrary configuration of external momenta, viz.
\beq \label{eq:flow4gen}
& & \pk\Gamma^{(4)}_\kp(p_1,p_2,p_3,p_4)=-\frac{1}{2}\int_r \pk R_\kp(r)\,G_\kp^2(r)\,\Gamma^{(6)}_\kp(r,-r,p_1,p_2,p_3,p_4) 
\nn
& & \hspace{1.0cm}+\,\int_r \Gamma^{(4)}_\kp(p_1,p_2,r,-p_1-p_2-r)\,G_\kp(r+p_1+p_2)G_\kp^2(r)\pk R_\kp(r)\,\Gamma^{(4)}_\kp(r,-p_1-p_2-r,p_3,p_4)\nn
& & \hspace{1.0cm}+\,\int_r \Gamma^{(4)}_\kp(p_1,p_3,r,-p_1-p_3-r)\,G_\kp(r+p_1+p_3)G_\kp^2(r)\pk R_\kp(r)\,\Gamma^{(4)}_\kp(r,-p_1-p_3-r,p_2,p_4)\nn
& & \hspace{1.0cm}+\,\int_r \Gamma^{(4)}_\kp(p_1,p_4,r,-p_1-p_4-r)\,G_\kp(r+p_1+p_4)G_\kp^2(r)\pk R_\kp(r)\,\Gamma^{(4)}_\kp(r,-p_1-p_4-r,p_3,p_2).\nn
\eeq
 Equation~(\ref{eq:flow4gen}) is now explicitly crossing symmetric, ensuring in a trivial way that the crossing symmetry of the $n$-point functions is preserved along the flow. As we have already mentioned, the same property does not hold for the flow formulation of the 2PI equations that is discussed in the main text (see the discussion at the end of Sect.~\ref{sec:gamma4flow}). However, we shall see that the flow formulation allows for a simple extension of this equation that preserves crossing symmetry through approximations. 

In the present discussion, we do not immediately restrict ourselves to translational invariant system and assume the propagator to be a function of two momentum variables, $G(p_1,p_2)$. By taking two derivatives of the Luttinger-Ward functional with respect to $G(p,q)$, one obtains the kernel\footnote{We absorb two factors of $(2\pi)^d$ in the definition of $\delta/\delta G(p,q)$.}
\beq\label{eq:Lambda2}
{\cal I}(p_1,p_2;p_3,p_4)=4\frac{\delta^2\Phi[G]}{\delta G(p_1,p_2)\delta G(p_3,p_4)}.
\eeq
This can be used to obtain the complete four-point function from the Bethe-Salpeter equation
\beq
\Gamma^{(4)}(p_1,p_2;p_3,p_4) & = & {\cal I}(p_1,p_2;p_3,p_4)-\frac{1}{2}\int_{q,k,r,s}\Gamma^{(4)}(p_1,p_2;q,r) G(-q,-k)G(-r,-s){\cal I}(-k,-s;p_3,p_4)\nn
& = & {\cal I}(p_1,p_2;p_3,p_4)-\frac{1}{2}\int_{q,k,r,s}{\cal I}(p_1,p_2;q,r) G(-q,-k)G(-r,-s)\Gamma^{(4)}(-k,-s;p_3,p_4)\,.\nn
\eeq
Here it is understood that $G$, ${\cal I}$ and $\Gamma^{(4)}$ are full Fourier transforms (see the discussion around Eq.~(\ref{gamman}), where full and reduced Fourier transforms are introduced). Restricting now the propagators to be those of a translation invariant system, $G(p,q)=\delta^{(d)}(p+q)G(p)$, and extracting similar $\delta$-functions from ${\cal I}$ and $\Gamma^{(4)}$, we arrive at the following equation for the reduced Fourier transforms (we keep the same notation for the reduced and full Fourier transforms, as done throughout the paper)
\beq
\Gamma^{(4)}(p_1,p_2;p_3,p_4) & = & {\cal I}(p_1,p_2;p_3,p_4)\nonumber\\
& - & \frac{1}{2}\int_{q}\Gamma^{(4)}(p_1,p_2;q,-p_1-p_2-q) G(q)G(p_1+p_2+q){\cal I}(-q,p_1+p_2+q;p_3,p_4)\nn
& = & {\cal I}(p_1,p_2;p_3,p_4)\nonumber\\
& - & \frac{1}{2}\int_{q}{\cal I}(p_1,p_2;q,-p_1-p_2-q) G(q)G(p_1+p_2+q)\Gamma^{(4)}(-q,p_1+p_2+q;p_3,p_4)\nn
\eeq
where it is understood that $p_4=-(p_1+p_2+p_3)$ and we have used $G(-q)=G(q)$. In the case where $p_1=p=-p_2$ and $p_3=q=-p_4$, we recover Eq.~(\ref{BS1}).\footnote{In order to verify  that the kernels coincide, we recall that the functional $\Phi$ in (\ref{eq:Lambda}) is not exactly the same as the one in (\ref{eq:Lambda2}). The former is the restriction of the latter to propagators of the form $G(p,q)=\delta^{(d)}(p+q)G(p)$ and there is also a volume factor $V$ that has been factored out. For the sake of clarity, we momentarily denote the reduced functional $\phi[g]=\Phi[G=g\delta]$. Now
\beq
\frac{\delta\phi}{\delta G(p)} & = & \frac{1}{V}\int_{q,r}\frac{\delta(\delta^{(d)}(q+r)G(q))}{\delta G(p)}\left.\frac{\delta\Phi}{\delta G(q,r)}\right|_{G(q,r)=\delta^{(d)}(q+r)G(q)}\nn
& = & \frac{1}{V}\int_{q,r}\delta^{(d)}(q+r)\delta^{(d)}(p-q)\left.\frac{\delta\Phi}{\delta G(q,r)}\right|_{G(q,r)=\delta^{(d)}(q+r)G(q)}=\frac{1}{V}\frac{\delta\Phi}{\delta G(p,-p)\,.}
\eeq
and similarly $\delta^2\phi/\delta g(p)\delta g(q)=(1/V)\delta^2\phi/\delta G(p,-p)\delta G(q,-q)$. If we explicitly factor out the momentum conservation delta from $\delta^2\phi/\delta G(p,-p)\delta G(q,-q)$ while keeping the same notation for the reduced object, that is $\delta^2\phi/\delta G(p,-p)\delta G(q,-q)\to \delta^{(d)}(p-p+q-q)\delta^2\phi/\delta G(p,-p)\delta G(q,-q)$ we arrive at $\delta^2\phi/\delta g(p)\delta g(q)=\delta^2\phi/\delta G(p,-p)\delta G(q,-q)$ and so the kernels identify indeed.}

We can proceed identically in the presence of a regulator $\kappa$. Then, after taking a $\kappa$-derivative and slightly generalizing the analysis in Sect.~\ref{sec:gamma4flow}, we arrive at
\begin{eqnarray}\label{eq:BSgen}
& & \partial_\kappa\Gamma^{(4)}_\kappa(p_1,p_2;p_3,p_4)=\partial_\kappa{\cal I}_\kappa(p_1,p_2;p_3,p_4)\nonumber\\
& & \hspace{1.0cm}-\,\frac{1}{2}\int_q\Gamma^{(4)}_\kappa(p_1,p_2;-q,q-p_1-p_2)\,\partial_\kappa [G_\kappa(q)G_\kappa(p_1+p_2-q)]\,\Gamma^{(4)}_\kappa(q,p_1+p_2-q,p_3,p_4)\nonumber\\
& & \hspace{1.0cm}-\,\frac{1}{2}\int_q\partial_\kappa{\cal I}_\kappa(p_1,p_2,-q,q-p_1-p_2)\,G_\kappa(q)G_\kappa(p_1+p_2-q)\,\Gamma^{(4)}_\kappa(q,p_1+p_2-q,p_3,p_4)\nonumber\\
& & \hspace{1.0cm}-\,\frac{1}{2}\int_q\Gamma^{(4)}_\kappa(p_1,p_2,-q,q-p_1-p_2)\,G_\kappa(q)G_\kappa(p_1+p_2-q)\,\partial_\kappa {\cal I}_\kappa(q,p_1+p_2-1;p_3,p_4)\nonumber\\
& & \hspace{1.0cm}+\,\frac{1}{4}\int_q\int_k \Gamma^{(4)}_\kappa(p_1,p_2;-q,p_1+p_2-q)\,G_\kappa(q)G_\kappa(p_1+p_2-q)\,\partial_\kappa\nonumber\\
& & \hspace{2.0cm}\times\,{\cal I}_\kappa(q,p_1+p_2-q;k,p_3+p_4-k)\,G_\kappa(k)G_\kappa(p_3+p_4-k)\,\Gamma^{(4)}_\kp(-k,k-p_3-p_4;p_3,p_4).\nn
\end{eqnarray} 
This equation is quite different from (\ref{eq:flow4}). In particular, even though its solution in the absence of approximations obeys crossing symmetry, this symmetry is not manifest in the equation itself, and it generally gets lost in the presence of approximations. It is in this sense that the flow equation (\ref{eq:flowGamma4gen}), which is only a particular case of Eq.~(\ref{eq:BSgen}), is not compatible with crossing symmetry. This of course relates to the discussion at the end of Sect.~\ref{sec:2PI} of how the different channels are non-equivalently resummed by the Bethe-Salpeter equation.

To cope with this issue, we write a  symmetrize version of Eq.~(\ref{eq:BSgen}), viz.
\begin{eqnarray}\label{eq:BSsym}
& & \partial_\kappa\Gamma^{(4)}_\kappa(p_1,p_2,p_3,p_4)=\frac{1}{3}\Bigg[\partial_\kappa{\cal I}_\kappa(p_1,p_2;p_3,p_4)\nonumber\\
& & \hspace{0.7cm}-\,\frac{1}{6}\int_q\Gamma^{(4)}_\kappa(p_1,p_2;-q,q-p_1-p_2)\,\partial_\kappa [G_\kappa(q)G_\kappa(p_1+p_2-q)]\,\Gamma^{(4)}_\kappa(q,p_1+p_2-q,p_3,p_4)\nonumber\\
& & \hspace{0.7cm}-\,\frac{1}{6}\int_q\partial_\kappa{\cal I}_\kappa(p_1,p_2,-q,q-p_1-p_2)\,G_\kappa(q)G_\kappa(p_1+p_2-q)\,\Gamma^{(4)}_\kappa(q,p_1+p_2-q,p_3,p_4)\nonumber\\
& & \hspace{0.7cm}-\,\frac{1}{6}\int_q\Gamma^{(4)}_\kappa(p_1,p_2,-q,q-p_1-p_2)\,G_\kappa(q)G_\kappa(p_1+p_2-q)\,\partial_\kappa {\cal I}_\kappa(q,p_1+p_2-1;p_3,p_4)\nonumber\\
& & \hspace{0.7cm}+\,\frac{1}{12}\int_q\int_k \Gamma^{(4)}_\kappa(p_1,p_2;-q,p_1+p_2-q)\,G_\kappa(q)G_\kappa(p_1+p_2-q)\nonumber\\
& & \hspace{1.7cm}\times\,\partial_\kappa{\cal I}_\kappa(q,p_1+p_2-q;k,p_3+p_4-k)\,G_\kappa(k)G_\kappa(p_3+p_4-k)\,\Gamma^{(4)}_\kp(-k,k-p_3-p_4;p_3,p_4)\nn
& & \hspace{0.7cm}+\,(p_2\leftrightarrow p_3)+(p_2\leftrightarrow p_4)\Bigg].
\end{eqnarray} 
This equation is, by construction, crossing-symmetric. It is obeyed by the exact four-point function. When coupled to Eq.~(\ref{eq:BS2b}), and upon truncations of the $\Phi[G]$ functional, it generates a new expansion scheme for both the two- and the four-point function, where crossing symmetry is explicitly implemented.\footnote{The inclusion of channels beyond the standard $\Phi$-derivable approximation has been discussed in Ref.~\cite{Fukushima:2012xw} within a particular approximation in the context of magnetic catalysis. The present discussion makes these considerations more systematic.}

We can even go a step further. We have illustrated above that subdivergent contributions that appear in the flow of ${\cal I}_\kappa$ could always be written in terms of the  loop approximations $\Gamma^{(4)}_{\kappa,L}$ to the four-point function. These approximations are always crossing symmetric. However they differ from the four-point function that enters Eq.~(\ref{eq:BS2b}). We can solve this issue by replacing any occurrence of $\Gamma^{(4)}_{\kappa,L}$ by the function $\Gamma^{(4)}_\kappa$ that solves Eq.~(\ref{eq:BSsym}). For instance, at four-loop order in $\Phi$, Eq.~(\ref{eq:dkIk}) would be replaced by
\beq\label{eq:dkIk3}
\partial_\kappa{\cal I}_\kappa(q,p) & = & -\int_r \Gamma^{(4)}_\kp(r,q,p)\,\pk G_\kp(r)\,G_\kp(r+q+p)\,\Gamma^{(4)}_\kp(r,q,p)+(q\rightarrow -q)\nonumber\\
& + & 2\int_s \pk G_\kp(s)\int_r \Gamma^{(4)}_\kp(r,q,p) G_\kp(r)\,G_\kp(r+q+p)\,\nonumber\\
& & \hspace{1.0cm}\times\,\Gamma^{(4)}_\kp(-r,-s,-p)\,G_\kp(r+s+p)\,\Gamma^{(4)}_\kp(r+p+q,s,-q)+(q\rightarrow -q)\,.
\eeq
By combining Eqs.~(\ref{eq:BS2b}), (\ref{eq:BSsym}) and (\ref{eq:dkIk3}), we have now an ``improved'' four-loop $\Phi$-derivable truncation\footnote{Of course, as one increases the truncation order of $\Phi[G]$ new terms appear in (\ref{eq:dkIk3}) but they can alway be determined in a systematic way.} where not only crossing symmetry is manifest, but also, where only one version of the four-point function runs along the flow. This is a simplification since it eliminates the need to deal with various versions of the four-point function.

Of course, in doing what we do here, we depart from the strict correspondence with the diagrammatic approach. In particular the solution of the above flow equations is likely sensitive to the choice of regulator. So there is a trade-of between extending $\Phi$-derivable approximations in a crossing symmetric way, as we propose to do here, and the reintroduction of a regulator dependence. We mention, however, that, since the new scheme is still based on a skeleton expansion, we expect this regulator dependence to be controlled, at least formally, by the number of loops kept in the expansion.

\subsection{Other possible truncations of the fRG hierarchy using skeleton expansions}
 The analysis of the present paper has revealed the role of various skeleton expansions, beyond the skeleton expansion of the Luttinger-Ward functional at the core of $\Phi$-derivable approximations. These other skeleton expansions suggest other possibilities to truncate the fRG hierarchy which we now briefly review.

One possibility is to use directly the two-skeleton expansion of $\Gamma^{(4)}_L$. Thus one may replace the exact flow equation (\ref{eq:BS2b}) by
\beq
\pk \Sigma_\kappa(p)=-\frac{1}{2}\int_q \,\pk R_\kappa(q)\,G^2_\kp(q)\,\Gamma^{(4)}_{L,\kp}(q,p)\,,
\eeq
This is clearly a systematically improvable approximation. It respects crossing symmetry since all the channels enter the four-point function $\Gamma^{(4)}_{L,\kp}(q,p)$. These $\Gamma^{(4)}_{L,\kp}(q,p)$ can be obtained as renormalized diagrams or as solutions of the flow equations (\ref{eq:Ggreat}).

Yet another truncation consists in using the expansion of the six-point function in terms of four-skeletons, as discussed at the end of \ref{app:great}. We write
\beq
\pk \Sigma_\kappa(p)=-\frac{1}{2}\int_q \,\pk R_\kappa(q)\,G^2_\kp(q)\,\Gamma_\kappa^{(4)}(q,p)
\eeq
with
\beq
& & \pk\Gamma^{(4)}_\kp(p_1,p_2,p_3,p_4)=-\frac{1}{2}\int_r \pk R_\kp(r)\,G_\kp^2(r)\,\Gamma^{(6)}_{L,\kp}(r,-r,p_1,p_2,p_3,p_4) 
\nn
& & \hspace{1.0cm}+\,\int_r \Gamma^{(4)}_\kp(p_1,p_2,r,-p_1-p_2-r)\,G_\kp(r+p_1+p_2)G_\kp^2(r)\pk R_\kp(r)\,\Gamma^{(4)}_\kp(r,-p_1-p_2-r,p_3,p_4)\nn
& & \hspace{1.0cm}+\,\int_r \Gamma^{(4)}_\kp(p_1,p_3,r,-p_1-p_3-r)\,G_\kp(r+p_1+p_3)G_\kp^2(r)\pk R_\kp(r)\,\Gamma^{(4)}_\kp(r,-p_1-p_3-r,p_2,p_4)\nn
& & \hspace{1.0cm}+\,\int_r \Gamma^{(4)}_\kp(p_1,p_4,r,-p_1-p_4-r)\,G_\kp(r+p_1+p_4)G_\kp^2(r)\pk R_\kp(r)\,\Gamma^{(4)}_\kp(r,-p_1-p_4-r,p_3,p_2)\nn
\eeq
with $\Gamma^{(6)}_{L,\kp}$ an explicitly finite expression in terms of $G_\kappa$ and $\Gamma^{(4)}_\kappa$. Again, this truncation is systematically improvable by adding more terms in the expansion of $\Gamma^{(6)}_\kappa$ in terms of four-skeletons.

 \section{Conclusions}
In the present work we have studied a particular truncation of the flow equations for the 1PI effective action for a scalar $\varphi^4$ theory in four dimensions in the symmetric phase, extending the fRG-2PI approach initiated in \cite{Blaizot:2010zx}. This truncation exploits the relation that exists between the four-point function and the two-point function in the 2PI formalism based on the Luttinger-Ward functional.  It provides an  exact formulation of the 2PI equations  in terms of flow equations.  The solution of these flow equations remains independent of the choice of the regulator that controls the flows within the so-called $\Phi$-derivable approximations based on a selection of a finite set of skeletons contributing to the LW functional. This suggests specific approximation schemes in which skeletons are ordered according to their number of loops, and  one selects skeletons up to a given number of loops. Once such an approximation is made, the two-point and four-point functions that are obtained are of course only approximate. However the independence of the solution of the flow equations on the choice of the regulator persists. 

One important benefit of writing the 2PI equations in terms of flow equations is that the functional RG provides much insight on the renormalization, which constitutes a major part of the present paper. We have shown that the flow equations that govern a given $\Phi$-derivable approximation are finite, that is, they are independent of any ultraviolet cutoff that may need to be introduced at intermediate stages. Furthermore, they can be made independent of the parameters of the bare theory. It follows that the renormalization can be achieved by following essentially the same route as in the standard 1PI fRG, i.e. without introducing counterterms: the choice of counterterms is replaced by a suitable choice of initial conditions. The flow equations also clarify the mechanisms of elimination of divergences, in particular they take care automatically of the subdivergences, and resolve the subtle issue of hidden subdivergences that complicates the diagrammatic approach to renormalization. All in all, the flow equations provide a complete understanding of the all-order renormalization of $\Phi$-derivable approximations, and clarify a number of issues concerning higher orders that were left unsettled in previous studies. 
 
The formulation of the 2PI equations in terms of flow equations not only provides an efficient tool to understand the formal aspects of  the renormalization of $\Phi$-derivable approximations, it also leads to various practical implementations.  It offers furthermore a flexibility that allows us to extend the $\Phi$-derivable approximations beyond the strict diagrammatic domain where they are usually defined. We have provided examples of such extensions, as well as suggestions for new possible truncations of the fRG equations. Finally, we note that the  strategy that has been developed in this paper could also be tried with other non-perturbative approaches defined in terms of diagrams, such as higher ($n$PI) effective actions or Dyson-Schwinger equations. \\

\noindent{\bf Acknowledgements}
We thank N.~Wink for discussions. This work is supported by EMMI, the
BMBF grant 05P18VHFCA. It is part of and supported by the DFG
Collaborative Research Centre SFB 1225 (ISOQUANT) as well as by the
DFG under Germany's Excellence Strategy EXC - 2181/1 - 390900948 (the
Heidelberg Excellence Cluster STRUCTURES). JPB is thankful for the warm hospitality of the ITP Heidelberg which he visited as a Jensen professor during early stages of this work. UR wishes to thank both the ITP Heidelberg and the ECT* for support during the initial stages of this work.

\appendix

\section{Illustrative example: the $2$-loop approximation }\label{sec:twoloop}
In this appendix, we illustrate some of the formal developments of the main text with the simplest non trivial example, that of the two-loop $\Phi$-skeleton (left diagram in Fig.~\ref{fig:skeletonsPhi}). The self-energy is momentum independent and so is the four-point function $\Gamma^{(4)}(p,q)=\Gamma^{(4)}(0,0)$, with  ${\cal I}(p,q)=\lambda_{\rm b}$. 
To alleviate the notation, we set $\lambda =\Gamma^{(4)}(0,0)$, and call $m^2$ the solution of the gap equation, to be identified with the physical (or renormalized) mass.  

\subsection{General considerations}
We start by considering the gap equation, the BS equation, and their solutions. This will allow us in particular to recall well known features of the $\varphi^4$ field theory in four dimensions. 

In the two-loop approximation, the gap equation and the BS equation take simple forms. The gap equation reads
  \beq\label{gap2l0}
 m^2=m_{\rm b}^2+\Sigma(m)=m_{\rm b}^2+\frac{\lambda_{\rm b}}{2}I(m),\quad I(m)\equiv \int_{q<\Lambda_{\rm uv}} \frac{1}{q^2+m^2},
 \eeq
and the BS equation (\ref{BS1}) can be written as
\beq\label{BS2l0}
 \frac{1}{\lambda}=\frac{1}{\lambda_{\rm b}}+\frac{1}{2}J(m),
\qquad J(m)\equiv \int_{q<\Lambda_{\rm uv}} \frac{1}{(q^2+m^2)^2}\,.
\eeq
 These two equations provide the values of the physical mass $m$,  and the four-point function $\lambda$, as a function of  $\lambda_{\rm b}$, $m_{\rm b}$ and the ultraviolet cutoff $\Lambda_{\rm uv}$.  At weak coupling, and for $\Lambda_{\rm uv}\gg m$, we have approximately 
 \beq\label{gapuv}
 m^2-m_{\rm b}^2\approx \lambda_{\rm b} \Lambda_{\rm uv}^2/(32\pi^2),\qquad  \frac{1}{\lambda}\approx\frac{1}{\lambda_{\rm b}}+\frac{1}{16\pi^2}\left( -1+\ln\frac{\Lambda_{\rm uv}^2}{m^2}\right).
 \eeq
The last relation shows that, at fixed $\lambda_{\rm b}$, $\lambda$ is a decreasing function of $\Lambda_{\rm uv}$ (eventually going to zero as $\Lambda_{\rm uv}$ goes to infinity). 
Alternatively, Eqs.~(\ref{gapuv})  specify how the parameters of the Lagrangian, $\lambda_{\rm b}$ and $m_{\rm b}$, need to change when one varies $\Lambda_{\rm uv}$, so as to maintain  $\lambda$ and $m$ at  their physical values.  Consider for instance the derivative of $\lambda_{\rm b}$  with respect to $\Lambda_{\rm uv}$, at fixed $\lambda$. One gets\footnote{Note that the BS equation captures only 1/3 of the one-loop beta-function, as only a single channel out of three is being taken into account.} 
\beq\label{betaoneloop} 
\Lambda_{\rm uv}\frac{\del }{\del \Lambda_{\rm uv}}
 \left( \frac{1}{\lambda_{\rm b}} \right)\approx -\frac{1}{32\pi^2}. 
\eeq
 The negative sign is indicative of the presence of a so-called ``Landau pole'': $\lambda_{\rm b}$
 blows up for a finite value $\Lambda_{\rm L}$ of $\Lambda_{\rm uv}$. To see that, we integrate Eq.~(\ref{betaoneloop}) and get, with $\Lambda_0$ some reference scale, 
 \beq
 \lambda_{\rm b}(\Lambda_{\rm uv})=\frac{\lambda_{\rm b}(\Lambda_0)}{1-\frac{\lambda_{\rm b}(\Lambda_0)}{32\pi^2}\log\frac{\Lambda_{\rm uv}}{\Lambda_0}}.
 \eeq
 This shows that  $\lambda_{\rm b}(\Lambda_{\rm uv})$ blows up when $\Lambda_{\rm uv}=\Lambda_{\rm L}$, with
 \beq\label{Landau}
 \Lambda_{\rm L}=\Lambda_0 \exp\left(\frac{32\pi^2}{\lambda_{\rm b}(\Lambda_0)}  \right).
 \eeq 
 As well known, the presence of this Landau pole renders delicate the discussions concerning the renormalization, and in particular the limit of infinite cutoff $\Lambda_{\rm uv}\to\infty$. We  always assume in this paper that we work in a regime of sufficiently weak coupling (that is $\lambda_{\rm b}(\Lambda_0)\ll 1$), so that we can allow $\Lambda_{\rm uv}$ to become large while staying far below $\Lambda_L$, i.e., $\Lambda_0\ll\Lambda_{\rm uv}\ll \Lambda_{\rm L}$.

 \subsection{Flow equations}
 Consider now the analogs of Eqs.~(\ref{gap2l0}) and (\ref{BS2l0})
 obtained when a regulator is added to all the propagators, as in Eq.~(\ref{renormgapeqn}). These read
 respectively 
\beq\label{gap2lkappa}
 m_\kappa^2=m_{\rm b}^2+\frac{\lambda_{\rm b}}{2}\int_{q<\Lambda_{\rm uv}} G_\kappa(q) ,\qquad
 G_\kappa(q)=\frac{1}{q^2+m_\kappa^2+R_\kappa(q)}, 
\eeq 
and
 \beq\label{BS2lk}
 \frac{1}{\lambda_\kappa}=\frac{1}{\lambda_{\rm b}}+\frac{1}{2}\int_{q<\Lambda_{\rm uv}}G^2_\kappa(q),
 \eeq 
 where we have set $\lambda_\kappa=\Gamma_\kappa^{(4)}(0,0)$. The presence of the regulator does not alter significantly the general structure of the equations. In particular, for $\kappa=0$ one recovers trivially the solution written above.

We may however proceed as in Sect.~\ref{sec:contactRG} and obtain $m_\kappa$ and $\lambda_\kappa$  as solutions of flow equations. 
The flow equation for $m_\kappa$  is obtained simply by taking the derivative of Eq.~(\ref{gap2lkappa}). One gets      
\beq
\partial_\kappa m_\kappa^2=- \frac{\lambda_{\rm b}}{2}  \int_{q<\Lambda_{\rm uv}}  G^2_\kappa(q)\left(\partial_\kappa m_\kappa^2+\partial_\kappa R_\kappa  \right),
\eeq    
or, using Eq.~(\ref{BS2lk}), 
\beq\label{eqnform2}
\partial_\kappa m_\kappa^2=-\frac{1}{2} \lambda_\kappa \int_{q<\Lambda_{\rm uv}} (\partial_\kappa R_\kappa)\,  G_\kappa^2(q)\,.
\eeq
This is indeed the expected flow equation for the two-point function (see Eq.~(\ref{eq:BS2b})). Note that the flow of the mass and that of the coupling constant are naturally coupled. 
The flow equation for  $\lambda_\kappa$, can be obtained similarly by differentiating Eq.~(\ref{BS2lk}) with respect to $\kappa$. It reads
\beq\label{eqnforg}
\partial_\kappa\left ( \frac{1}{\lambda_\kappa} \right)=\frac{1}{2}\partial_\kappa\int_{q<\Lambda_{\rm uv}} G^2_\kappa(q),\qquad \partial_\kappa \lambda_\kappa =-\frac{\lambda_\kappa^2}{2}\partial_\kappa\int_{q<\Lambda_{\rm uv}} G^2_\kappa(q).
\eeq
  The writing on the right corresponds indeed to Eq.~(\ref{eq:flowGamma4gen}),  where only the first line contributes since $\del_\kappa {\cal I}_\kappa=0$ in the present case.

It remains to determine the initial conditions for these flow equations (see the discussion in Sect.~\ref{sec:summ}). We want to choose these at a large value $\kappa=\Lambda$, in such a way as to  recover the standard 2PI results at $\kappa=0$. This easily done since we  know the explicit solution, given in Eqs.~(\ref{gap2lkappa}) and (\ref{BS2lk}) above:  it is sufficient to analyze the behavior of this solution as $\kappa=\Lambda$ becomes large. If we use a sharp infrared regulator,  the answer is immediate: when $\Lambda=\Lambda_{\rm uv}$, $\Sigma_\Lambda=0$ so that $m_\Lambda=m_{\rm b}$, and $\lambda_\Lambda=\lambda_{\rm b}$. For a smooth regulator, e.g. $R_\kappa(q)=(\kappa^2-q^2)\theta(\kappa-q)$, and $\Lambda>\Lambda_{\rm uv}$, one gets instead
\beq
\frac{1}{\lambda_\Lambda}=\frac{1}{\lambda_{\rm b}}+\frac{1}{64\pi^2} \frac{\Lambda_{\rm uv}^4}{(\Lambda^2+M_\Lambda^2)^2},\qquad \Sigma_\Lambda=\frac{\lambda_{\rm b}}{64\pi^2}\frac{\Lambda_{\rm uv}^4}{\Lambda^2+M_\Lambda^2},
\eeq
where $m_\Lambda^2=m^2+\Sigma_\Lambda$.
 The dependence on $\Lambda$ reflects the fact that, with a smooth regulator,  the fluctuations (carrying momenta below the ultraviolet cutoff $\Lambda_{\rm uv}$) are only gradually suppressed as $\Lambda$ grows;  they are entirely suppressed only as $\Lambda\to\infty$: when $\Lambda/\Lambda_{\rm uv}\to \infty$, $\lambda_\Lambda\to\lambda_{\rm b}$ and $\Sigma_\Lambda/\Lambda_{\rm uv}^2\to 0$, the deviations from these limits being powers of $\Lambda_{\rm uv}/\Lambda$. 
 
 \subsection{Renormalization and hidden sub-divergences}\label{twolooprenorm}
We now examine the renormalization with counterterms, as outlined in Sect.~\ref{sec:2PIren}. In the 2-loop approximation, there is no field renormalization ($Z=1$) and  therefore  we have simply 
\beq\label{barerenrom}
\lambda_{\rm b}=\lambda+\delta \lambda,\qquad  m_{\rm b}^2=m^2+\delta m^2.
\eeq

\begin{figure}[htbp]
\begin{center}
\includegraphics[width=8cm]{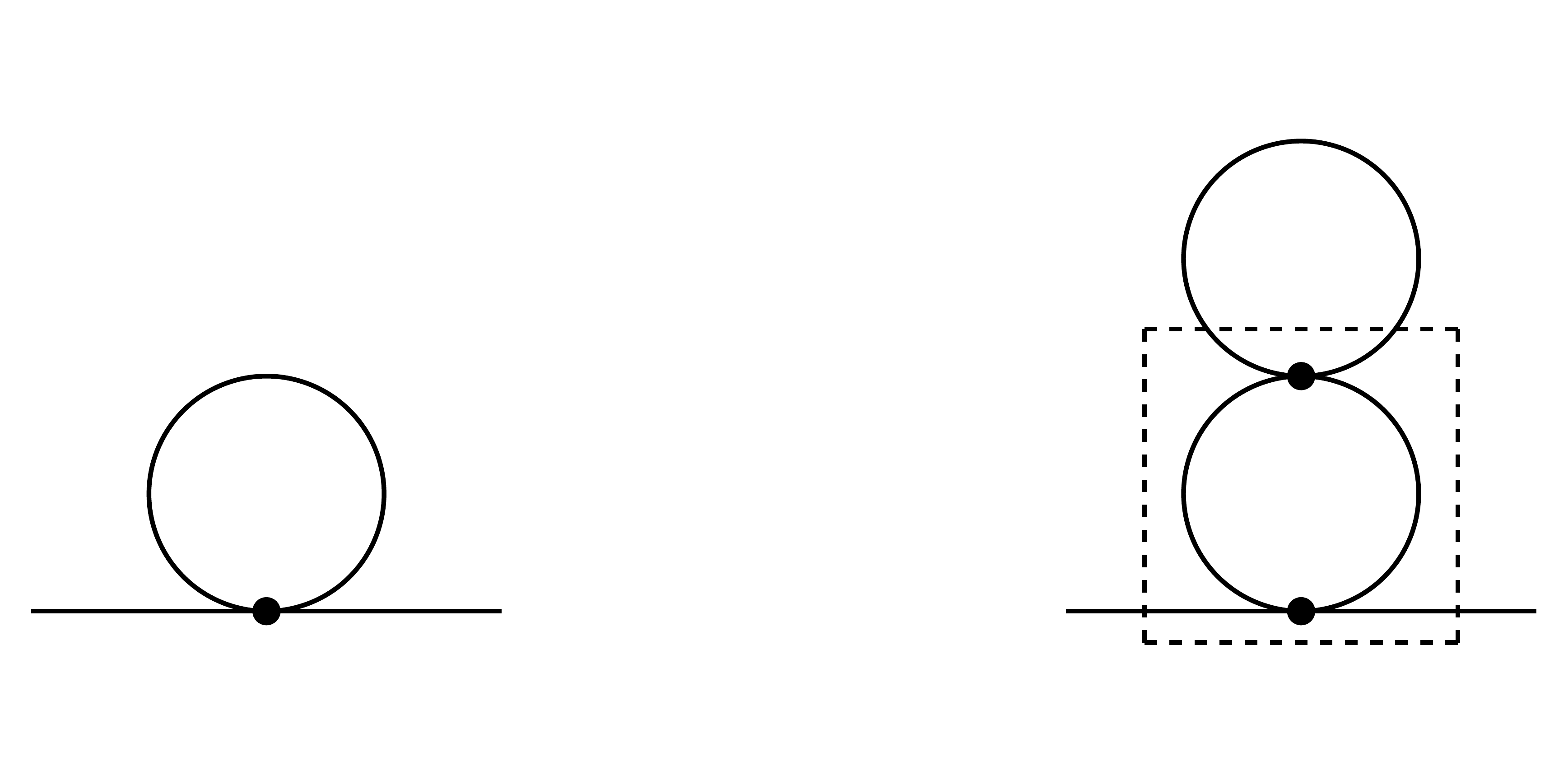}
\caption{The leading order contribution to the self-energy (left) and it first perturbative correction (right). The subdiagram isolated by the dashed line is divergent. This divergence corresponds to a subdivergence of the leading order self-energy when the latter is evaluated with a self-consistent propagator. Such a subdivergence is eliminated by a renormalization of the coupling constant. }\label{fig:subdivetadpole}
\end{center}
\end{figure}

With the self-energy given in Eq.~(\ref{gap2l0}), the gap equation reads
 \beq\label{gap2l0b}
 m^2=m_{\rm b}^2+\frac{\lambda_{\rm b}}{2}I(m)=m^2+\delta m^2+\frac{\lambda_{\rm b}}{2}I(m),
 \eeq
 where we have used Eq.~(\ref{renormcond})  to identify the  renormalized mass $m$ with the solution of the gap equation. The one-loop self-energy is divergent when $\Lambda_{\rm uv}\to \infty$. This divergence may be absorbed in the mass counterterm
 \beq\label{deltamnaif}
 \delta m^2=-\frac{\lambda_{\rm b}}{2}I(m).
 \eeq
  However, 
 by doing this simple subtraction, one ignores the fact that there are subdivergences hidden in $I(m)$, and failing to eliminate those properly may lead to difficulties. This is the case in particular in finite temperature calculations.  Such subdivergences appear explicitly when one performs a perturbative analysis, i.e., expand the self-consistent propagator in powers of the coupling, and they are related to the renormalization of the coupling constant (see Fig.~\ref{fig:subdivetadpole} for an explicit example at two-loop perturbative order).  

 We  show now that such subdivergences are eliminated when one simultaneously renormalize the self-energy and the BS equation. For better illustration, we  consider a calculation at finite temperature and  write the relevant integrals as follows
\beq
I(M)=I_0(M)+I_T(M),\nn
J(M)=J_0(M)+J_T(M)
\eeq
where $I_0(M)$ and $I_T(M)$ denote respectively  the zero temperature  and the finite temperature contributions to  $I(M)$, and similarly for 
$J_0$ and $J_T$ ($J=-\del I/\del M^2$). At one-loop order this separation is not ambiguous. 
 We denote by $M$ the solution of the finite temperature gap equation, the renormalized mass $m$ being given by the solution of this equation at zero temperature. Similarly, we call $\Gamma^{(4)}(M)$ the solution of the BS equation, with the renormalized coupling given by $\lambda=\Gamma^{(4)}(m)$ at $T=0$. We have
 \beq
M^2=m^2+\delta m^2+\frac{1}{2}(\lambda+\delta\lambda)I(M),\qquad \frac{1}{\Gamma^{(4)}(M)}=\frac{1}{\lambda+\delta\lambda}+\frac{1}{2}J(M).
\eeq
At this point, one may impose the renormalization conditions (at $T=0$) and determine the counterterms. We get, 
\beq\label{deltalambda2l}
\delta\lambda=\frac{\lambda}{2}J_{0}(m)\,\frac{\lambda}{1 -\frac{\lambda}{2} J_{0}(m)},\qquad \delta m^2=-\frac{\lambda+\delta\lambda}{2} I_0(m).
\eeq
Using this expression for the mass counterterm, we rewrite the gap equation as follows
\beq
M^2-m^2=\frac{1}{2}(\lambda+\delta\lambda)\left[I(M)-I_0(m)\right].
\eeq
Now, we have
\beq
I(M)&=&I_0(m)+\left( M^2-m^2 \right)\left.\frac{\del I_0}{\del M^2}\right|_{M=m}+C(M,m)\nn
&=& I_0(m)-\left( M^2-m^2 \right)J_0(m)+C(M,m)
\eeq
where $C(M,m)$ is a finite quantity, with $C(M,M)=I_T(M)$. Note how the quadratic divergence drops in the difference $I(M)-I_0(m)$, leaving a logarithmic divergence that cancels against that of $J_0(m)$ (the divergent intermediate loop in Fig.~\ref{fig:subdivetadpole}). By using the explicit expression of $\delta \lambda$ in terms of $J_0(m)$ given above, Eq.~(\ref{deltalambda2l}), one finally obtains the following simple form of the gap equation
\beq
M^2-m^2=\frac{\lambda}{2}C(M,m).
\eeq
At zero temperature, $M=m$ is obviously solution (since then $C(m,m)=0$). At finite temperature the equation remains finite. 
Observe that the elimination of the subdivergences has been crucial to obtain this simple result. This was achieved by  considering  the ``internal structure'' of the self-consistent propagators in order to exhibit the hidden subdivergences. 
As we shall see in the next section, handling subdivergences is made much easier with the flow equations. 

Before moving to the next section, we mention that one could also impose the renormalization conditions directly at finite temperature. In this case 
\beq
\delta m^2=-\frac{\lambda+\delta\lambda}{2} I(m)
\eeq
and the gap equation becomes $M^2=m^2$ which is the simple statement that the renormalized mass is chosen to coincide with the solution of the gap equation. Here, it may seem that the four-point function plays no role in the gap equation. This is however an illusion since, in this scheme, the renormalized mass is temperature dependent, and, when inquiring how it depends on the temperature the four-point function reemerges. Indeed, since the bare mass $m_{\rm b}^2=m^2+\delta m^2$ is temperature independent, on has
\beq
0=\frac{\rmd m_{\rm b}^2}{\rmd T}=\frac{\rmd m^2}{\rmd T}-\frac{\lambda+\delta\lambda}{2}\frac{\rmd I}{\rmd T}\,,
\eeq
where we used that $\lambda_{\rm b}=\lambda+\delta\lambda$ is also temperature independent. Now the temperature dependence of $I$ is either explicit, through $I_T$, or implicit through $m^2$. Then, one finds
\beq
0=\frac{\rmd m^2}{\rmd T}-\frac{\lambda+\delta\lambda}{2}\left[-J(m^2)\frac{\rmd m^2}{\rmd T}+\frac{\partial I}{\partial T}\right]\,,
\eeq
where we used that $\partial I/\partial m^2=-J(m^2)$. Solving for ${\rmd m^2}/{\rmd T}$, one find eventually
\beq
\frac{\rmd m^2}{\rmd T}=\frac{\Gamma^{(4)}(m)}{2}\frac{\partial I}{\partial T}\,,
\eeq
which involves the four-point function, as announced.

\subsection{Renormalization with flow equations}\label{sec:renromtwoloop}
In order to illustrate the developments in Sects.~\ref{sec:42} and \ref{sec:2PIren} we write the flow equations for $m_\kappa^2$ and $\lambda_\kappa$ as follows (here, for simplicity, we switch back to the zero temperature case)
\beq\label{flowformal2}
\del_\kappa m_\kappa^2 = {\cal F}_\kappa^{(2)}(m_\kappa^2,\lambda_\kappa),\qquad \del_\kappa \lambda_\kappa = {\cal F}_\kappa^{(4)}(m_\kappa^2,\lambda_\kappa),
\eeq
and recall the important properties that the flows ${\cal F}_\kappa^{(n)}$ are finite. It is convenient however, for integrating the equations, to keep an ultraviolet cutoff. We shall then be able to illustrate the strategies followed respectively in Sects.~\ref{sec:2PIren}  and \ref{sec:42}.  

In the first case, that of Sect.~\ref{sec:2PIren}, one writes the solution of the flow equations formally as 
\beq
m^2_\kappa=m_{\rm b}^2+ \int_\infty^\kappa \rmd\kappa' \left.{\cal F}^{(2)}_{\kappa'}(m^2_{\kappa'},\lambda_{\kappa'})\right|_{\Lambda_{\rm uv}},\qquad  \lambda_\kappa=\lambda_{\rm b} + \int_\infty^\kappa \rmd\kappa' \left.{\cal F}^{(4)}_{\kappa'}(m^2_{\kappa'},\lambda_{\kappa'})\right|_{\Lambda_{\rm uv}}\,,  
\eeq
where we have made explicit the dependence of the flows on the ultraviolet cutoff. The initial condition is set at the scale $\Lambda\gg \Lambda_{\rm uv}$, actually $\Lambda\to\infty$, and we used  $m_{\Lambda\to\infty}^2=m_{\rm b}^2$ and $\lambda_{\Lambda\to\infty}= \lambda_{\rm b}$.  By replacing the bare parameters by their expressions (\ref{barerenrom}) in terms of renormalized ones and counterterms, one recovers the formal expressions (\ref{eq:ct1}) and (\ref{eq:ct3}) of the counterterms (note that $\delta\lambda$ here is the same as what is called $\delta\tilde\lambda$ in Sect.~\ref{sec:2PIren}):
\beq
\delta m^2=-\int_\infty^0 \rmd\kappa' \left.{\cal F}^{(2)}_{\kappa'}(m^2_{\kappa'},\lambda_{\kappa'})\right|_{\Lambda_{\rm uv}},\qquad \delta \lambda=-\int_\infty^0 \rmd\kappa' \left.{\cal F}^{(4)}_{\kappa'}(m^2_{\kappa'},\lambda_{\kappa'})\right|_{\Lambda_{\rm uv}}\,.
\eeq
Explicit expressions are easily obtained by integrating the flow equation Eq.~(\ref{eqnforg}) for $\lambda_\kappa$  between the scale $\Lambda$ and the scale $\kappa$. One gets
\beq\label{GammaLambda}
\frac{1}{\lambda_\kappa}-\frac{1}{\lambda_\Lambda}=\int_\Lambda^\kappa \rmd\kappa'\int_{q<\Lambda_{\rm uv}} G_{\kappa'}(q)\del_{\kappa'}G_{\kappa'}(q), 
\eeq
or, letting $\Lambda\to \infty$, and setting $\lambda_{\kappa=0}=\lambda$, 
\beq\label{GammaLambda3}
\frac{1}{\lambda}-\frac{1}{\lambda_{\rm b}}=\int_\infty^0\rmd\kappa'\int_{q<\Lambda_{\rm uv}} G_{\kappa'}(q)\del_{\kappa'}G_{\kappa'}(q)=\frac{1}{2}\int_{q<\Lambda_{\rm uv}}G^2(q),  
\eeq
where we have used $G_{\Lambda\to\infty}=0$. The expression of the counterterms, already given in Eq.~(\ref{deltalambda2l}), follows by replacing $\lambda_{\rm b}$ by $\lambda +\delta \lambda$. One may proceed similarly for the mass. By using directly the solution given in Eq.~(\ref{gap2lkappa}), one gets
\beq\label{mkaplam}
    \frac{m_\kappa^2-m_\Lambda^2}{\lambda_{\rm b}}=\frac{1}{2}\int_{q<\Lambda_{\rm uv}} \left[G_\kappa(q)-G_\Lambda(q)  \right].
   \eeq
  Letting $\Lambda\to \infty$, using $m^2_{\Lambda\to\infty}=m_{\rm b}^2=m^2+\delta m^2$, and setting $m^2_{\kappa=0}=m^2$, one immediately recovers the expression (\ref{deltalambda2l}) of the mass counterterm. In fact, it is interesting to rewrite this equation (\ref{mkaplam}) in terms of the renormalized parameters
\beq\label{gap2l2b}
 \frac{m_\kappa^2}{\lambda}=\frac{m^2}{\lambda}+\frac{1}{2}\int_{q<\Lambda_{\rm uv}}
 \left[ G_\kappa(q)-G(q) +(m_\kappa^2-m^2) G^2(q) \right]. \eeq
One recognizes in this integral the pattern of the elimination of  divergences already discussed at the end of Sect.~\ref{twolooprenorm}, with the last term proportional to $G^2$ cancelling the logarithmic divergence in the difference $G_\kappa-G$, and  leaving a potential quadratic divergence that disappears in the difference. This simple example illustrates the efficiency of the flow equation in dealing with (hidden) subdivergences. 

We  now return to the flow equations (\ref{flowformal2}), and integrate them from 0 to $\Lambda$, and $\Lambda$ is now kept finite. For $\lambda_\kappa$ the result is written in Eq.~(\ref{GammaLambda}), which we can rewrite as follows
\beq\label{GammaLambda2}
\frac{1}{\lambda_\kappa}-\frac{1}{\lambda_\Lambda}=\frac{1}{2}\int_{q<\Lambda_{\rm uv}}\left[ G^2_\kappa(q)-G_\Lambda^2(q)   \right]. 
\eeq
The integral is now finite, which results from the fact that the integration over the scale parameter $\kappa$ is limited to the finite interval $[0,\Lambda]$. The limit $\Lambda_{\rm uv}\to \infty$ can then be trivially taken. One sees that $\lambda_\Lambda$ is the quantity that is related by the flow to the renormalized coupling $\lambda=\lambda_{\kappa=0}$. In other word this is the proper initial condition on the flow, when this is initialized at $\Lambda<\Lambda_{\rm uv}$. Note how  $\lambda_\kappa$ remains independent of $\Lambda$: as one varies $\Lambda$, one changes the amount of fluctuations that are integrated out (in the integral in the right-hand side), and that change is exactly compensated by a change in the value of $\lambda_\Lambda$.  If $\Lambda$ is large enough, the flow of $\lambda_\Lambda$ is simple, $\Lambda \del_\Lambda \lambda_\Lambda\simeq -1/(32 \pi^2)$. This specifies how $\lambda_\Lambda$ has to change as one modifies the initial scale $\Lambda$ so as to leave invariant the flow of $\lambda_\kappa$  for $\kappa<\Lambda$, and in particular  the renormalized coupling $\lambda=\lambda_{\kappa=0}$. Similar considerations apply of course for the dependence of the bare parameters or the counterterms on $\Lambda_{\rm uv}$, as given for instance by Eqs.~(\ref{gapuv}) for the bare parameters, or Eqs.~(\ref{deltalambda2l})  for  the counterterms 

Similar manipulations can be done for the mass. One obtains easily the following equation,    
\beq\label{gap2l2a}
 \frac{m_\kappa^2}{\lambda_\Lambda}=\frac{m_\Lambda^2}{\lambda_\Lambda}+\frac{1}{2}\int_{q<\Lambda_{\rm uv}} \left[ G_\kappa(q)-G_\Lambda(q) +(m_\kappa^2-m_\Lambda^2) G_\Lambda^2(q) \right], \eeq
 in terms of the parameters at scale $\Lambda$. Here again we can set $\Lambda_{\rm uv}\to \infty$. By choosing $\lambda_\Lambda$ and $m_\Lambda$, the renormalized parameters at the scale $\Lambda$, as initial conditions  for the coupled flow equations for  $\Gamma_\kappa^{(4)}$ and $m_\kappa$, one eliminates the need to determine the counterterms of the standard renormalization procedure. Renormalization here amounts to appropriate subtractions at the scale $\Lambda$, as shown for instance in Eqs.~(\ref{GammaLambda}) and (\ref{gap2l2a}). Once such subtractions are made at the initial scale, the coupled equations take care automatically of all potential four-point subdivergences.

\section{Four point insertions in $\del_\kappa {\cal I}_\kappa$}  \label{app:great}
\begin{figure}[htbp]
\begin{center}
\includegraphics[width=12cm]{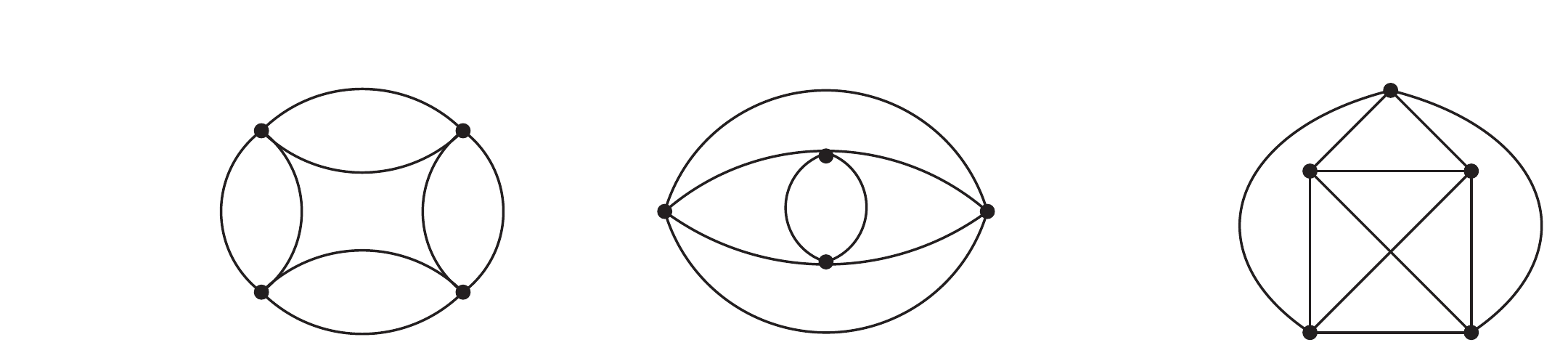}
\caption{Some five and six-loop diagrams that contribute to $\Phi$, with various topologies.}\label{fig:fourfivePhi}
\end{center}
\end{figure}

\begin{figure}[htbp]
\begin{center}
\includegraphics[width=12cm]{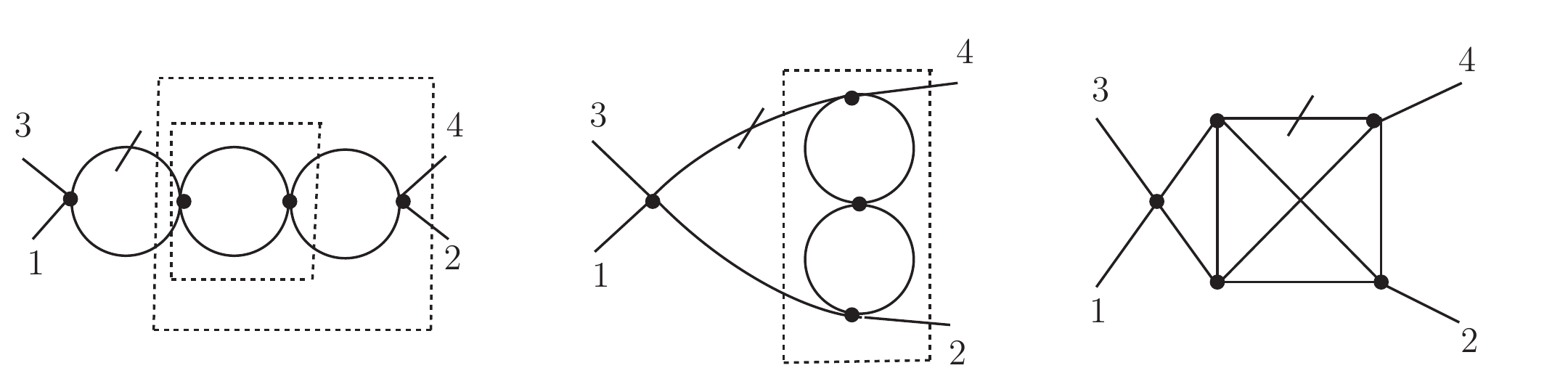}
\caption{Some diagrams that contribute to ${\cal I}_{12;34}$, and that can be deduced from those of $\Phi$ in Fig.~\ref{fig:fourfivePhi} by functional differentiation.  The lines that carry a slash are the lines that are opened when differentiating ${\cal I}$, leading effectively to diagrams for six-point functions. The dotted lines isolate four-insertions.  The middle bubble in the left diagram is an insertion embedded in the larger insertion made by  the two most right bubbles which constitute a maximal insertion for $\del {\cal I}$. The diagram on the right is irreducible, i.e., it contains only trivial four-insertions.}\label{fig:fourfiveI}
\end{center}
\end{figure}

In this section we generalize the analysis presented in Sect.~\ref{sec:flowI} to arbitrary diagrams contributing to the flow of ${\cal I}_\kappa$. The diagrams contributing to ${\cal I}_\kappa$  are obtained from the skeleton diagrams of $\Phi$ by opening two lines, and the derivative $\del_\kappa$ entails opening one extra line, thereby making the diagrams of  $\del_\kappa {\cal I}_\kappa$ effectively those of a six-point function, $\delta {\cal I}/\delta G$. Examples of diagrams contributing to $\Phi$ and ${\cal I}$, and exhibiting various topologies, are given below (Figs.~\ref{fig:fourfivePhi} and \ref{fig:fourfiveI}) at five and six-loop orders for $\Phi$. Our goal in this Appendix is to analyze the four-point insertions in the diagrams contributing to $\del_\kappa {\cal I}_\kappa$ and show how these can be resummed into complete four-point functions.  At the end of this Appendix, we mention possible  generalizations of this analysis to other classes of diagrams as well as to other $n$-point functions. 

Let us first specify what we mean by a four-point insertion, which we shall refer to simply as a four-insertion. Consider a diagram ${\cal D}$ contributing to $\del_\kappa{\cal I}_\kappa$, noted $\del{\cal I}$ for short. We draw this, as in Fig.~\ref{fig:fourfiveI}, as a diagram contributing to ${\cal I}$, with one line carrying a slash, which is the line to be opened to yield the corresponding diagram of $\delta {\cal I}/\delta G$.  The external lines of the diagrams are labelled by a number, specifying the channel in which ${\cal I}$ is irreducible (see \ref{sec:relation} for more details). Thus, for instance, the diagrams in Fig.~\ref{fig:fourfiveI} are irreducible in the channel (12;34). The external lines are considered parts of the diagram.   A  four-insertion  is any 1PI four-point subgraph of ${\cal D}$, possibly including some of the external lines of ${\cal D}$, that is obtained by cutting up to four internal lines of ${\cal D}$,  such  that, after the cut, the diagram splits into two disconnected pieces.  Including the external lines in the process  allows us in particular to consider as insertions the tree-level vertices (as well as insertions containing vertices attached to these external lines). The four-insertions that isolate a single tree-level vertex will be called trivial. A diagram in which only trivial four-insertions can be identified will be said to be irreducible. Examples are displayed in Fig.~\ref{fig:fourfiveI}. 

\begin{figure}[htbp]
\begin{center}
\includegraphics[width=12cm]{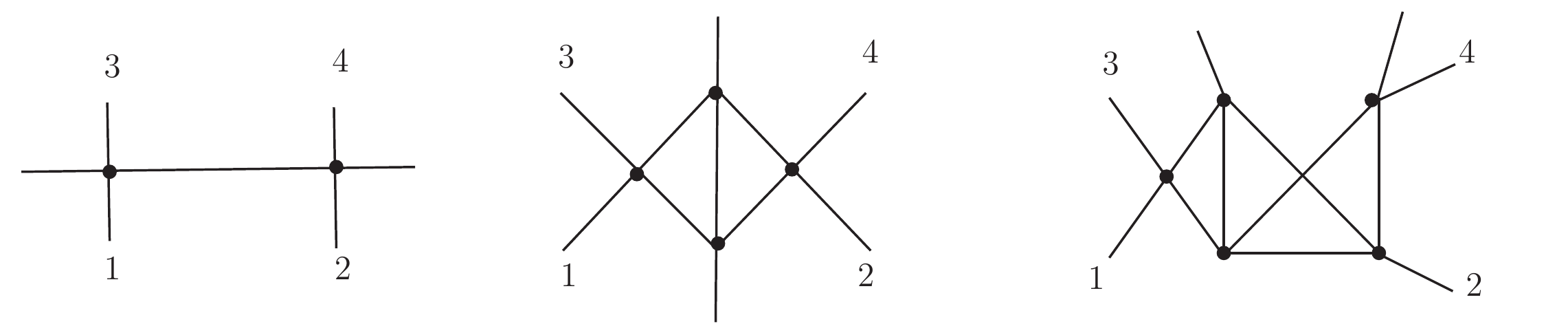}
\caption{Some of the irreducible diagrams that contribute to $\del{\cal I}$ that are deduced from the three diagrams of Fig.~\ref{fig:fourfivePhi} by the procedure indicated in the text. These diagrams contain only trivial four-insertions. The left diagram is clearly one-line reducible: it is split into two parts when the line that joins the two vertices is cut. }\label{fig:fourfiveS}
\end{center}
\end{figure}

 Given a diagram ${\cal D}$, it is straightforward to make the list of all its four-insertions. Elements of this list that have the same topology, but involve distinct elements of ${\cal D}$, are to be considered as distinct.  Among the four-insertions of ${\cal D}$ some can be imbedded into larger four-insertions (see Fig.~\ref{fig:fourfiveI} for an example). We call {\it maximal insertion} a four-insertion which is not itself a four-insertion in a larger one. 
  We shall argue in this appendix that these maximal insertions can be identified without ambiguity in any diagram  ${\cal D}$ contributing to $\del {\cal I}$. It is then possible to 
  substitute them by trivial ones, i.e., by tree-level vertices. In so doing,  one transforms ${\cal D}$ into an irreducible diagram, in the sense defined above, which we shall refer to as a \textit{four-skeleton}.  Examples of four-skeletons  contributing to $\del{\cal I}$ are given in Fig.~\ref{fig:fourfiveS}. That the maximal four-insertions can be identified unambiguously means  that a given diagram of $\del{\cal I}$ has a unique four-skeleton. It follows that  
       one can generate all the diagrams of $\del {\cal I}$ by replacing  the vertices of the four-skeletons by the complete four-point function, which is our ultimate goal (see  Sect.~\ref{sec:flowI}). This replacement does not change the overall irreducibility properties of $\del {\cal I}$: a four-skeleton of $\del{\cal I}_{12;34}$ is irreducible in the channel $(12;34)$ (it contains no four-insertion with lines 1,2 or 3,4 as external legs), and this property subsists  after the replacement of the trivial vertices by full four-point functions.  It is also clear that all the diagrams of $\del {\cal I}$ can be generated in this way, since each diagram of $\del {\cal I}$ has a unique skeleton. Furthermore, the argument that we shall develop shortly, showing that maximal four-insertions cannot have common elements, indicates that symmetry factors factorize into symmetry factors attached to the skeletons, and those of the individual four-point functions that sit at the vertices of the four-skeletons. This insures that each diagram is properly counted. An explicit verification by direct evaluation of the symmetry factors will be given elsewhere \cite{Urko_overlap}.
    
  The issues that we are considering in this appendix bear similarities with those addressed at the beginning of this paper about 2PI, or more generally $n$PI, approximations. Among typical such issues,  let us recall, for instance, that vacuum diagrams do not, in general, have unique  two-skeletons, and the substitution of the bare propagators by dressed ones in the Luttinger-Ward functional generates an over-counting of the original diagrams.   By contrast,  the two-skeletons of the two-point function (or higher $n$-point functions) can be identified without ambiguity and it is then possible  to generate, without double counting,  all the diagrams of the two-point function by substituting in its two-skeletons   the bare propagators with the full ones. However, the notion of maximal four-insertion, and the fact that we consider external lines in the definition of the four-insertions, are important elements that make the present analysis deviate somewhat  from more standard diagrammatic analysis in $n$PI formalisms (see e.g. \cite{DeDominicis:1964b}). Note for instance that the analysis of this appendix will apply to $\del{\cal I}$ but not to  ${\cal I}$. For one thing, with our definition, each diagram of ${\cal I}$ is a (maximal) four-insertion, so that ${\cal I}$ has a unique skeleton, the tree-level vertex. But  the replacement of this vertex by a full four-point function would not be legitimate since it would not preserve the irreducibility of ${\cal I}$. It is easy to see that no such constraint remains on the four-insertions of $\del{\cal I}$.

The proof that maximal four-insertions in the diagrams of $\del{\cal I}$ can be non ambiguously identified relies on the property that, in such diagrams, any two maximal four-insertions cannot have any element in common, that is, they do not overlap. In order to establish the latter property we shall consider first the case of a two-point function for which such overlaps are possible, and we shall identify how  two insertions ${\cal M}_1$ and ${\cal M}_2$ (not necessarily maximal) can overlap. Then we shall return to the diagrams of $\del {\cal I}$  and,  by using a reductio ad absurdum argument, we will show that no overlap of two four-insertions can occur there  if we assume ${\cal M}_1$ and ${\cal M}_2$ to be maximal.

  Consider then a diagram ${\cal D}$, in which two insertions ${\cal M}_1$ and ${\cal M}_2$ have been identified.  The insertion ${\cal M}_1$ (a similar discussion can be made for ${\cal M}_2$) has been obtained by cutting a quadruplet of lines in the diagram, including possibly some of the external lines of ${\cal D}$. We shall call these four lines the external legs of the insertion: these are composed, as just said,  of cut internal lines of ${\cal D}$ and possibly some external lines of ${\cal D}$.  In ${\cal M}_1$, there could also exist sets of internal lines such that, once cut,  the insertion splits into two disconnected parts. We shall  call {\it connecting lines}  such internal lines.  
 \begin{figure}[h]
\begin{center}
\includegraphics[angle=0,scale=0.75]{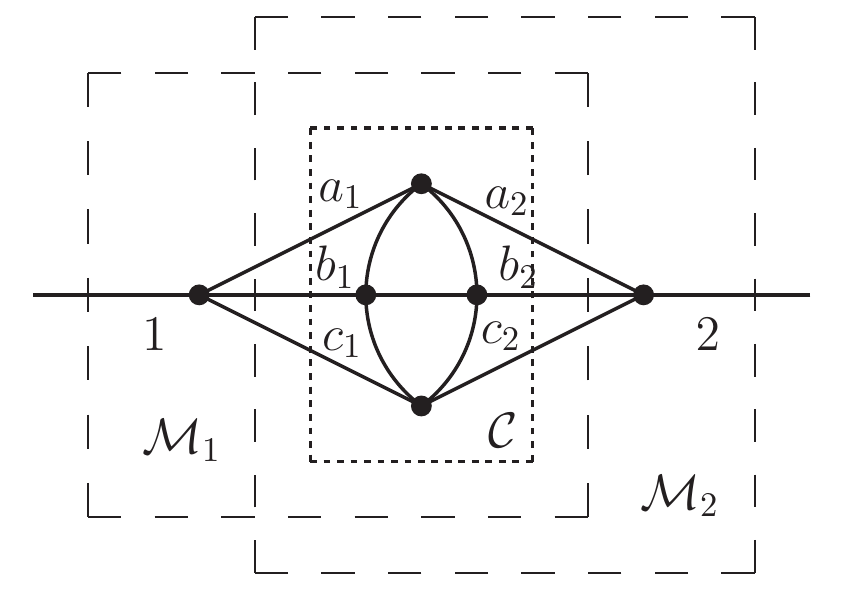}
\caption{Example of two overlapping (non maximal) four-insertions ${\cal M}_1$ and ${\cal M}_2$,  in a diagram ${\cal D}$ with two external lines ($n=2$).}\label{fig:max_overlap_2A}
\end{center}
\end{figure}
 To make things more concrete consider for instance the diagram of Fig.~\ref{fig:max_overlap_2A}, that contributes to a two-point function ($n=2$). 
   In Fig.~\ref{fig:max_overlap_2A},  ${\cal M}_1$ is isolated from ${\cal D}$ by cutting the lines $a_2,b_2,c_2$. Together with the external line labelled 1, these lines constitute the external legs of ${\cal M}_1$. By cutting the lines $a_1,b_1,c_1$ one splits ${\cal M}_1$ into two disconnected pieces, one of them containing the vertex attached to the line 1, the other part being a six point function with $a_1,b_1,c_1$ and $a_2,b_2,c_2$ as external legs.  The lines $a_1,b_1,c_1$ are connecting lines of ${\cal M}_1$. Similarly, the insertion called ${\cal M}_2$, has $a_1,b_1,c_1$ and 2 as external legs, and $a_2, b_2, c_2$ as connecting lines. ${\cal M}_2$ and ${\cal M}_1$ share a common subgraph, the six-point function labelled ${\cal C}$, already identified as part of ${\cal M}_1$. 
 The structure that we see emerging on this simple example is that depicted more generally in Fig.~\ref{fig:generic} below. It reveals one possible type of overlap between four-insertions which, however, are here non maximal. 

\begin{figure}
\begin{center}
\includegraphics[angle=0,scale=0.5]{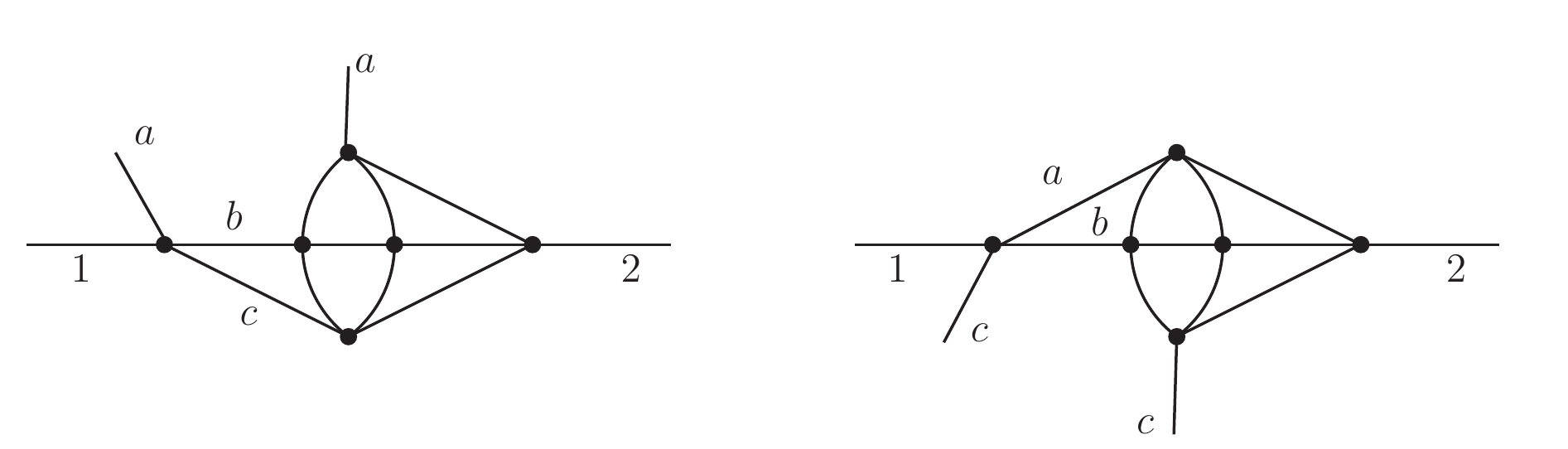}
\caption{Example of two overlapping maximal four-insertions in a diagram with $n=2$. The two diagrams represent two distinct four-insertions that have all their vertices, and all their lines but two  in common.}\label{fig:max_overlap_2B}
\end{center}
\end{figure}

There is  another possibility of overlap, which is  illustrated  in Fig.~\ref{fig:max_overlap_2B}.
  The two displayed four-insertions there are now maximal, and overlapping. They share all their vertices and their lines, except  the line $a$, opened in the left one and  $c$ opened in the right one. There are no connecting lines. The two lines $a$ and $c$  play a special role: they are not lines of ${\cal C}$, the set of common elements,  but they join two vertices that belong to ${\cal C}$. For this reason, such lines will be called {\it returning lines}. Each of these lines belongs to one of the four-insertions but not to the other. Typically, opening such a line in an $n$-point function generates a $n+2$ point function.  Thus, returning lines may appear as parts of the external legs of a four-insertion, but they play no role in isolating the insertion from the rest of the diagram. For a more general discussion on how connecting and returning lines describe the insertion of a graph into another graph, see \cite{Urko_overlap}.

With these elements in mind, we  now return to the diagram displayed in Fig.~\ref{fig:generic}, which is a part of a bigger diagran ${\cal D}$. This exhibits two overlapping insertions ${\cal M}_1$ and ${\cal M}_2$ and has $n$ external lines (we keep here $n\geq 4$  as a free parameter). We denote by ${\cal C}$ the part common to  ${\cal M}_1$ and ${\cal M}_2$. We assume that ${\cal M}_1$ contains $m_1>0$ lines connecting ${\cal C}$ and ${\cal M}_1/{\cal C}$ , where ${\cal M}_1/{\cal C}$ denotes the complementary part of ${\cal C}$ in ${\cal M}_1$,\footnote{${\cal M}_1/{\cal C}$ is the set of all the vertices of ${\cal M}_1$ that are not in ${\cal C}$ together with the lines attached to them. The $m_1$ connecting lines are not counted as part of the external lines of ${\cal M}_1/{\cal C}$.} and similarly for ${\cal M}_2$.  In order to isolate ${\cal M}_1$ in ${\cal D}$, one has to cut a number of lines:  these include the connecting lines of ${\cal M}_2$, as well as possibly internal lines of ${\cal D}$. That is, the lines whose number is denoted by  $n_{\cal C}$  may contain external legs of ${\cal D}$  as well as cut internal lines of ${\cal D}$ that connect a vertex of ${\cal C}$ to the rest of the diagram. These $n_{\cal C}$ lines  are common to both ${\cal M}_1$ and ${\cal M}_2$.  Similarly,  the $n_{{\cal M}_1/{\cal C}}$ external lines of ${\cal M}_1/{\cal C}$ may contain external lines of ${\cal D}$ and cut internal lines of ${\cal D}$.  The total number of external lines of the four-insertion ${\cal M}_1$, $ n_{{\cal M}_1}$,  is therefore $n_{{\cal M}_1}=n_{{\cal M}_1/{\cal C}}+n_{\cal C}+m_2$. The same reasoning holds for ${\cal M}_2$ so that we have
\beq\label{nM12}
4 = n_{{\cal M}_1}=n_{{\cal M}_1/{\cal C}}+n_{{\cal C}}+m_2
= n_{{\cal M}_2}=n_{{\cal M}_2/{\cal C}}+n_{{\cal C}}+m_1\,,
\eeq
where we have used the fact that both ${\cal M}_1$ and ${\cal M}_2$ are four-insertions, so that $n_{{\cal M}_1}=n_{{\cal M}_2}=4$.  
\begin{figure}[h]
\begin{center}
\includegraphics[angle=0,scale=0.65]{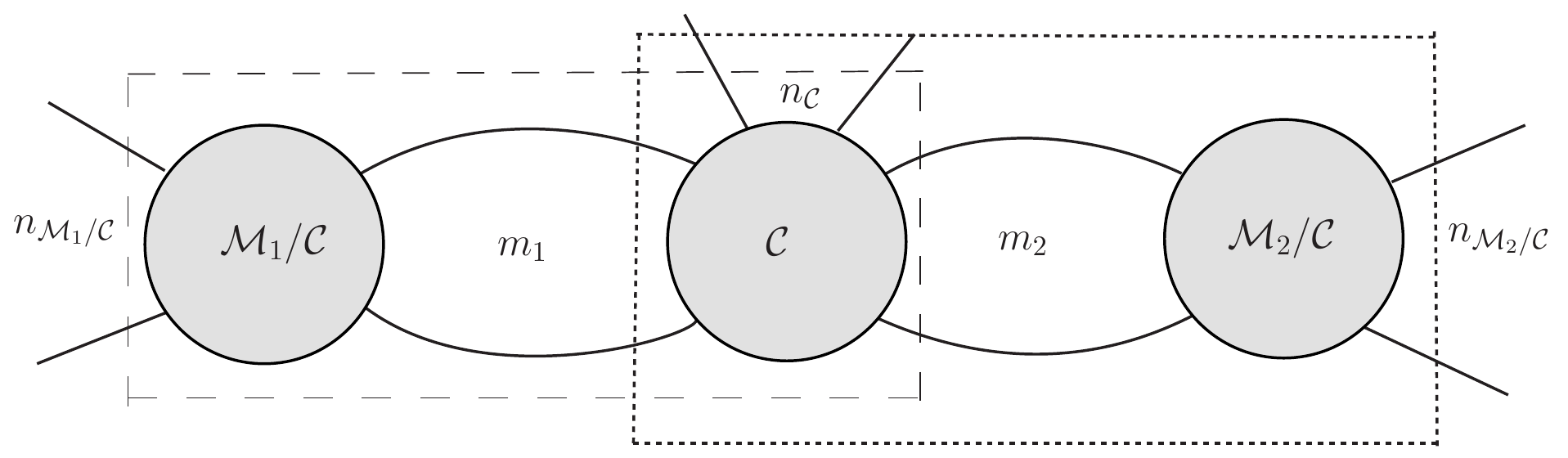}
\caption{Generic configuration with an overlap of two maximal vertex insertions.}\label{fig:generic}
\end{center}
\end{figure}

At this point, we note  that  ${\cal M}_1\cup{\cal M}_2$, the graph that contains all the elements of ${\cal M}_1$ and ${\cal M}_2$,  can be viewed as a 1PI subgraph of ${\cal D}$ with $n_{{\cal M}_1/{\cal C}}+n_{{\cal M}_2/{\cal C}}+n_{\cal C}$ external legs. This number is necessary greater than 2 since we started from a diagram ${\cal D}$ that  does not contain any self-energy insertion and our assumption that ${\cal D}$ has at least four external legs exclude the case $n=2$.  Since it is not possible to isolate three-point functions in ${\cal D}$, we must have then $n_{{\cal M}_1/{\cal C}}+n_{{\cal M}_2/{\cal C}}+n_{{\cal C}}\geq 4$. The case $n_{{\cal M}_1/{\cal C}}+n_{{\cal M}_2/{\cal C}}+n_{{\cal C}}=4$ is excluded because it would imply that both ${\cal M}_1$ and ${\cal M}_2$ are insertions of   ${\cal M}_1\cup {\cal M}_2$, which is impossible since ${\cal M}_1$ and ${\cal M}_2$ are maximal. We conclude therefore that  $n_{{\cal M}_1/{\cal C}}+n_{{\cal M}_2/{\cal C}}+n_{{\cal C}}> 4$
which, with the help of  Eq.~(\ref{nM12}) we can rewrite as 
\beq\label{constraintnC}
m_1+m_2+n_{{\cal C}}<4.
\eeq
But $m_1\ge 2$ and $m_2\ge 2$ since ${\cal M}_1$ and ${\cal M}_2$ are 1PI subgraphs, so that 
$
 m_1+m_2+n_{\cal C}\ge 4,
$
in  contradiction with Eq.~(\ref{constraintnC}). 
We conclude that two four-insertions of ${\cal D}$ cannot share a common subgraph ${\cal C}$ with the structure exhibited in Fig.~\ref{fig:generic}. 

The acute reader may have noticed that we did not take into account the possibility that returning lines could be attached to ${\cal C}$. That is, we have ignored the possibility that the dotted box on the right of Fig.~\ref{fig:generic} cuts one or various lines that start at a vertex of ${\cal C}$ and return to a vertex of ${\cal C}$ while not being cut by the dashed box on the left. One could also consider similar returning  lines that are cut by the left box but not by the right one. In fact it is easy to argue that such cases cannot occur here. Assume indeed that there is, for instance, one returning line contributing to ${\cal M}_1$. It would contribute to two external lines of ${\cal M}_1$, meaning that the insertion that has been isolated from ${\cal D}$ before opening the returning line is a two-point function. But this is excluded by our assumptions that ${\cal D}$ is 2PI and $n\geq 4$. 

Now, even though this possibility is excluded in the present case, it is nevertheless interesting to push the reasoning by keeping open the possibility of returning lines, in particular in view of potential generalizations of the present analysis \cite{Urko_overlap}. To do so, let us assume that in addition to the lines indicated in Fig.~\ref{fig:generic} there are $r_1$ returning lines in ${\cal M}_1$ and $r_2$ returning lines in ${\cal M}_2$. It is easy to see that the mere effect of their presence is to change $m_1\to m_1+2r_1$ and similarly for $m_2$. The inequality (\ref{constraintnC}) gets modified by the same substitution and leads to the same contradiction, and hence the same conclusion. 

We need now examine the two possibilities that are not covered by the generic case that we have just considered and which assumed that both $m_1$ and $m_2$ were different from zero. The first situation is that in which the two four-insertions ${\cal M}_1$ and ${\cal M}_2$ share all their vertices. This corresponds to $m_1=m_2=0$. If we ignore possible returning lines, it is clear that ${\cal M}_1={\cal M}_2$ which is just the trivial case of overlap between a four-insertion and itself. With the possibility of returning lines taken into account,  Eq.~(\ref{nM12}) gets replaced by $4=n_{\cal C}+2r_1=n_{\cal C}+2r_2$ , so that  either $r_1=r_2=1$ and $n_{\cal C}=2$ or $r_1=r_2=2$ and $n_{\cal C}=0$. In the first case, ${\cal M}_1\cup {\cal M}_2$ is a two-point function and in the second case a zero-point function, but this is just impossible since we excluded these possibilities.

The second situation is that where the common part ${\cal C}$ shares all its vertices with say ${\cal M}_1$,  but where there are still connecting lines within ${\cal M}_2$. This correspond to the case $m_1\ne 0, m_2=0$ (clearly, the case $m_1= 0, m_2\ne 0$ can be treated similarly). If we ignore possible returning lines, it is clear that ${\cal M}_1\subset  {\cal M}_2$  which is not possible since we assumed ${\cal M}_1$ to be maximal. If we include the possibility of returning lines, following the same reasoning as above, we arrive at $m_2+2r_1 +2r_2+n_{\cal C}< 4$. But since $m_2\ge 2$ and at least one of the $r_i$ is different from zero, we have $m_2+2r_1 +2r_2+n_{\cal C}\ge  4$ which again leads to a contradiction.

This concludes our proof that in the diagrams contributing to $\del {\cal I}$ there is no ambiguity in isolating four-vertex insertions, and consequently in identifying the four-skeletons. This proof extends in fact to a wider class of diagrams. Clearly, it applies to any connected two-skeleton diagram that has a number $n\ge 4$ of external lines (this is why we kept $n\ge 4$ as a free parameter in the previous discussion), and by extension, to any disconnected diagram made of such pieces. One example is provided by the diagrams that enter the 2PI  $n$-point functions  $\delta^m {\cal I}/\delta G^m$, with $m>1$ ($m=1$ corresponding to $\del{\cal I}$). These correspond to the sum of $n$-point diagrams that are 2PI with respect to the cuts that leave the legs originating from a given derivative $\delta/\delta G$ on the same side of the cut. Their connected parts involve skeleton graphs with more than four legs. Another case that fits the picture is  $\Gamma^{(p)}$ for $p\geq 4$ once written in terms of two-skeleton diagrams (which it is always possible to do unambiguously), and also $\delta^m \Gamma^{(p)}/\delta G^m $. Note that both $\delta \Gamma^{(4)}/\delta G$ and $\delta {\cal I}/\delta G$, enter the flow equations that are considered in Sect.~\ref{sec:flowI}, once convoluted with $\del_\kappa G_\kappa$. Of course the substitution of the  bare vertices by the full four-point function requires that there  is no  restriction on the type of four-insertions that can appear (such as the restriction of irreducibility already mentioned for ${\cal I}$), a property that needs to be checked for any infinite class of diagrams that one wants to analyse.\\

The previous discussion concerned four-point insertions in connected two-skeleton diagrams with $n\ge 4$ external lines (or disconnected combination of those), the case for which we have shown that the four-point insertions do not overlap. In a diagram with more than six external lines, we may encounter overlapping maximal insertions with more than six external lines. To see that, we shall extend the discussion of the generic situation depicted in Fig.~\ref{fig:generic} to the case of a diagram with $n\ge 6$ external lines built on four-skeletons, with two overlapping insertions ${\cal M}_1$ and ${\cal M}_2$. For simplicity we restrict  the discussion to the case where the considered insertions have 6 external lines, and the assumption of a four-skeleton eliminates returning lines.     The same reasoning as that leading to Eq.~(\ref{nM12}) yields now
\beq\label{eq:sat}
 n_{{\cal M}_1}=n_{{\cal M}_1/{\cal C}}+n_{{\cal C}}+m_2
= n_{{\cal M}_2}=n_{{\cal M}_2/{\cal C}}+n_{{\cal C}}+m_1=
6\,.
\eeq
\begin{figure}[h]
\begin{center}
\includegraphics[angle=0,scale=0.45]{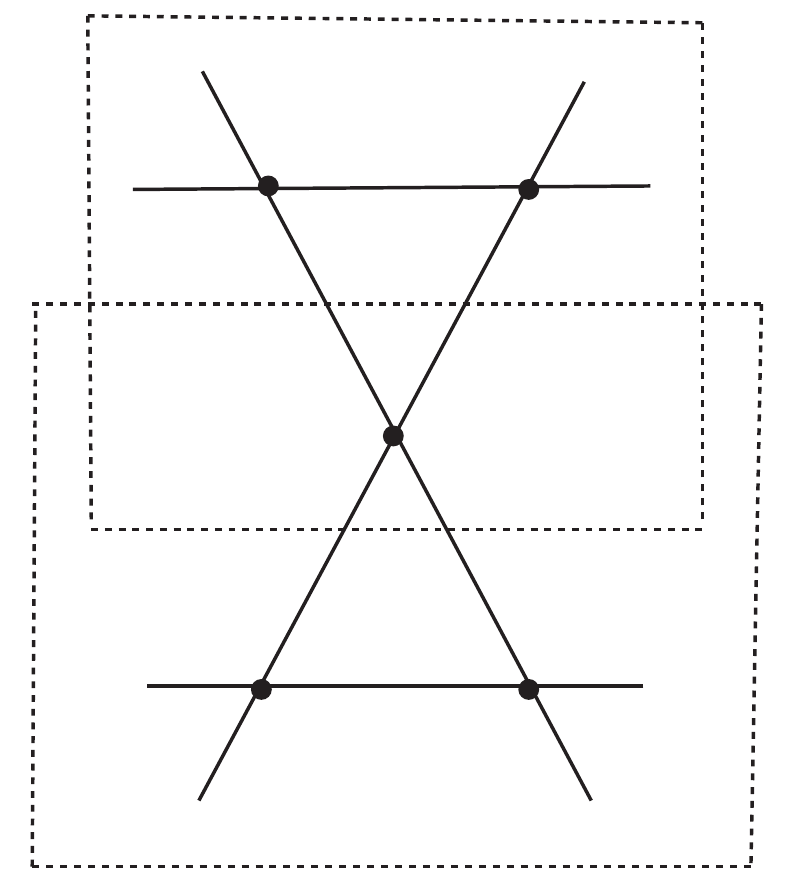}
\caption{Example of overlapping six-vertex insertion within a $8$-point function. }\label{fig:six_overlap}
\end{center}
\end{figure}
Moreover, the union of ${\cal M}_1$ and ${\cal M}_2$ is a 1PI $(n_{{\cal M}_1/{\cal C}}+n_{{\cal M}_2/{\cal C}}+n_{\cal C})$-point vertex function whose number of legs cannot be less than $4$ since we have assumed that ${\cal D}$ is a four-skeleton with more than six legs. This number  of legs must be even, so it cannot be 5.  It cannot be  $6$ either  since this would mean that ${\cal M}_1\cup {\cal M}_2$ is a six-point vertex that contains the maximal six-point vertices ${\cal M}_1$ and ${\cal M}_2$. It follows that
\beq
n_{{\cal C}}+n_{{\cal M}_1/{\cal C}}+n_{{\cal M}_2/{\cal C}}>6\,.
\eeq 
Combining this with (\ref{eq:sat}) together with the fact that $m_i\geq 2$ (since ${\cal M}_i$ is 1PI), one arrives at
\beq
4\leq m_1+m_2+n_{{\cal C}}<6
\eeq
There is indeed room for overlap, with for instance the solution $m_1=m_2=2$ and $n_{{\cal C}}=0$. An  illustration is given  in Fig.~\ref{fig:six_overlap} showing a contribution to the eight-point function with two overlapping maximal six-point vertex insertions.

\section{A simple relation between $\Gamma^{(4)}_{L,\kappa}$ and ${\cal I}_{L,\kappa}$}\label{sec:relation}
Following an analysis similar to that in Ref.~\cite{DeDominicis:1964b}, we present here a simple relation between the four point function $\Gamma^{(4)}$ and its irreducible parts in the three independent channels. 
We  consider here the general situation where the propagator is a function of two positions or two momenta, and we write it simply  as $G_{ij}$ with $G_{ij}=G(x_i,x_j)$ or $G_{ij}=G(p_i,p_j)$. The kernel ${\cal I}$ is then a function of four variables
\beq
{\cal I}_{12;34}\equiv 4\frac{\delta^2\Phi[G]}{\delta G_{12}\delta G_{34}}\,,
\eeq
where the semi-colon indicates the channel in which ${\cal I}$ is irreducible. The diagrams contributing to ${\cal I}_{12;34}$ cannot be split into two disconnected pieces containing respectively the pairs $(1,2)$ and $(3,4)$ by cutting two of its internal lines: they are irreducible in the channel $(12;34)$.  
${\cal I}_{12;34}$ can be used to construct the four-point function $\Gamma^{(4)}_{1234}$ from the Bethe-Salpeter equation
\beq\label{eq:BS10}
\Gamma^{(4)}_{1234} & \!\!\!=\!\!\! & {\cal I}_{12;34}-\frac{1}{2}\,{\cal I}_{12;ij}\,G_{ik}G_{jl}\,\Gamma^{(4)}_{kl34}\nn
& \!\!\!=\!\!\! & {\cal I}_{12;34}-\frac{1}{2}\,\Gamma^{(4)}_{12ij}\,G_{ik}G_{jl}\,{\cal I}_{kl;a34}\,.
\eeq 
Note that $\Gamma^{(4)}_{1234}$ is crossing symmetric, in contrast to the kernel ${\cal I}_{12;34}$, which only obeys the following properties ${\cal I}_{12;34}={\cal I}_{34;12}$ as follows from its definition, and ${\cal I}_{12;34}={\cal I}_{21;34}={\cal I}_{12;43}$ as follows from the symmetry $G_{ij}=G_{ji}$.  But, for instance, it is not invariant under the exchange of $2$ and $3$. 

We shall write the BS equation in the following way
\beq\label{eq:B1}
\Gamma^{(4)}_{1234}={\cal I}_{12;34}+\Delta\Gamma^{(4)}_{12;34},
\eeq
separating the contributions of the diagrams that are irreducible in the channel $(12;34)$ from those which are not. 
One can write similar relations for the other two channels:
\beq\label{eq:BS10b}
\Gamma^{(4)}_{1234}&\!\!\!\!=\!\!\!\!&{\cal I}_{13;24}+\Delta\Gamma^{(4)}_{13;24},\\
\Gamma^{(4)}_{1234}&\!\!\!\!=\!\!\!\!&{\cal I}_{14;23}+\Delta\Gamma^{(4)}_{14;23}.\label{eq:B3}
\eeq 
In fact, when only quartic interactions are present and the field expectation value is assumed to vanish, any diagram contributing to $\Gamma^{(4)}_{1234}$ can be reducible in only one channel.\footnote{Indeed, consider a diagram that admits a cut that leaves $(1,2)$ and $(3,4)$ on each side. Then the diagram writes necessarily $X_{12ij}G_{ik}G_{jl}Y_{kl34}$. Now, if we assume that there is a second possible cut that leaves $(1,3)$ and $(2,4)$ on each side, then,  because $1$ and $2$ cannot be on the same side of the cut, and similarly for $3$ and $4$, the only possibility is that the diagram writes $X_{1ii'} G_{i'j'} X'_{2jj'} G_{ik}G_{jl}Y_{kk'3}G_{k'l'}Y'_{ll'4}$ which involves three-point functions $X$, $X'$, $Y$ and $Y'$. However, in a quartic theory with vanishing field expectation value, three-point functions vanish and there are no such diagrams.} The decompositions above are therefore non ambiguous. 

Among the diagrams that contribute to ${\cal I}_{12;34}$, which are irreducible in the channel $(12;34)$, there are diagrams that are also irreducible in the other two channels. We call $\bar {\cal I}_{1234}$ the sum of the diagrams that are irreducible in all channels (including the elementary vertex). Now, the diagrams that contribute to ${\cal I}_{12;34}$ are either fully irreducible, or reducible in either the channel $(13;24)$ or $(14;23)$. The latter contributions are respectively  $\Delta\Gamma^{(4)}_{13;24}$ and $\Delta\Gamma^{(4)}_{14;23}$, so that 
\beq
{\cal I}_{12;34}=\bar {\cal I}_{1234}+\Delta\Gamma^{(4)}_{13;24}+\Delta\Gamma^{(4)}_{14;23}.
\eeq
 It follows that
\beq
\Gamma^{(4)}_{1234}&=&{\cal I}_{12;34}+\Delta\Gamma^{(4)}_{12;34}\\
&=& \bar{\cal I}_{1234}+\Delta\Gamma^{(4)}_{13;24}+\Delta\Gamma^{(4)}_{14;23}+\Delta\Gamma^{(4)}_{12;34}.
\eeq
From this relation and (\ref{eq:B1})-(\ref{eq:B3}), a simple calculation yields the following identity
\beq\label{eq:id}
\Gamma^{(4)}_{1234}= \frac{1}{2}\Big[{\cal I}_{12;34}+{\cal I}_{13;24}+{\cal I}_{14;23}-\bar {\cal I}_{1234}\Big].
\eeq
This identity, valid at any loop order, allows one  to construct $\Gamma^{(4)}_L$ from ${\cal I}_L$. It also shows that $\bar {\cal I}_{1234}$ is crossing-symmetric as we anticipated by our choice of notation.

\section{$\partial_\kappa {\cal I}_\kappa$ and power counting}\label{sec:dI}
Let us briefly recall how Weinberg's theorem can be used to determine the large momentum behavior of a given $n$-point function $\Gamma^{(n)}(p_1,\dots,p_n)$  as some (not necessarily all) of its external momenta grow large. This involves the identification  of subgraphs attached to the large external momenta, and such that all internal lines of these subgraphs carry large momenta. Because of this, one can expand the propagators in powers of the small scales present in the subgraph (such as the mass or the regulator), or the (small) external momenta. The leading asymptotic behavior is then determined by power counting applied to the subgraph in question. This leading behavior does not depend on the small scales, which however remain present in the rest of the diagram and may contribute as a multiplicative factor.

More concretely, consider for instance a diagram contributing to $\Gamma^{(4)}_\kappa(p,q)$ and suppose that  $p$ is ``large'', i.e., $p\gg q,\kappa$. It is easily seen that the subgraphs attached to $p$ that have a maximal superficial degree of divergence are four-point subgraphs. These behave logarithmically at large $p$ (by power counting). Similar results hold in the regimes $q\gg p,\kappa$ and $p,q\gg\kappa$.  Returning to the case $p\gg q,\kappa$, we note that a typical diagram contributing to $\Gamma^{(4)}_\kappa(p,q)$ that possesses four-point subgraphs attached to $p$ will yield contributions of the form  $\Gamma^{(4)}_\kappa(p,q)\sim \ln p\times r_\kappa(q)$, as discussed above\footnote{By `$\ln p$' we mean a function that grows logarithmically. This could include powers of logarithms.}. It follows in particular that for $p\gg q,\kappa$,  $\partial_\kappa\Gamma^{(4)}_\kappa(p,q)\sim \ln p\times \partial_\kappa r_\kappa(q) $. In other words, $\partial_\kappa\Gamma^{(4)}_\kappa(p,q)$ also counts as $0$ in the power counting for $p\gg q,\kappa$. This result extends to the regimes $q\gg p,\kappa$ and $p,q\gg\kappa$. 

The situation is different for $\partial_\kappa {\cal I}_\kappa(p,q)$. Indeed, although ${\cal I}_\kappa(p,q)$ may also contain subgraphs contributing to the four-point function, its s-channel two-particle irreducibility implies that the only four-point subgraph that one can attach to the external legs carrying the momentum $p$ is the diagram itself. It follows that here  $r_\kappa(q)=1$, and the logarithmic asymptotic behavior is independent of $\kappa$. As a result,  $\partial_\kappa {\cal I}_\kappa(p,q)$ is suppressed by at least one unit (in fact two) as compared to  $\partial_\kappa\Gamma^{(4)}_\kappa(p,q)$ and finally counts as $-2$ in the power counting. Similar remarks apply to the regimes $q\gg p,\kappa$ and $p,q\gg\kappa$. 

The same result can be obtained directly from the flow equation (\ref{eq:chain_rule}). The two-particle irreducibility of $\delta {\cal I}/\delta G$ means that the leading asymptotic behavior of $\partial_\kappa {\cal I}_\kappa(p,q)$ comes from the regime where all momenta in the integral are large and is thus given by the superficial degree of divergence of that integral. Using that $\partial_\kappa G_\kappa$ counts as $-4$ and $\delta {\cal I}/\delta G$ as $-2$ (since ${\cal I}$ counts as $0$ but $\delta/\delta G$ kills one integral and one propagator), we find the superficial degree of divergence $\delta_{\del{\cal I}}=4-4-2=-2$, in agreement with the result above.

\section{Renormalized loop skeleton expansion}\label{sec:rsexp}
In this section, we show that the renormalized solution $\Gamma^{(4)}_{L,\kp}$ of the flow equations (\ref{eq:Ggreat}) with initial conditions such that the conditions (\ref{eq:cond}) and (\ref{eq:cons}) are fulfilled has the polynomial form (\ref{eq:polynomial}). To this purpose we write $\Gamma^{(4)}_{L,\kp}$ as in Eq.~(\ref{eq:exp_fin}) and show that $\Delta\Gamma^{(4)}_{L,\kappa}$ is proportional to $\lambda^{L+1}$.

 We start by considering $\Delta\Gamma^{(4)}_{L=0,\kappa}(p_i)=\Gamma^{(4)}_{L=0,\kp}(p_i)$, without any prejudice on its original diagrammatic structure. Since its flow vanishes, see Eq.~(\ref{eq:trivial_flow}), the value of $\Gamma^{(4)}_{L=0,\kp}(p_i)$ does not depend on $\kappa$. Furthermore, since $\Gamma^{(4)}_{L=0,\Lambda}(p_i)$ does not depend on $p_i$, $\Gamma^{(4)}_{L=0,\kp}(p_i)$ cannot depend on $p_i$ either. The constant value of  $\Gamma^{(4)}_{L=0,\kappa=0}(p_i=0)$ is fixed  from the conditions (\ref{eq:cond}) and (\ref{eq:cons}):
\beq
\Gamma^{(4)}_{L=0,\kp}(p_i)=\lambda\,.
\eeq
Let us now consider a generic $\Delta\Gamma^{(4)}_{L,\kappa(p_i)}$. We proceed recursively assuming that  $\Delta\Gamma^{(4)}_{L',\kappa}(p_i)$ has been shown to be proportional to $\lambda^{L'+1}$ for $L'<L$, and show that the property extends to $L'=L$. Writing
\beq
\Gamma^{(4)}_{L,\kp}(p_i)=\lambda_{L,\Lambda}+\int_\Lambda^\kappa d\kappa' \partial_{\kappa'}\Gamma^{(4)}_{L,\kp'}(p_i),
\eeq
and imposing the conditions (\ref{eq:cond})-(\ref{eq:cons}), we find 
\beq
\lambda_{L,\Lambda}=\lambda-\int_\Lambda^0 d\kappa' \partial_{\kappa'}\Gamma^{(4)}_{L,\kp'}(p_i=0),
\eeq
so that 
\beq
\Gamma^{(4)}_{L,\kp}(p_i)=\lambda-\int_\Lambda^0 d\kappa' \partial_{\kappa'}\Gamma^{(4)}_{L,\kp'}(p_i=0)+\int_\Lambda^\kappa d\kappa' \partial_{\kappa'}\Gamma^{(4)}_{L,\kp'}(p_i)\,.
\eeq
By subtracting a similar equation with $L$ replaced by $L-1$, we obtain
\beq\label{eq:flow_Delta}
\Delta\Gamma^{(4)}_{L,\kp}(p_i)=-\int_\Lambda^0 d\kappa' \partial_{\kappa'}\Delta\Gamma^{(4)}_{L,\kp'}(p_i=0)+\int_\Lambda^\kappa d\kappa' \partial_{\kappa'}\Delta\Gamma^{(4)}_{L,\kp'}(p_i)\,.
\eeq

We return now to Eq.~(\ref{eq:Ggreat}), and we perform the substitution indicated in  Eq.~(\ref{eq:exp_fin}). We note then  that each term on the right-hand side of Eq.~(\ref{eq:Ggreat}) now involves $\Delta\Gamma^{(4}_{L',\kp}$ with $L'<L$. We call $\bar L$ the number of loops in a given term of Eq.~(\ref{eq:Ggreat}); for instance the first term has $\bar L=1$ loop, the second term $\bar L=2$ loops and so on. From the definition of the operator $[\_]_{\{L\}}$ given after Eq.~(\ref{eq:exp_fin}), we are instructed to keep in the expansion of terms of order $\bar L$, all the terms that are such that $L=\bar L+\sum L'$. Using our recurrence assumption, each term scales with $\lambda$ as $\lambda^{\sum L'+V}$, with $V$ the number of explicit vertices of the considered term. It is easily seen that $V=\bar L+1$, from which it follows finally that $\partial_\kappa\Delta\Gamma^{(4)}_{L,\kp}$ is proportional to $\lambda^{\sum L'+\bar L+1}=\lambda^{L+1}$. In view of  Eq.~(\ref{eq:flow_Delta}), this conclusion extends to $\Delta\Gamma^{(4)}_{L,\kp}$ itself.

\section{Remarks on the 2PI $n$-point functions}\label{App:npoint2PI}
In this Appendix we analyze the flow equations for the 2PI $n$-point functions. The first two equations, that for the self-energy $\Sigma_\kappa(p)$ and that for the irreducible kernel ${\cal I}_\kappa (p,q)$, have already been given in the main text, Eqs.~(\ref{delSigma}) and (\ref{eq:chain_rule}) respectively. It is convenient to rewrite the equation for the self-energy in terms of the two-point function, i.e.,  
\beq
\partial_\kappa \Gamma^{(2)}_\kappa(p)=\frac{1}{2}\int_q \partial_\kappa G_\kappa(q)\,{\cal I}_\kappa(q,p)\,,\label{eq:start}
\eeq
with $G_\kappa^{-1}(q)=\Gamma^{(2)}_{L-1,\kappa}(q)+R_\kappa(q)$ and where ${\cal I}_\kappa$  needs to be seen at this stage as a sum of skeleton diagrams in the bare theory. To remove  this reference to the bare theory, we may obtain  ${\cal I}_\kappa$ from the integration of a flow equation. This is easily obtained by noticing that the $\kappa$ dependence of ${\cal I}_\kappa(q,p)$ originates solely from the propagator $G_\kappa$. We get (see Eq.~(\ref{eq:chain_rule}))
\beq
\partial_\kappa {\cal I}_\kappa(q,p)=\int_{r_1} \partial_\kappa G_\kappa(r_1)\,\left.\frac{\delta {\cal I}(q,p)}{\delta G(r_1)}\right|_{G=G_\kappa}\,.\label{eq:first}
\eeq
Again, since  $\delta {\cal I}/\delta G$ is to be seen as a sum of skeleton diagrams in the bare theory, we repeat the previous step and obtain $\delta {\cal I}/\delta G$ from the integration of a flow equation 
\beq
\partial_\kappa \left.\frac{\delta {\cal I}(q,p)}{\delta G(r_1)}\right|_{G=G_\kappa}=\int_{r_2} \partial_\kappa G_\kappa(r_2)\,\left.\frac{\delta^2 {\cal I}(q,p)}{\delta G(r_2)\delta G(r_1)}\right|_{G=G_\kappa}\,.
\eeq
We can continue this procedure until we reach $\delta^{m+1} {\cal I}(q,p)/\delta G(r_{m+1})\cdots\delta G(r_1)=0$ with $m$ equal to the maximal number of propagators in the diagrams of ${\cal I}$, in the considered $\Phi$-derivable approximation. In this case, the flow of $\delta^{m} {\cal I}(q,p)/\delta G(r_{m})\cdots\delta G(r_1)$ vanishes and the tower of flow equations terminates\footnote{For instance, in the $L$-loop approximation, the diagrams that contribute to $\Phi$ contain up to $2L-2$ propagators and the sequence of 2PI $n$-point functions that can be constructed terminates with 2PI $(4L-4)$-point functions.} One then arrives at a reformulation of $\Phi$-derivable approximations as a system of flow equations for the quantities $\Gamma^{(2)}_\kappa(p)$, ${\cal I}_\kappa(q,p)$ and $\delta^{k} {\cal I}(q,p)/\delta G(r_{k})\cdots\delta G(r_1)|_{G=G_\kappa}$, with $1\leq k\leq m$. These equations are (\ref{eq:start}), (\ref{eq:first}) and
\beq
\partial_\kappa \left.\frac{\delta^k {\cal I}(q,p)}{\delta G(r_k)\cdots\delta G(r_1)}\right|_{G=G_\kappa}=\int_{r_{k+1}} \partial_\kappa G_\kappa(r_{k+1})\,\left.\frac{\delta^{k+1} {\cal I}(q,p)}{\delta G(r_{k+1})\cdots\delta G(r_1)}\right|_{G=G_\kappa}\,,\quad 1\leq k \leq m\,.\label{eq:tower}
\eeq
As we have seen, these equations rely on the fact that the Luttinger-Ward functional $\Phi$ has no explicit dependence on $\kappa$, all the dependence on $\kappa$ being carried by the propagator.

\begin{figure}[h]
\begin{center}
\includegraphics[angle=0,width=15cm]{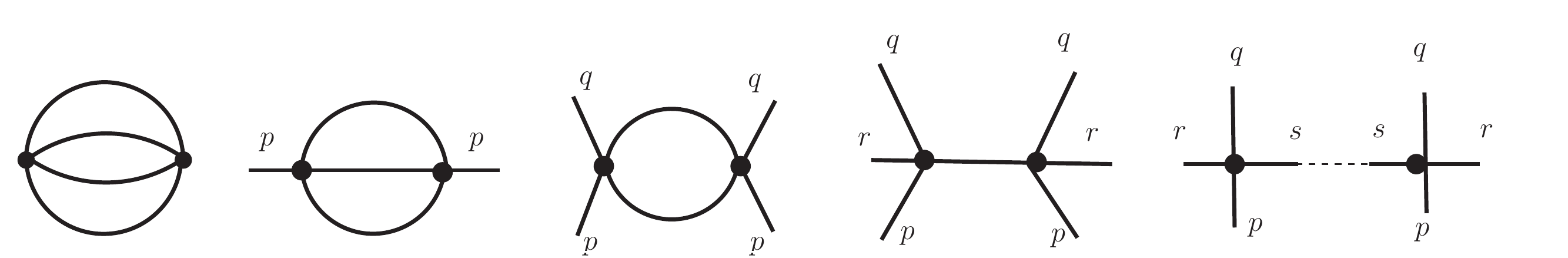}
\caption{The various 2PI $n$-point functions that appear in the three-loop approximation to $\Phi$. From left to right: $\Phi$, $\Sigma$, ${\cal I}$, ${\cal J}$and ${\cal K}$.  }\label{fig:AppB}
\end{center}
\end{figure}

As an illustration, consider  the three-loop approximation for $\Phi$ (see Fig.~\ref{fig:AppB}). The irreducible kernel ${\cal I}_\kappa$ obeys
\beq\label{eq:chain_rule2}
\partial_\kappa {\cal I}_\kappa(q,p)=\int_r \partial_\kappa G_\kappa(r){\cal J}_\kappa(p,q,r),\qquad {\cal J}_\kappa(p,q,r)\equiv \frac{\delta{\cal I}_\kappa(q,p)}{\delta G_\kappa(r)}.
\eeq
where ${\cal J}_\kappa(p,q,r)$ is a six-point function with a tree structure: a propagator connecting two vertices. A further functional derivative yields an eight-point function ${\cal K}_\kappa(q,p,r,s)$
\beq
 \del_\kappa{\cal J}_\kappa(q,p,r)=\int_s\del_\kappa G_\kappa(s) {\cal K}_\kappa(q,p,r,s),\qquad {\cal K}_\kappa(q,p,r,s)\equiv \frac{\delta{\cal J}_\kappa(q,p,r)}{\delta G_\kappa(s)},
\eeq
composed of two disconnected vertices.
Clearly, in the three-loop $\Phi$-derivable approximation, ${\cal K}_\kappa(q,p,r,s)$ is the last 2PI $n$-point function that can be constructed and we have $\partial_\kappa {\cal K}_\kappa=0$.

While at a given loop order, the sequence of 2PI $n$-point functions is finite and provides therefore a possible practical scheme to determine $\del_\kappa {\cal I}_\kappa$, there are two features that make this procedure somewhat unsatisfactory, as we now explain. The first one is easily dealt with, while the second one points to a conceptual issue that motivated the strategy adopted  in Sect.~\ref{sec:flowI}.

 Let us consider the first issue. From the 2PI nature of the derivatives $\delta^k {\cal I}(q,p)/\delta G(r_k)\cdots\delta G(r_1)$, it follows that there are no four-point functions attached directly to the two legs associated to a given derivative $\delta/\delta G(r_i)$. From Weinberg theorem, it follows that $\delta^k {\cal I}(q,p)/\delta G(r_k)\cdots\delta G(r_1)$ counts as a strictly negative contribution in the power counting of the integral in Eq.~(\ref{eq:tower}). Since $\partial_\kappa G_\kappa(r_{k+1})$ counts as $-4$, the integral is finite. In contrast, Eq.~(\ref{eq:first}) is not finite by power couting since the superficial degree of divergence is $\delta=4-4+0=0$. There is however a simple solution to this problem. All  one needs to do is  replace Eqs.~(\ref{eq:start}) and (\ref{eq:first}) by
\beq
\partial_\kappa \Gamma^{(2)}_\kappa(p)=\frac{1}{2}\int_q \partial_\kappa R_\kappa(q)\,\Gamma^{(4)}_\kappa(q,p)\,,\label{eq:new_start}
\eeq
and 
\begin{eqnarray}\label{eq:new_first}
\partial_\kappa\Gamma^{(4)}_\kappa(p,q)=\partial_\kappa{\cal I}_\kappa(p,q) & - & \frac{1}{2}\int_r\Gamma^{(4)}_\kappa(p,r)\,\partial_\kappa G_\kappa^2(r)\,\Gamma^{(4)}_\kappa(r,q)\nonumber\\
& - & \frac{1}{2}\int_r\partial_\kappa{\cal I}_\kappa(p,r)\,G_\kappa^2(r)\,\Gamma^{(4)}_\kappa(r,q)\nonumber\\
& - & \frac{1}{2}\int_r\Gamma^{(4)}_\kappa(p,r)\,G_\kappa^2(r)\,\partial_\kappa {\cal I}_\kappa(r,q)\nonumber\\
& + & \frac{1}{4}\int_r\int_s \Gamma^{(4)}_\kappa(p,r)\,G_\kappa^2(r)\,\partial_\kappa {\cal I}_\kappa(r,s)\,G_\kappa^2(s)\,\Gamma^{(4)}_\kp(s,q). 
\end{eqnarray} 
These equations have been shown to be finite by power counting in the main text. We then arrive at a system of finite flow equations, Eqs.~(\ref{eq:new_start}), (\ref{eq:new_first}) and (\ref{eq:tower}) for the functions $\Gamma^{(2)}_\kappa(p)$, $\Gamma^{(4)}_\kappa(q,p)$ and $\delta^{k} {\cal I}(q,p)/\delta G(r_{k})\cdots\delta G(r_1)$, with $1\leq k\leq m$. We note that ${\cal I}_\kappa$ does not appear in this set since it enters the equations only through $\partial_\kappa {\cal I}_\kappa$, and its flow equation never needs to be integrated.

The second unsatisfactory feature concerns the initialization of the system of equations. Because $\Gamma^{(2)}_\kappa$ and $\Gamma^{(4)}_\kappa$ are 1PI functions, their initial conditions are simple and given in Eq.~(\ref{eq:init3}). 
On the other hand, the functions $\delta^{k} {\cal I}(q,p)/\delta G(r_{k})\cdots\delta G(r_1)$'s are not 1PI functions. They can contain in particular disconnected pieces involving delta functions in momentum space. The developments in \ref{app:great} allow us to clarify the structure of the initial conditions to a large extent. In fact, the functions $\delta^{k} {\cal I}(q,p)/\delta G(r_{k})\cdots\delta G(r_1)$ enjoy the same properties as $\delta {\cal I}/\delta G(r)$: they are expressible as skeleton diagrams in which tree-level vertices are replaced by the exact $\Gamma^{(4)}$. In a given $\Phi$-derivable approximation, one truncates at a given loop order, in which case the exact $\Gamma^{(4)}$ needs to be replaced by $\Gamma^{(4)}_L$, as appropriate.\footnote{If ${\cal I}$ is made of diagrams up to $L-2$ loops, $\delta^{k} {\cal I}(q,p)/\delta G(r_{k})\cdots\delta G(r_1)$ should be seen as made of diagrams up to $L-2-k$ loops, provided one counts the possible delta functions that may appear, as negative loops.} Among the  skeleton diagrams, there are diagrams where all propagators have been cut and there remain only the tree-level vertices multiplied by appropriate delta functions. After replacing the tree-level vertices by four-point functions, one obtains products of $\Gamma^{(4)}_L$'s (expanded to the relevant loop order) multiplied by delta functions, which survive when $\kappa$ is taken large an lead to products of the initial conditions $\lambda_{L,\Lambda}$ (again, expanded to the relevant loop order). The other diagrams, involve, in addition, some $\Gamma^{(4)}_L$'s connected by propagators. These diagrams are suppressed at large $\kappa$ because, once expressed in terms of the $\Gamma^{(4)}_L$'s, all their possible loops have a negative superficial degree of divergence.

To make things more concrete, let us take a few examples. At three-loop order, the tower of flow equations involves $\delta {\cal I}/\delta G$ and $\delta^2 {\cal I}/\delta G^2$ whose diagrammatic contributions are shown as the last two diagrams of Fig.~16 and are obviously expressed in terms of $\Gamma^{(4)}_{L=0}$.  In $\delta {\cal I}/\delta G$, the two $\Gamma^{(4)}_{L=0}$ are connected by a propagator which suppresses the contribution at large $\kappa$ . One may then initialize $\delta {\cal I}/\delta G$ to $0$. On the other hand, no propagator appears in $\delta^2 {\cal I}/\delta G^2$ together with the two vertices, so this quantity has a non-trivial initialization: 
\beq
\left.\frac{\delta^2 {\cal I}(q,p)}{\delta G(r_2)\delta G(r_1)}\right|_{G=G_\Lambda}=-\lambda_{0,\Lambda}^2\delta(p+q+r_1+r_2)
\eeq
(recall that $\lambda_{0,\Lambda}$ is simply $\lambda$ with our choice of renormalization conditions). At four-loop order, the tower of flow equations involves $\delta {\cal I}/\delta G$, $\delta^2 {\cal I}/\delta G^2$, $\delta^3 {\cal I}/\delta G^3$, $\delta^4 {\cal I}/\delta G^4$. A simple analysis reveals that $\delta {\cal I}/\delta G$ and $\delta^3 {\cal I}/\delta G^3$ do not contain contributions involving only disconnected four-point functions. These objects are therefore suppressed at large $\kappa$ and need to be initialized to $0$. On the other hand $\delta^2 {\cal I}/\delta G^2$ and $\delta^4{\cal I}/\delta G^4$ contain such contributions and thus require a non-trivial initialization. We have for instance
\beq
\frac{\delta^2 {\cal I}(q,p)}{\delta G(r_2)\delta G(r_1)}=-\Big[\big[\Gamma^{(4)}_{L=1}(p,q,r_1)\big]^2\Big]_{1}\delta(p+q+r_1+r_2)
\eeq
so the initial condition is
\beq
\left.\frac{\delta^2 {\cal I}(q,p)}{\delta G(r_2)\delta G(r_1)}\right|_{G=G_\Lambda}=-\Big[\lambda_{1,\Lambda}^2\Big]_{1}\delta(p+q+r_1+r_2)\,.
\eeq
One can similarly deduce the initial condition for $\delta^4{\cal I}(q,p)/\delta G(r_2)\delta G(r_2)\delta G(q)\delta G(p)|_{G=G_\kappa}$. We note that the four-point functions $\Gamma^{(4)}_{L}$ that were found to play a major role in our main discussion, also appear here in the determination of the initial conditions for some of the flow equations of the 2PI $n$-point functions. Therefore, these  functions cannot be ignored, and need to be properly treated.

Let us finally mention that while the hierarchy of equations (\ref{eq:start}), (\ref{eq:first}) and (\ref{eq:tower}) is identical to that discussed in Ref.~\cite{Carrington:2017lry}, our analysis deviates from that presented in that reference. According to Ref.~\cite{Carrington:2017lry}, only a subset of the flow equations needs to be used. More precisely, one must go up in the hierarchy until one finds the first (so-called terminal) 2PI $n$-point function that fulfills certain ``consistency conditions'', as defined in \cite{Carrington:2017lry}, that ensure that the corresponding flow equation can be integrated exactly in terms of diagrams. These consistency conditions, which are not the same as those discussed in the main text, see Eq.~(\ref{eq:cons}), express the regularity of the 2PI $n$-point functions $X^{(n)}(p_i)$ in the zero-momentum limit, that is $\Delta X^{(n)}(p_i)\equiv X^{(n)}(p_i)-X^{(n)}(p_i=0)$ should approach $0$ as $p_i\to 0$.  According to the authors of \cite{Carrington:2017lry} such limits may lead to a $0\times\infty$ indetermination due to the possible presence of subdivergences. We find no trace of such problematic limits. For one thing, the present theory has a Landau pole which requires the presence of an explicit UV cut-off. This ensures that $\Delta X^{(n)}(p_i)$ approaches $0$ as $p_i\to 0$. It could happen that this limit is approached only for extremely tiny values of $p_i$ if subintegrals remain strongly sensitive to $\Lambda_{\rm uv}$. However, our finite flow equations all generate the appropriate subtractions at the scale $\Lambda$ that ensure that no strong dependence on $\Lambda_{\rm uv}$ is present and $\Delta X^{(n)}(p_i)$ approaches $0$ smoothly in a reasonable range of $p_i$ near 0. 

In our approach, we have introduced a hierarchy of flow equations beyond the equation for $\partial_\kappa {\cal I}_\kappa$, but these are written in terms of the $\Gamma^{(4)}_{L,\kp}$'s rather than the 2PI $n$-point functions. As explained in the main text, the main purpose of this hierarchy is to clarify the renormalization, see Sects.~\ref{sec:42} and \ref{sec:2PIren}. As discussed in Sect.~\ref{sec:practical}, for practical applications, we can directly evaluate $\partial_\kappa {\cal I}_\kappa$ in terms of the diagrammatically renormalized $\Gamma^{(4)}_{L,\kp}$'s.\\

\bibliographystyle{unsrt}
\bibliography{2PIbib}

\end{document}